\documentclass[reprint,superscriptaddress,aps,prl,]{revtex4-2}

\usepackage{graphicx}
\usepackage{float}
\usepackage{dcolumn}
\usepackage{bm}
\usepackage{bbm}
\usepackage{braket}
\usepackage{xcolor}
\usepackage{physics}
\usepackage{hyperref}
\usepackage{leftindex}

\renewcommand\({\begin{equation}}	
\renewcommand\){\end{equation}}
\newcommand{\al}[1]{\begin{aligned}#1\end{aligned}}

\makeatletter
\DeclareFontFamily{OMX}{MnSymbolE}{}
\DeclareSymbolFont{MnLargeSymbols}{OMX}{MnSymbolE}{m}{n}
\SetSymbolFont{MnLargeSymbols}{bold}{OMX}{MnSymbolE}{b}{n}
\DeclareFontShape{OMX}{MnSymbolE}{m}{n}{
    <-6>  MnSymbolE5
   <6-7>  MnSymbolE6
   <7-8>  MnSymbolE7
   <8-9>  MnSymbolE8
   <9-10> MnSymbolE9
  <10-12> MnSymbolE10
  <12->   MnSymbolE12
}{}
\DeclareFontShape{OMX}{MnSymbolE}{b}{n}{
    <-6>  MnSymbolE-Bold5
   <6-7>  MnSymbolE-Bold6
   <7-8>  MnSymbolE-Bold7
   <8-9>  MnSymbolE-Bold8
   <9-10> MnSymbolE-Bold9
  <10-12> MnSymbolE-Bold10
  <12->   MnSymbolE-Bold12
}{}

\let\llangle\@undefined
\let\rrangle\@undefined
\DeclareMathDelimiter{\llangle}{\mathopen}%
                     {MnLargeSymbols}{'164}{MnLargeSymbols}{'164}
\DeclareMathDelimiter{\rrangle}{\mathclose}%
                     {MnLargeSymbols}{'171}{MnLargeSymbols}{'171}
\makeatother

\begin{document}

\title{\texorpdfstring{Resilience of Entangling Gates to Spatiotemporally Correlated Noise}{Resilience of Entangling Gates to Spatiotemporally Correlated Noise}}

\author{Xinru Tang}
\affiliation{Department of Physics, University of Basel, Klingelbergstrasse 82, 4056 Basel, Switzerland}
\author{Sander Driessen}
\affiliation{Department of Physics, University of Basel, Klingelbergstrasse 82, 4056 Basel, Switzerland}
\affiliation{Department of Applied Physics, Eindhoven University of Technology, Eindhoven 5612 AZ, The Netherlands}
\author{Ji Zou}
\email{ji.zou@kfupm.edu.sa}
\affiliation{Physics Department, King Fahd University of Petroleum and Minerals, 31261, Dhahran, Saudi Arabia}
\affiliation{Quantum Center, KFUPM, Dhahran, Saudi Arabia}
\author{Stefano Bosco}
\affiliation{QuTech, Delft University of Technology, Lorentzweg 1, 2628 CJ Delft, The Netherlands}
\author{Jelena Klinovaja} 
\affiliation{Department of Physics, University of Basel, Klingelbergstrasse 82, 4056 Basel, Switzerland}
\author{Daniel Loss} 
\affiliation{Department of Physics, University of Basel, Klingelbergstrasse 82, 4056 Basel, Switzerland}
\affiliation{Physics Department, King Fahd University of Petroleum and Minerals, 31261, Dhahran, Saudi Arabia}
\affiliation{Quantum Center, KFUPM, Dhahran, Saudi Arabia}
\affiliation{RDIA Chair in Quantum Computing}

\date{\today}

\begin{abstract}

As quantum processors scale, spatially and temporally correlated noise is emerging as a limiting factor in the fidelity of multiqubit gates. Here we show that this makes the choice of entangling operation a first-order design decision: for two-qubit gates in semiconductor spin qubits, the CZ and $\sqrt{\mathrm{iSWAP}}$ gates exhibit a pronounced resilience to correlated noise relative to the SWAP gate, even though correlated fluctuations of the entangling interaction can degrade gate performance to a level comparable to that of local noise. We trace this hierarchy to the redistribution of noise-sensitive channels away from zero frequency and show it to hold for noise spectra dominated by low-frequency weight, including \mbox{$1/f$} and Lorentzian. These results follow from a general expression, derived here, for the average gate infidelity under arbitrary spatiotemporal noise correlations, valid for generic multiqubit systems and system-environment coupling structures. The infidelity takes the form of a spectral overlap between the noise correlation matrix and a gate-dependent filter function, in which the coherent dynamics of the gate itself shapes the spectral window through which correlated fluctuations enter. The framework converts measured noise correlations into quantitative error budgets and provides a concrete design principle for selecting entangling operations in scalable quantum architectures.
\end{abstract}

\maketitle

\textit{Introduction}|Noise is the fundamental obstacle to scalable and fault-tolerant quantum computation \cite{Preskill2018quantumcomputingin,palma1996quantum, SCHLOSSHAUER20191}. Over the past two decades, substantial theoretical and experimental effort has been devoted to characterizing and suppressing decoherence in qubits, largely under the assumption of spatially local and approximately Markovian error channels.  Various mitigation strategies have been developed within this framework, including operation at decoherence-free sweet spots \cite{SweetSpots4ChargeNoise,FullyTunableHyperfineInteractions,HoleSpinQubitsinThinCurvedQuantumWells,ReducedSensitivitytoChargeNoise,Asingleholespinwithenhancedcoherence,LOCCNet,PhysRevLett.116.116801}, quantum error correction codes \cite{PerfectQECC,TheoryofQEC,FTQCvsGaussiannoise,Preskill2012SufficientCO}, dynamical decoupling protocols \cite{DDofOPS,FTQDD,DD4multiqubit,PhysRevB.77.174509,Nguyen2025ColorIC,1d4l-73x6,PhysRevLett.98.100504,Vezvaee2026}, and optimal control techniques \cite{QCbyanOptimalControlAlgorithm,OptimalcontrolgeometryandQC,HybridApproachtoQuantumOptimalControl,hansen2022implementation,PhysRevA.104.062415,Bosco2023PhaseDrivingHS,Rimbach-Russ_2023}. These approaches have significantly improved qubit coherence and gate fidelities across multiple platforms. 

However, as quantum processors scale  in size, density, and connectivity, spatially correlated and non-Markovian error mechanisms become unavoidable \cite{DemonstrationofnonMarkovianprocesscharacterisation,Mechanicallyinducedcorrelatederrorsonsuperconducting,k2lx-vfqp,PhysRevB.107.174511}, directly challenging the independent error paradigm that underlies most existing mitigation strategies. Such correlations arise generically from  common electromagnetic or phononic environments \cite{PhysRevB.77.155420,PhysRevA.65.032326}, residual cross talk \cite{corsstalk,PhysRevB.105.085414,PhysRevB.104.045420}, collective calibration drifts \cite{Rao2025TowardsAT}, spin qubit shuttling \cite{mokeev2024modelingdecoherencefidelityenhancement,78sg-ttth,vanRiggelen-Doelman2024,krzywda2026coherence,PhysRevB.88.075301}, and fluctuations in shared classical control hardware \cite{Mechanicallyinducedcorrelatederrorsonsuperconducting}. Recent experiments in spin qubit systems and other quantum computing platforms have directly measured correlated error channels whose magnitude is even comparable to that of local noise \cite{Yoneda2022,AriasPRA2023,RojasArias2025InferringCN,RojasArias2024TheOO,Donnelly2025,superconductingqubits}. Such findings demonstrate that correlated noise is not just a higher-order correction within otherwise independent error models but rather a key feature of realistic error processes that can fundamentally alter error accumulation and impact the scalability and reliability of fault-tolerant architectures \cite{QECGoogle,QECcorr2004,topoQECcorr2018,Nickerson2019analysingcorrelated,QECinspatcorr2005,Breakdownofsurfacecode,GoogleAI,Chen2021,Spatiallycorrelated}.

\begin{figure}[t!]
    \centering
\includegraphics[width=.84\linewidth]{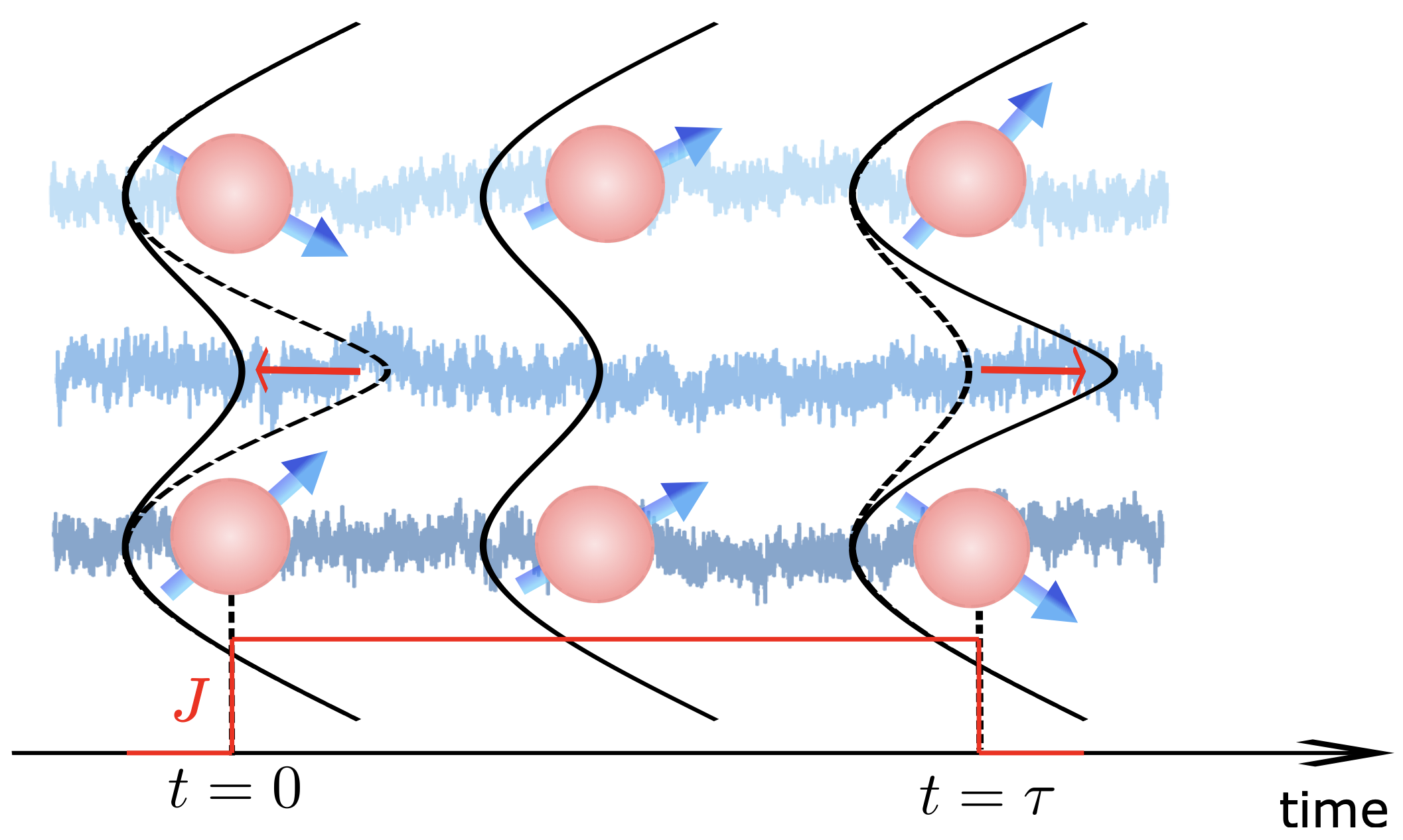}
    \caption{\textbf{Schematic of the system-environment coupling model.} At $t=0$, the qubit–qubit interaction $H_{\text{int}}$ with coupling strength $J$ is switched on and remains active until $t=\tau$. Each qubit is coupled to a common bath or noise source exhibiting spatial and temporal correlations, leading to fluctuations in the qubit energies and interaction strength (colored traces).}
    \label{fig:1}
\end{figure}

In this Letter, we show that spatiotemporally correlated noise renders multiqubit gate fidelity strongly gate-dependent, turning the choice of entangling operation into a design tool. To this end, we establish a framework to quantitatively characterize multiqubit gate operations in the presence of both spatially and temporally correlated noise. We derive an expression for the average gate fidelity with an effective filter function~\cite{PhysRevLett.87.270405,PhysRevB.77.174509,PhysRevLett.109.020501,PhysRevLett.113.250501,Green_2013,PhysRevA.95.022121,PhysRevLett.127.170403,PhysRevResearch.3.043047} that incorporates arbitrary spatial correlations and temporal spectra to leading order in weak stationary noise, going beyond standard independent error models. Our analysis  quantifies the impact of different correlation structures and identifies the dominant contributions governing multiqubit-gate error accumulation. In particular, we find that certain classes of correlated noise can degrade gate fidelity to a level comparable to, or even exceeding, that induced by local noise, while other noise correlations produce only a negligible effect.

Applying this framework to two-qubit gates in spin qubit systems, we explicitly show that correlated noise couples nontrivially to the gate-generating interaction \cite{QuantumcomputationwithQD,Coupledquantumdotsasquantumgates,QCwithColdTrappedIons} and therefore produces strongly gate-dependent error behavior. Interestingly, we find that, for low-frequency noise, including \mbox{$1/f$} and Lorentzian spectra, controlled-phase (CZ) gate (equivalent to controlled-NOT (CNOT) up to single-qubit operations) and the $\sqrt{\text{iSWAP}}$ gate exhibit enhanced robustness relative to the SWAP gate.  Furthermore, we identify the underlying mechanism and establish its generality across low-frequency noise spectra. In general, by tuning the  interaction to realize distinct operations, one effectively reshapes how correlated fluctuations are filtered by the coherent dynamics.

\textit{Model}|We consider an $N$-qubit system subjected to a generic noisy environment, governed by $H=H_{S}+H_{E}+H_{\text{SE}},$
where $H_{S}$, $H_{E}$, and $H_{\text{SE}}$ denote the Hamiltonians of the system, environment, and their coupling, respectively. We take the Hamiltonian of the interacting $N$-qubit system to be $H_{S}=\sum_{i}\Delta_{i}\sigma_{i}^{z}/2+H_{\text{int}}$. We set $\hbar=1$, such that $\Delta_i$ are the qubit frequencies and $H_{\text{int}}$ describes a generic inter-qubit coupling, which enables multiqubit gates. 
 As illustrated in Fig.~\ref{fig:1}, we focus on the scenario where the qubit-qubit interaction $H_{\text{int}}$ with coupling strength $J$ is switched on  at $t=0$, remains active for the duration $\tau$ to implement the gate, and is then switched off at $t=\tau$; we assume instantaneous (rectangular) switching of $H_{\text{int}}$ throughout. We note that, with large spin-orbit interaction, this approach enables the realization of a broad class of two-qubit gates in a single step \cite{SOenabledrealizationofarbitrary2qubitgates}.  We quantify noise-induced infidelity relative to the exact noiseless propagator generated by $H_S$.
 
In any realistic platform, however, the qubits inevitably couple to their environment  via $H_{\text{SE}}=\sum_{\alpha}\hat{A
}_{\alpha}\otimes\hat{B}_{\alpha}$, where $\alpha$ labels distinct  channels. The operators $\hat{A}_{\alpha}$ are Hermitian and act on the qubit Hilbert space. They  represent either single-qubit operators, $\hat{A} \propto \sigma_{i}^{a}$, or multiqubit operators, $\hat{A} \propto \sigma_{i}^{a}\sigma_{j}^{b}\sigma_{k}^{c}\cdots$, where $i,j,k$ label the qubits and $a,b,c \in \{x, y, z\}$ denote spin components. On the other hand, $\hat{B}_{\alpha}$ are also Hermitian but act on the environment. Physically, they  describe environmental degrees of freedom such as fluctuating gate electric fields, local charge or spin defects, nuclear spins, and phonon modes. Coupling to such a complex environment leads to imperfect gate operations, producing not only independent qubit errors but also errors that are correlated across different qubits and in time. The goal of this work is to understand how different types of correlated noise affect the performance of multiqubit gates, in regimes where other control errors are negligible due to standard mitigation techniques such as pulse shaping and dynamical decoupling.

\textit{General formula for quantum gate fidelity}|To this end, we evaluate the average gate fidelity $\overline{\mathcal{F}}$~\cite{AGF,Abad} of a generic multiqubit operation generated by $H_{\text{int}}$ in the presence of correlated noise:
\begin{equation}
 \overline{\mathcal{F}} \equiv \int d\psi \bra{\psi}\mathcal{U}^\dag_g \mathcal{E}_{\text{lab}}(\ket{\psi}\bra{\psi})\mathcal{U}_g\ket{\psi},
\end{equation}
where the integral is over all pure initial states $\ket{\psi}$ and $\mathcal{U}_g=e^{-iH_S \tau}$ denotes the ideal (target) quantum gate, while $\mathcal{E}_{lab}$ is the quantum channel describing the noisy implementation in the laboratory frame. It is related to the interaction-picture error channel by $\mathcal{E}_{\text{lab}}(\rho)=\mathcal{U}_g\mathcal{E}_{I}(\rho)\mathcal{U}_g^{\dagger}$.

The starting point is the von Neumann equation that describes the density matrix of the combined system in the interaction picture, $\dot{\rho}_{\text{tot}}(t)=\mathcal{L}(t)\rho_{\text{tot}}(t)$, with the Liouville superoperator $\mathcal{L}(t)=-i[H_{\text{SE}}(t),\star]$. Assuming that at $t=0$ the $N$-qubit system is in the state $\ket{\psi}$ and the environment is in a thermal state $\rho_E$, the noisy quantum channel in the interaction picture takes the form:
\(  \mathcal{E}_I(\ket{\psi}\bra{\psi}) = \Tr_E [\Phi_\tau (\ket{\psi}\bra{\psi}\otimes \rho_E)],  \)
with the quantum map $\Phi_\tau=\mathcal{T}\exp{\int_0^\tau ds\,\mathcal{L}(s) }$ and time-ordering operator $\mathcal{T}$. Here, $\Tr_E$ stands for the trace over the environment Hilbert space. We work to second order in the system-environment coupling. A sufficient condition
for a bounded coupling Hamiltonian is
 $||H_{\text{SE}}||_{\infty}\tau \ll 1$, where $||A||_{\infty}\equiv \sup_{\|\ket{\psi}\|=1} \|A\ket{\psi}\|$ denotes the operator norm~\cite{SM,PhysRevA.78.012355}. 
After algebraic manipulations detailed in the SM \cite{SM}, we obtain the average gate infidelity $\overline{\mathcal{I}}=1-\overline{\mathcal{F}}$:
\(\al{
\overline{\mathcal{I}}= \sum_{\alpha,\beta}\int^\tau_0&\!\!\!dt_1\!\int^{t_1}_0\!\!\!dt_2\, S_{\alpha\beta}(t_1-t_2) \Bigg\{ \frac{\Tr_S[\hat A_\alpha(t_1)\hat A_\beta(t_2)] }{d+1} \\
&-\frac{\Tr_S[\hat A_\alpha(t_1)]\Tr_S[\hat A_\beta(t_2)]}{d(d+1)}\Bigg\} +\text{H.c.},
} \label{Eq:timedomain} \)
 where $\Tr_S$ denotes the trace over the $N$-qubit Hilbert space of dimension $d=2^N$. Here, we introduced the noise correlation function $S_{\alpha\beta}(t_1-t_2)=\tr_{E}[\rho_{E}\hat{B}_{\alpha}(t_1)\hat{B}_{\beta}(t_2)]\equiv \langle \hat{B}_{\alpha}(t_1)\hat{B}_{\beta}(t_2)\rangle$, assuming stationary noise statistics, whose nontrivial temporal structure encodes the non-Markovian effects. The Hilbert–Schmidt inner product $\Tr_S[\hat A_\alpha(t_1)\hat A_\beta(t_2)]$ measures the overlap between system operators at different times in operator space. 

Since the system operators $\hat{A}_{\alpha}$ are composed of Pauli matrices, we have $\Tr_S\hat{A}_{\alpha}=0$ and then the average gate infidelity can  be recast in the frequency-domain
filter-function form~\cite{PhysRevLett.87.270405,PhysRevB.77.174509,PhysRevLett.109.020501,PhysRevLett.113.250501,Green_2013,PhysRevA.95.022121,PhysRevLett.127.170403,PhysRevResearch.3.043047}:
\begin{align}
    \overline{\mathcal{I}}\!=\!\frac{1}{d+1}\sum_{\alpha,\beta}\int_{-\infty}^{\infty}\frac{d\omega}{2\pi}S^C_{\alpha\beta}(\omega)F_{\alpha\beta}(\omega, \tau),\label{mainres}
\end{align}
 where $S^C_{\alpha\beta}(\omega)$ represents the classical component of the noise. The effective filter function is given by the convolution:
\begin{align}
F_{\alpha\beta}(\omega,\tau)
  = \int\!\frac{d\tilde{\omega}}{2\pi} K_{\alpha\beta}(\tilde{\omega})
     F_{\mathrm{FID}}(\omega+\tilde{\omega},\tau).
\label{eq:filter_function}
\end{align}
Here, $K_{\alpha\beta}({\omega})
  = \int dt\, e^{i{\omega}t}\,\mathrm{Tr}_S[\hat{A}_\alpha(t)\hat{A}_\beta]$ describes the  spectral response of the qubit system and 
$F_{\mathrm{FID}}(\omega,\tau)
=4\sin^2(\omega\tau/2)/\omega^2$ is the free-induction-decay (FID) kernel.  We note that the gate-generating interaction shapes the spectral response
$K_{\alpha\beta}$ and therefore determines how the effective
filter samples the noise spectrum.
  To make this gate dependence explicit, we use the spectral
decomposition $H_S=\sum_n E_n\ket{n}\bra{n}$ and obtain
\( F_{\alpha\beta}(\omega,\tau)=\sum_{nm}\hat{A}_{\alpha}^{nm}\hat{A}_{\beta}^{mn} F_{\text{FID}}(\omega-\omega_{nm},\tau).\label{filterfunction}\)
Each shifted kernel
$F_{\mathrm{FID}}(\omega-\omega_{nm},\tau)$
is peaked at $\omega=\omega_{nm}=E_n-E_m$,
with peak value $\tau^2$.
The corresponding channel therefore samples noise most strongly
near its transition frequency.
The contribution of each channel is weighted by
$\hat{A}_{\alpha}^{nm}\hat{A}_{\beta}^{mn}$, where
$\hat{A}_{\alpha}^{nm}
\equiv\bra{n}\hat{A}_{\alpha}\ket{m}$.
The individual matrix elements set the selection rules and
coupling strengths for noise-induced transitions, in analogy
with dipole matrix elements in light--matter interaction.
The overall gate infidelity is thus governed by the spectral
overlap between these weighted, shifted filters and the
environmental noise spectrum.

We point out that, while the noise correlation $S_{\alpha\beta}(t_1-t_2)$ generally contains both quantum and classical components of the environmental fluctuations, only the classical part, $S^C_{\alpha\beta}(t)= \langle \{\hat{B}_\alpha(t),\hat{B}_\beta\}\rangle/2$,
contributes to the average gate infidelity, as seen in Eq.~\eqref{mainres}. Here, $S^C(\omega)= \int dt\, e^{i\omega t} S^C(t)$ defines the classical noise spectrum that is directly measurable in experiments.

We also stress that Eq.~\eqref{mainres} applies to general multiqubit gates implemented through direct interaction switching under generic noise. As an example, consider pairwise interactions described by
$H_S=\sum_i \Delta_i \sigma_i^z/2+\sum_{i>j}\sum_{a,b}J^{ab}_{ij}\sigma_i^a\sigma_j^b/4$. The qubit system couples to the environment through operators $\hat{A}_{\alpha}=\sigma_i^z$ and $\sigma_i^a\sigma_j^b$, leading to fluctuations in the qubit frequencies $\delta\Delta_i$ and in the
inter-qubit couplings $\delta J^{ab}_{ij}$, represented by environment operators $\hat{B}_{\alpha}=\delta\Delta_{i}/2$ and $\delta J^{ab}_{ij}/4$, respectively. Three classes of noise spectra are particularly relevant in experiments: (i) \textit{frequency noise} $S_{\Delta_i\Delta_j}(\tau)=\langle \hat{B}_{\Delta_i}(t+\tau)\hat{B}_{\Delta_j}(t)\rangle$,
producing local errors for $i=j$ and spatially correlated errors
for $i\neq j$; (ii) \textit{interaction noise}
$S_{JJ}(\tau)
 =\langle \hat{B}_{ J^{ab}_{ij}}(t+\tau)\hat{B}_{ J^{cd}_{kl}}(t)\rangle$,
describing the noise in the entangling interaction which fluctuates in time and across different qubit pairs; and
(iii) \textit{cross correlations}
$S_{\Delta J}(\tau)
 =\langle \hat{B}_{\Delta_i}(t+\tau)\hat{B}_{J^{ab}_{ij}}(t)\rangle$,
capturing correlated fluctuations between the inter-qubit coupling and the qubit frequency. We note that the contribution of spatially correlated frequency noise to the infidelity can be negative; however, it is bounded by that of local noise, ensuring that the total infidelity remains positive \cite{SM}. Recent experiments reveal that these different noise correlations exhibit complicated temporal structures in general and, importantly, can be comparable in magnitude~\cite{Yoneda2022,AriasPRA2023,RojasArias2025InferringCN,RojasArias2024TheOO,Donnelly2025,superconductingqubits,exchangecrosstalk}. Their contributions to the average gate infidelity are fully captured by
Eq.~\eqref{mainres}. 

For instance, in the white-noise limit where
$S_{\alpha\beta}(\omega)=S_{\alpha\beta}$ are frequency-independent
constants, the infidelity reduces to:
\begin{align}
    \overline{\mathcal{I}}=\frac{\tau}{d+1}\sum_{\alpha,\beta}S_{\alpha\beta}\Tr_{S}[\hat{A}_{\alpha}\hat{A}_{\beta}],\label{Eq:memorylesslimit}
\end{align}
 which is linear in gate time $\tau$, a hallmark of Markovian decoherence. It is clear from Eq.~\eqref{Eq:memorylesslimit} that, in this memoryless limit $S_{\alpha\beta}(t)\propto\delta(t)$, the overlap of different noise channels in operator space $\mathrm{Tr}_S(\hat A_\alpha \hat A_\beta)$ determines which fluctuations contribute to the gate infidelity.  For the frequency noise $S_{\Delta_i\Delta_j}$, the overlap $\mathrm{Tr}_S(\sigma_i^z\sigma_j^z)$ vanishes for $i\neq j$, so only on-site fluctuations lead to the usual single-qubit dephasing, while spatially correlated frequency noise does not contribute in this limit. Similarly, cross correlations $S_{\Delta J}$ involve orthogonal operators and therefore vanish. In contrast, noise in the inter-qubit couplings $S_{JJ}$ contributes directly because the corresponding operators satisfy $\mathrm{Tr}_S(\sigma_i^a\sigma_j^b\sigma_i^a\sigma_j^b)= 2^N$, producing simultaneous errors on both qubits and contributing to entangling-gate decoherence in proportion to their spectral strengths.

\textit{Spin qubit gates under correlated $1/f$ noise}|We now investigate the effects of correlated noise on two-qubit gates in semiconductor quantum dots, one of the leading platforms for scalable quantum computing \cite{QuantumcomputationwithQD,RevModPhys.95.025003,Chatterjee2021,RevModPhys.79.1217,Vandersypen2017}. In these systems, two qubits are coupled via a Heisenberg exchange interaction \cite{QuantumcomputationwithQD,Coupledquantumdotsasquantumgates,DiVincenzo2000,doi:10.1126/science.1116955}. An effective spin-orbit interaction (SOI) is usually introduced to enable electrical qubit control, either through a micromagnet-induced field gradient in electron spin qubits or via the intrinsic SOI, for example in Ge hole spin qubits. This SOI misaligns the spin quantization axes by an angle $\theta$, yielding an anisotropic coupling~\cite{ExchangeanisotropiesinmicrowavedrivenSTqubits,Anisotropicexchangeinteractionoftwoholespinqubits}:
\begin{equation}
H_{S}=\frac{\Delta_{1}}{2}\sigma_{1}^{z}+\frac{\Delta_{2}}{2}\sigma_{2}^{z}+\frac{J}{4}[\Vec{\sigma}_{1}R_{y}(-\theta)\Vec{\sigma}_{2}].
\end{equation}
Here $\Delta_i$ are the qubit frequencies and $J$ parametrizes the anisotropic exchange, with $R_{y}(-\theta)$ a rotation around the $y$-axis. In the regime $\Delta_1,\Delta_2\gg J$, two key parameters characterize this Hamiltonian: the normalized detuning $\delta\equiv|\Delta_1-\Delta_2|/J$ and the  angle $\theta$. Together they determine the structure of the effective interaction and the resulting two-qubit gate.

\begin{figure*}[t]
\centering
\includegraphics[width=0.9\textwidth]{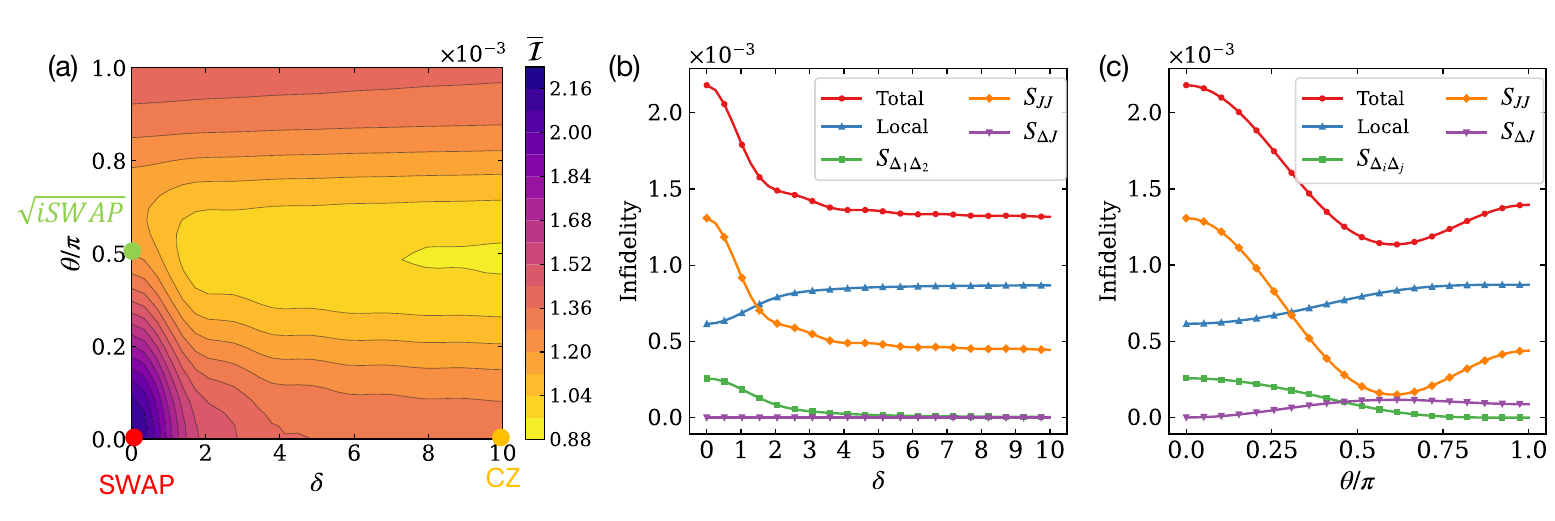}
\caption{\textbf{Infidelity of two-qubit gates under $1/f$ noise.} (a) Total infidelity across the $(\delta,\theta)$ plane. (b) Line cuts of the infidelity at $\theta = 0$, showing dependence of infidelity induced by each type of noise on detuning. 
(c) Line cuts at $\delta = 0$, showing infidelity induced by each type of noise as a function of rotation angle $\theta$. Parameters used in the plots: $J=1$, $\Delta_{1}=10$, $\Delta_{2}=\Delta_{1}+\delta \cross J$, $\tau=\pi/J$, $T=2\pi\cross 100\tau$, $\eta=0.01$. }
\label{fig:2}
\end{figure*}
In theory there are typically three operating regimes \cite{SM}: (i) $\delta=0$, $\theta=0$, corresponding to the isotropic Heisenberg exchange and the SWAP gate; (ii) $\delta\gg1$, where large detuning suppresses the transverse coupling and yields an effective Ising interaction, enabling the CZ gate~\cite{CPhase}; and (iii) $\delta=0$, $\theta=\pi/2$, where the fully anisotropic limit produces, after a rotating-wave approximation, an effective interaction $\frac{J}{8}(\sigma_{1}^{x}\sigma_{2}^{x}+\sigma_{1}^{y}\sigma_{2}^{y})$ that realizes  $\sqrt{\mathrm{iSWAP}}$ up to single-qubit rotations (see SM~\cite{SM}). 

The qubit system couples to its environment through the operators
$\hat{A}_\alpha \in \{\sigma_i^z,\, \vec{\sigma}_1 R_y(-\theta)\vec{\sigma}_2\}$,
corresponding to fluctuations in the qubit frequencies $\delta\Delta_i$ and in the
anisotropic exchange interaction $\delta J$, represented by environment operators $\hat{B}_{\alpha}=\delta\Delta_{i}/2$ and $\delta J/4$, respectively. 
In solid-state spin qubits, a dominant source of such fluctuations
is charge noise, which originates microscopically from ensembles of
two-level fluctuators
(TLFs)~\cite{Lor21f,RojasArias2025InferringCN,Yoneda2018,PhysRevLett.121.076801,PhysRevB.110.235305,Muller_2019}.
The statistical properties of these fluctuators determine both the
spectral shape and the spatial correlations of the noise entering
Eq.~\eqref{mainres}.
The most commonly observed form is $1/f$ noise \cite{1fnoiseplanarge,
Interactingtwolevelsystemsasasourceof1fchargenoiseinQD,
1fnoiseinsemiconductorsarisingfromtheheterogeneousdetrappingprocess,
Decoherenceinducedbyasparsebathoftwolevelfluctuators,RevModPhys.53.497,Elsayed2024,10.1063/1.4954700,RevModPhys.86.361}, characterized by
the spectrum
$S(\omega) = 2\pi  T^{-2+\eta}/|\omega|^{1-\eta}$, where $\eta$ is a small positive regularization parameter that
controls the low-frequency divergence, and $T$ is a reference time scale that sets the overall noise strength. We take all noise correlations to
share the same spectrum and amplitude as an illustrative fully correlated benchmark, $S^C_{\alpha\beta}(\omega)=S(\omega)$~\cite{Yoneda2022,AriasPRA2023,RojasArias2025InferringCN,RojasArias2024TheOO,Donnelly2025,superconductingqubits,exchangecrosstalk}.
We show that, for a generic two-spin system under
this spectrum, the average gate infidelity is
\begin{align}\label{Eq:1fInfidelity}
\overline{\mathcal{I}} = \frac{4}{5}
  \sum_{\alpha\beta}\sum_{nm}
  \hat{A}_{\alpha}^{nm}\hat{A}_{\beta}^{mn}
  \,\Gamma[\eta]\,\cos\!\left(\frac{\pi\eta}{2}\right)
  \left(\frac{\tau}{T}\right)^{\!2-\eta} \notag\\
  \times
  \sum_{k=0}^{\infty} \frac{(-1)^k}{(2k)!}
  \frac{(\omega_{nm}\tau)^{2k}}
       {[\eta-(2k\!+\!1)][\eta-(2k\!+\!2)]},
\end{align}
where $\Gamma$ is the Euler gamma function and the transition amplitudes $\hat{A}_{\alpha}^{nm}\hat{A}_{\beta}^{mn}$ set the weight of each channel at frequency $\omega_{nm}$. The nonlinear scaling in $\tau$  reflects the non-Markovian character of $1/f$ noise, in contrast to the behavior of white noise. 

Figure~\ref{fig:2}(a) maps the total infidelity across the $(\delta,\theta)$ parameter plane, exposing an intriguing
gate dependence. The SWAP gate [red dot in Fig.~\ref{fig:2}(a)] 
exhibits the
largest infidelity, forming a hot spot in the landscape, while both the CZ gate [orange dot \cite{CZcomment}] and the $\sqrt{\mathrm{iSWAP}}$ gate [green dot] lie in regions of  reduced error.
This hierarchy originates from the structure of the transition amplitudes $\hat{A}_{\alpha}^{nm}\hat{A}_{\beta}^{mn}$: the SWAP gate activates a larger number of noise-sensitive channels,
making it more vulnerable to the correlated fluctuations considered here than either the CZ or $\sqrt{\mathrm{iSWAP}}$ operation~\cite{Rimbach-Russ_2023}. We note that, among the three operations studied, this dividing line coincides with entangling character: SWAP, which maps product states onto product states and generates no entanglement, is precisely the vulnerable operation, while the entangling CZ and $\sqrt{\mathrm{iSWAP}}$ gates are the resilient ones. 

To elucidate the origin of this behavior, we examine one-dimensional cuts along the boundaries of the parameter plane. Figure~\ref{fig:2}(b) traces the evolution of individual noise
contributions as the system is tuned from the SWAP gate toward the CZ gate ($\theta=0$, increasing $\delta$).
Across all operating points, the dominant sources of infidelity are the local (on-site) noise and the interaction noise $S_{JJ}$; all other correlated channels remain comparatively small. At $\theta=0$ the cross-correlation $S_{\Delta J}$ vanishes identically, while the spatially correlated frequency noise $S_{\Delta_1\Delta_2}$ is suppressed once the large-detuning limit projects the exchange onto an effective $\sigma_1^z\sigma_2^z$ interaction.

The physical mechanism underlying the suppression of $S_{JJ}$ with increasing $\delta$ can be traced to the restructuring of the energy spectrum (see Fig. S1 in SM \cite{SM}). At the isotropic exchange point and with $\delta=0$, the eigenstates are the
singlet-triplet pair $\ket{S}$ and $\ket{T_0}$, separated by the exchange splitting $J$. 
As $\delta$ increases, the eigenstates rotate toward the computational basis $\ket{\uparrow\downarrow}$, $\ket{\downarrow\uparrow}$. While other dephasing channels remain unchanged, the zero-frequency dephasing channels of $\ket{S}$ and $\ket{T_0}$ induced by the $\{\sigma_1^x\sigma_2^x, \sigma_1^y\sigma_2^y\}$ terms become the transition between $\ket{\uparrow\downarrow}$ and $\ket{\downarrow\uparrow}$, acquiring a nonzero frequency $|\Delta_1-\Delta_2|$ \cite{SM}. 
The noise weight at this transition frequency is strongly suppressed by the $1/f$ spectrum, leading to a rapid decrease of the $S_{JJ}$ contribution
[Fig.~\ref{fig:2}(b)].

Meanwhile, the local noise is only mildly affected by this eigenstate rotation, increasing slightly as the singlet-triplet transition with frequency $J$ is converted into computational-basis dephasing at zero frequency.
The spatially correlated frequency noise $S_{\Delta_1\Delta_2}$ contributes weakly throughout \cite{frequencynoise}, since the relevant transition amplitudes enter with opposite signs and largely cancel. The effective filter functions for each gate are presented in Sec.~VI of the SM \cite{SM}. Taken together, these effects explain the reduced total infidelity of the CZ gate relative to the SWAP gate.

A similar suppression mechanism operates along the cut from the SWAP gate to the $\sqrt{\mathrm{iSWAP}}$ gate ($\delta=0$, increasing $\theta$), as shown in Fig.~\ref{fig:2}(c). The local noise again increases mildly with $\theta$, driven by the reduction of the $\ket{S}$-$\ket{T_0}$ splitting, which shifts the filter function toward zero frequency where $1/f$ noise is strongest. The interaction noise, however, displays a non-monotonic
dependence: it decreases initially, reaches a minimum beyond $\theta=\pi/2$, and then rises. To leading order in $J/\Delta$ at $\Delta_1=\Delta_2=\Delta$, the zero-frequency interaction-noise weight is proportional to $2+4\cos\theta+6\cos^2\theta$, whose minimum occurs at $\cos\theta=-1/3$ ($\theta\simeq0.608\pi$). Unlike the $\theta=0$ case, a finite rotation angle activates the cross-correlation channel $S_{\Delta J}$, but its contribution remains small due to the weak mixing between the $\ket{T_{\pm 1}}$ and $\ket{S,T_0}$ subspaces.

Both the large-detuning regime of the CZ gate and the large-rotation regime of the $\sqrt{\mathrm{iSWAP}}$ gate
act to shift noise-sensitive channels away from zero frequency, where the spectral weight of low-frequency noise is concentrated. This suppression mechanism is not specific to $1/f$ noise: it applies to any spectrum dominated by low-frequency weight, including Lorentzian (see below) and $1/f^{\gamma}$ spectra with $\gamma$ of order unity (for $\gamma\to0$ the spectrum becomes white and the gate dependence disappears, cf.~Eq.~\eqref{Eq:memorylesslimit}), all commonly
observed in solid-state qubit platforms \cite{RevModPhys.86.361,Muller_2019,RevModPhys.53.497}
. Selecting the operating point in the $(\delta,\theta)$ plane therefore reshapes the spectral window through which correlated fluctuations couple to the gate dynamics, yielding higher fidelities for the CZ and $\sqrt{\mathrm{iSWAP}}$ operations than for the SWAP gate across this class of low-frequency-dominated noise environments.

\textit{Spatially correlated Lorentzian noise}|Advances in materials engineering have substantially reduced TLF densities, yielding
devices in which only a few nearby fluctuators dominate the noise environment \cite{PaqueletWuetz2023,https://doi.org/10.1002/adma.202003361,PhysRevB.100.165305}. In this regime the noise spectrum takes a Lorentzian form, $S(\omega)= T_2^{*-1}/(1+\omega^2\tau_c^2)$, where $T_2^*$ is a reference time scale that sets the overall noise strength and $\tau_c$ is the correlation time of the dominant TLF. The Lorentzian introduces a characteristic frequency scale $\tau_c^{-1}$, absent in the scale-free $1/f$ case, that separates a white-noise plateau at
$\omega\ll\tau_c^{-1}$ from a $1/\omega^2$ roll-off at $\omega\gg\tau_c^{-1}$. 

\begin{figure}[h!]
    \centering
\includegraphics[width=\linewidth]{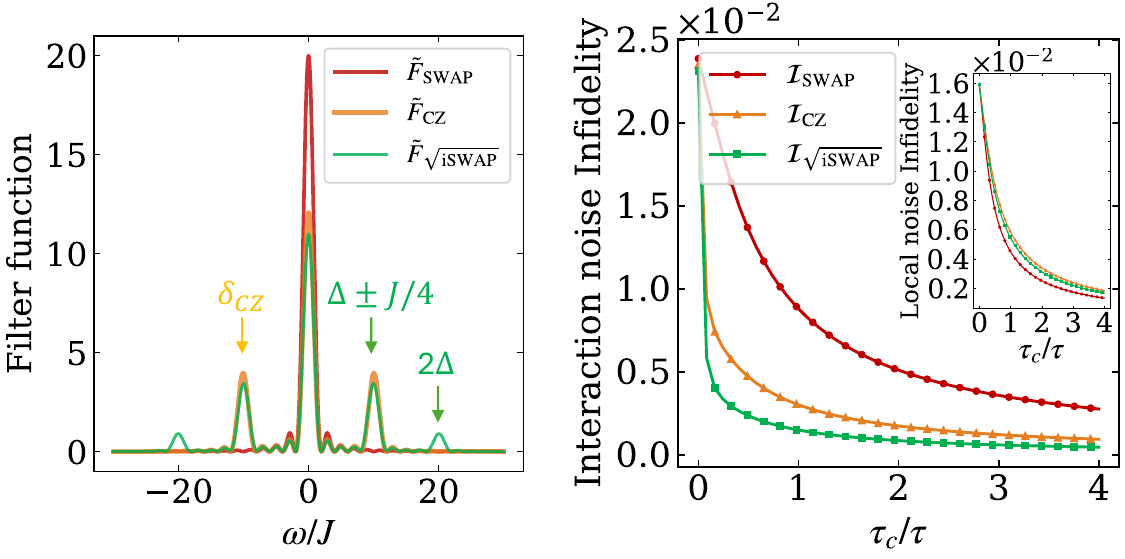}
    \caption{\textbf{Two-qubit gates under correlated Lorentzian noise.} Left panel: Normalized total filter functions $\tilde{F}=F/\tau^2$ at the three gate points. 
    Right panel: contributions from interaction noise as a function of the correlation time  $\tau_{c}$ at the SWAP gate point, CZ gate point, and $\sqrt{\text{iSWAP}}$ gate point. The inset shows the contributions from local noise as a function of the correlation time  $\tau_{c}$. 
    Parameters used in the plots: $J=1$, $\Delta_{1}=10$, $\Delta_{2}=\Delta_{1}+\delta \cross J$, $\tau=\pi/J$, $T_{2}^{*}=100\tau$. For the SWAP gate, $\delta=0,  \Delta_{1}=\Delta_{2}=\Delta$ and $\theta=0$; for the CZ gate, $\delta=\delta_{CZ}=10$ and $\theta=0$; and for the $\sqrt{\text{iSWAP}}$ gate, $\delta=0$, $\Delta_{1}=\Delta_{2}=\Delta$ and $\theta=\pi/2$.}
\label{fig:3}
\end{figure}

Since the Lorentzian spectrum is also low-frequency, the total infidelity across the $(\delta,\theta)$ plane follows the same hierarchy established for $1/f$ noise (see SM \cite{SM}), confirming the generality of the suppression mechanism described above. The time scale $\tau_c$ introduces additional features. Figure~\ref{fig:3}, left panel, displays the total filter functions at the
three operating points. The SWAP-gate filter is dominated by zero-frequency weight, whereas the CZ and $\sqrt{\mathrm{iSWAP}}$ filters carry significant spectral weight at higher frequencies (see SM for a detailed decomposition \cite{SM}).
This distinction has a direct consequence for the $\tau_c$ dependence of the infidelity, shown in Fig.~\ref{fig:3}, right panel. For both the interaction noise and the local noise (inset), the infidelity decreases monotonically with increasing $\tau_c$, as expected from the progressive suppression of high-frequency spectral weight.
The key observation, however, is that the decay rates differ markedly between gates and noise channels.
The interaction-noise contribution drops far more rapidly at the CZ and $\sqrt{\mathrm{iSWAP}}$ points than at the SWAP point, because the filter functions of these two gates sample noise at frequencies $\omega\gtrsim\tau_c^{-1}$ that fall within the Lorentzian roll-off. By contrast, the SWAP-gate filter is concentrated at zero frequency, where the Lorentzian spectrum remains flat and essentially $\tau_c$-independent. For local noise, the filter functions of all three gates are dominated by zero-frequency weight, so the infidelity decreases at a comparable rate regardless of the operating point [inset of Fig.~\ref{fig:3}, right panel].

\textit{Conclusion and outlook}|We have shown that spatiotemporally correlated noise makes multiqubit gate fidelity strongly gate-dependent: 
for the equal-amplitude, fully
correlated noise model studied here in semiconductor
spin qubits, the entangling CZ and $\sqrt{\mathrm{iSWAP}}$ gates are markedly more resilient to low-frequency correlated noise than the non-entangling SWAP gate, even though correlated fluctuations of the exchange interaction can degrade performance to a level comparable to local noise~\cite{anticorrelationcomment}. These results follow from a general expression for the average gate infidelity under arbitrary spatiotemporal noise correlations. The framework is not specific to spin qubits and applies to superconducting circuits, trapped ions, and neutral atoms wherever spatially correlated noise is present. Combined with microscopic noise modeling~\cite{modelingnoiseenvir,samadi2025statisticalstructurechargedisorder}, it enables platform-specific fidelity predictions and provides a concrete design principle for selecting entangling operations in scalable, fault-tolerant quantum processors.

\begin{acknowledgments}
\textit{Acknowledgments}|This work was supported by the Georg H. Endress Foundation, by the Swiss National Science Foundation, NCCR SPIN (grant no. 51NF40-225153), and by the Ibn Sina Global Scholarship. DL acknowledges the Deanship of Research at KFUPM and the Quantum Center for the support received under Grant no. CUP25102 and no. INQC2500,
respectively.
\end{acknowledgments}

\textit{Data availability}|The numerical code used to generate the figures in this work is available from the authors upon reasonable request. 
\nocite{thetheoryofopenquantumsystem,Haarmeasure,Clerk2008IntroductionTQ,Nielsen_Chuang_2010,iswapgate}
\bibliography{bib}

\begin{thebibliography}{112}%
\makeatletter
\providecommand \@ifxundefined [1]{%
 \@ifx{#1\undefined}
}%
\providecommand \@ifnum [1]{%
 \ifnum #1\expandafter \@firstoftwo
 \else \expandafter \@secondoftwo
 \fi
}%
\providecommand \@ifx [1]{%
 \ifx #1\expandafter \@firstoftwo
 \else \expandafter \@secondoftwo
 \fi
}%
\providecommand \natexlab [1]{#1}%
\providecommand \enquote  [1]{``#1''}%
\providecommand \bibnamefont  [1]{#1}%
\providecommand \bibfnamefont [1]{#1}%
\providecommand \citenamefont [1]{#1}%
\providecommand \href@noop [0]{\@secondoftwo}%
\providecommand \href [0]{\begingroup \@sanitize@url \@href}%
\providecommand \@href[1]{\@@startlink{#1}\@@href}%
\providecommand \@@href[1]{\endgroup#1\@@endlink}%
\providecommand \@sanitize@url [0]{\catcode `\\12\catcode `\$12\catcode `\&12\catcode `\#12\catcode `\^12\catcode `\_12\catcode `\%12\relax}%
\providecommand \@@startlink[1]{}%
\providecommand \@@endlink[0]{}%
\providecommand \url  [0]{\begingroup\@sanitize@url \@url }%
\providecommand \@url [1]{\endgroup\@href {#1}{\urlprefix }}%
\providecommand \urlprefix  [0]{URL }%
\providecommand \Eprint [0]{\href }%
\providecommand \doibase [0]{https://doi.org/}%
\providecommand \selectlanguage [0]{\@gobble}%
\providecommand \bibinfo  [0]{\@secondoftwo}%
\providecommand \bibfield  [0]{\@secondoftwo}%
\providecommand \translation [1]{[#1]}%
\providecommand \BibitemOpen [0]{}%
\providecommand \bibitemStop [0]{}%
\providecommand \bibitemNoStop [0]{.\EOS\space}%
\providecommand \EOS [0]{\spacefactor3000\relax}%
\providecommand \BibitemShut  [1]{\csname bibitem#1\endcsname}%
\let\auto@bib@innerbib\@empty
\bibitem [{\citenamefont {Preskill}(2018)}]{Preskill2018quantumcomputingin}%
  \BibitemOpen
  \bibfield  {author} {\bibinfo {author} {\bibfnamefont {J.}~\bibnamefont {Preskill}},\ }\bibfield  {title} {\bibinfo {title} {Quantum {C}omputing in the {NISQ} era and beyond},\ }\href {https://doi.org/10.22331/q-2018-08-06-79} {\bibfield  {journal} {\bibinfo  {journal} {{Quantum}}\ }\textbf {\bibinfo {volume} {2}},\ \bibinfo {pages} {79} (\bibinfo {year} {2018})}\BibitemShut {NoStop}%
\bibitem [{\citenamefont {Palma}\ \emph {et~al.}(1996)\citenamefont {Palma}, \citenamefont {Suominen},\ and\ \citenamefont {Ekert}}]{palma1996quantum}%
  \BibitemOpen
  \bibfield  {author} {\bibinfo {author} {\bibfnamefont {G.~M.}\ \bibnamefont {Palma}}, \bibinfo {author} {\bibfnamefont {K.-A.}\ \bibnamefont {Suominen}},\ and\ \bibinfo {author} {\bibfnamefont {A.}~\bibnamefont {Ekert}},\ }\bibfield  {title} {\bibinfo {title} {Quantum computers and dissipation},\ }\href@noop {} {\bibfield  {journal} {\bibinfo  {journal} {Proceedings of the Royal Society of London. Series A: Mathematical, Physical and Engineering Sciences}\ }\textbf {\bibinfo {volume} {452}},\ \bibinfo {pages} {567} (\bibinfo {year} {1996})}\BibitemShut {NoStop}%
\bibitem [{\citenamefont {Schlosshauer}(2019)}]{SCHLOSSHAUER20191}%
  \BibitemOpen
  \bibfield  {author} {\bibinfo {author} {\bibfnamefont {M.}~\bibnamefont {Schlosshauer}},\ }\bibfield  {title} {\bibinfo {title} {Quantum decoherence},\ }\href {https://doi.org/https://doi.org/10.1016/j.physrep.2019.10.001} {\bibfield  {journal} {\bibinfo  {journal} {Physics Reports}\ }\textbf {\bibinfo {volume} {831}},\ \bibinfo {pages} {1} (\bibinfo {year} {2019})},\ \bibinfo {note} {quantum decoherence}\BibitemShut {NoStop}%
\bibitem [{\citenamefont {Bosco}\ \emph {et~al.}(2021)\citenamefont {Bosco}, \citenamefont {Het\'enyi},\ and\ \citenamefont {Loss}}]{SweetSpots4ChargeNoise}%
  \BibitemOpen
  \bibfield  {author} {\bibinfo {author} {\bibfnamefont {S.}~\bibnamefont {Bosco}}, \bibinfo {author} {\bibfnamefont {B.}~\bibnamefont {Het\'enyi}},\ and\ \bibinfo {author} {\bibfnamefont {D.}~\bibnamefont {Loss}},\ }\bibfield  {title} {\bibinfo {title} {Hole spin qubits in $\mathrm{Si}$ finfets with fully tunable spin-orbit coupling and sweet spots for charge noise},\ }\href {https://doi.org/10.1103/PRXQuantum.2.010348} {\bibfield  {journal} {\bibinfo  {journal} {PRX Quantum}\ }\textbf {\bibinfo {volume} {2}},\ \bibinfo {pages} {010348} (\bibinfo {year} {2021})}\BibitemShut {NoStop}%
\bibitem [{\citenamefont {Bosco}\ and\ \citenamefont {Loss}(2021)}]{FullyTunableHyperfineInteractions}%
  \BibitemOpen
  \bibfield  {author} {\bibinfo {author} {\bibfnamefont {S.}~\bibnamefont {Bosco}}\ and\ \bibinfo {author} {\bibfnamefont {D.}~\bibnamefont {Loss}},\ }\bibfield  {title} {\bibinfo {title} {Fully tunable hyperfine interactions of hole spin qubits in {Si} and {Ge} quantum dots},\ }\href {https://doi.org/10.1103/PhysRevLett.127.190501} {\bibfield  {journal} {\bibinfo  {journal} {Phys. Rev. Lett.}\ }\textbf {\bibinfo {volume} {127}},\ \bibinfo {pages} {190501} (\bibinfo {year} {2021})}\BibitemShut {NoStop}%
\bibitem [{\citenamefont {Bosco}\ and\ \citenamefont {Loss}(2022)}]{HoleSpinQubitsinThinCurvedQuantumWells}%
  \BibitemOpen
  \bibfield  {author} {\bibinfo {author} {\bibfnamefont {S.}~\bibnamefont {Bosco}}\ and\ \bibinfo {author} {\bibfnamefont {D.}~\bibnamefont {Loss}},\ }\bibfield  {title} {\bibinfo {title} {Hole spin qubits in thin curved quantum wells},\ }\href {https://doi.org/10.1103/PhysRevApplied.18.044038} {\bibfield  {journal} {\bibinfo  {journal} {Phys. Rev. Appl.}\ }\textbf {\bibinfo {volume} {18}},\ \bibinfo {pages} {044038} (\bibinfo {year} {2022})}\BibitemShut {NoStop}%
\bibitem [{\citenamefont {Reed}\ \emph {et~al.}(2016)\citenamefont {Reed}, \citenamefont {Maune}, \citenamefont {Andrews}, \citenamefont {Borselli}, \citenamefont {Eng}, \citenamefont {Jura}, \citenamefont {Kiselev}, \citenamefont {Ladd}, \citenamefont {Merkel}, \citenamefont {Milosavljevic}, \citenamefont {Pritchett}, \citenamefont {Rakher}, \citenamefont {Ross}, \citenamefont {Schmitz}, \citenamefont {Smith}, \citenamefont {Wright}, \citenamefont {Gyure},\ and\ \citenamefont {Hunter}}]{ReducedSensitivitytoChargeNoise}%
  \BibitemOpen
  \bibfield  {author} {\bibinfo {author} {\bibfnamefont {M.~D.}\ \bibnamefont {Reed}}, \bibinfo {author} {\bibfnamefont {B.~M.}\ \bibnamefont {Maune}}, \bibinfo {author} {\bibfnamefont {R.~W.}\ \bibnamefont {Andrews}}, \bibinfo {author} {\bibfnamefont {M.~G.}\ \bibnamefont {Borselli}}, \bibinfo {author} {\bibfnamefont {K.}~\bibnamefont {Eng}}, \bibinfo {author} {\bibfnamefont {M.~P.}\ \bibnamefont {Jura}}, \bibinfo {author} {\bibfnamefont {A.~A.}\ \bibnamefont {Kiselev}}, \bibinfo {author} {\bibfnamefont {T.~D.}\ \bibnamefont {Ladd}}, \bibinfo {author} {\bibfnamefont {S.~T.}\ \bibnamefont {Merkel}}, \bibinfo {author} {\bibfnamefont {I.}~\bibnamefont {Milosavljevic}}, \bibinfo {author} {\bibfnamefont {E.~J.}\ \bibnamefont {Pritchett}}, \bibinfo {author} {\bibfnamefont {M.~T.}\ \bibnamefont {Rakher}}, \bibinfo {author} {\bibfnamefont {R.~S.}\ \bibnamefont {Ross}}, \bibinfo {author} {\bibfnamefont {A.~E.}\ \bibnamefont {Schmitz}}, \bibinfo {author} {\bibfnamefont {A.}~\bibnamefont {Smith}}, \bibinfo {author}
  {\bibfnamefont {J.~A.}\ \bibnamefont {Wright}}, \bibinfo {author} {\bibfnamefont {M.~F.}\ \bibnamefont {Gyure}},\ and\ \bibinfo {author} {\bibfnamefont {A.~T.}\ \bibnamefont {Hunter}},\ }\bibfield  {title} {\bibinfo {title} {Reduced sensitivity to charge noise in semiconductor spin qubits via symmetric operation},\ }\href {https://doi.org/10.1103/PhysRevLett.116.110402} {\bibfield  {journal} {\bibinfo  {journal} {Phys. Rev. Lett.}\ }\textbf {\bibinfo {volume} {116}},\ \bibinfo {pages} {110402} (\bibinfo {year} {2016})}\BibitemShut {NoStop}%
\bibitem [{\citenamefont {Piot}\ \emph {et~al.}(2022)\citenamefont {Piot}, \citenamefont {Brun}, \citenamefont {Schmitt}, \citenamefont {Zihlmann}, \citenamefont {Michal}, \citenamefont {Apra}, \citenamefont {Abadillo-Uriel}, \citenamefont {Jehl}, \citenamefont {Bertrand}, \citenamefont {Niebojewski}, \citenamefont {Hutin}, \citenamefont {Vinet}, \citenamefont {Urdampilleta}, \citenamefont {Meunier}, \citenamefont {Niquet}, \citenamefont {Maurand},\ and\ \citenamefont {De~Franceschi}}]{Asingleholespinwithenhancedcoherence}%
  \BibitemOpen
  \bibfield  {author} {\bibinfo {author} {\bibfnamefont {N.}~\bibnamefont {Piot}}, \bibinfo {author} {\bibfnamefont {B.}~\bibnamefont {Brun}}, \bibinfo {author} {\bibfnamefont {V.}~\bibnamefont {Schmitt}}, \bibinfo {author} {\bibfnamefont {S.}~\bibnamefont {Zihlmann}}, \bibinfo {author} {\bibfnamefont {V.~P.}\ \bibnamefont {Michal}}, \bibinfo {author} {\bibfnamefont {A.}~\bibnamefont {Apra}}, \bibinfo {author} {\bibfnamefont {J.~C.}\ \bibnamefont {Abadillo-Uriel}}, \bibinfo {author} {\bibfnamefont {X.}~\bibnamefont {Jehl}}, \bibinfo {author} {\bibfnamefont {B.}~\bibnamefont {Bertrand}}, \bibinfo {author} {\bibfnamefont {H.}~\bibnamefont {Niebojewski}}, \bibinfo {author} {\bibfnamefont {L.}~\bibnamefont {Hutin}}, \bibinfo {author} {\bibfnamefont {M.}~\bibnamefont {Vinet}}, \bibinfo {author} {\bibfnamefont {M.}~\bibnamefont {Urdampilleta}}, \bibinfo {author} {\bibfnamefont {T.}~\bibnamefont {Meunier}}, \bibinfo {author} {\bibfnamefont {Y.-M.}\ \bibnamefont {Niquet}}, \bibinfo {author} {\bibfnamefont
  {R.}~\bibnamefont {Maurand}},\ and\ \bibinfo {author} {\bibfnamefont {S.}~\bibnamefont {De~Franceschi}},\ }\bibfield  {title} {\bibinfo {title} {A single hole spin with enhanced coherence in natural silicon},\ }\href {https://doi.org/10.1038/s41565-022-01196-z} {\bibfield  {journal} {\bibinfo  {journal} {Nat. Nanotechnol.}\ }\textbf {\bibinfo {volume} {17}},\ \bibinfo {pages} {1072} (\bibinfo {year} {2022})}\BibitemShut {NoStop}%
\bibitem [{\citenamefont {Zhao}\ \emph {et~al.}(2021)\citenamefont {Zhao}, \citenamefont {Zhao}, \citenamefont {Wang}, \citenamefont {Song},\ and\ \citenamefont {Wang}}]{LOCCNet}%
  \BibitemOpen
  \bibfield  {author} {\bibinfo {author} {\bibfnamefont {X.}~\bibnamefont {Zhao}}, \bibinfo {author} {\bibfnamefont {B.}~\bibnamefont {Zhao}}, \bibinfo {author} {\bibfnamefont {Z.}~\bibnamefont {Wang}}, \bibinfo {author} {\bibfnamefont {Z.}~\bibnamefont {Song}},\ and\ \bibinfo {author} {\bibfnamefont {X.}~\bibnamefont {Wang}},\ }\bibfield  {title} {\bibinfo {title} {Practical distributed quantum information processing with loccnet},\ }\href {https://doi.org/10.1038/s41534-021-00496-x} {\bibfield  {journal} {\bibinfo  {journal} {npj Quantum Information}\ }\textbf {\bibinfo {volume} {7}},\ \bibinfo {pages} {159} (\bibinfo {year} {2021})}\BibitemShut {NoStop}%
\bibitem [{\citenamefont {Martins}\ \emph {et~al.}(2016)\citenamefont {Martins}, \citenamefont {Malinowski}, \citenamefont {Nissen}, \citenamefont {Barnes}, \citenamefont {Fallahi}, \citenamefont {Gardner}, \citenamefont {Manfra}, \citenamefont {Marcus},\ and\ \citenamefont {Kuemmeth}}]{PhysRevLett.116.116801}%
  \BibitemOpen
  \bibfield  {author} {\bibinfo {author} {\bibfnamefont {F.}~\bibnamefont {Martins}}, \bibinfo {author} {\bibfnamefont {F.~K.}\ \bibnamefont {Malinowski}}, \bibinfo {author} {\bibfnamefont {P.~D.}\ \bibnamefont {Nissen}}, \bibinfo {author} {\bibfnamefont {E.}~\bibnamefont {Barnes}}, \bibinfo {author} {\bibfnamefont {S.}~\bibnamefont {Fallahi}}, \bibinfo {author} {\bibfnamefont {G.~C.}\ \bibnamefont {Gardner}}, \bibinfo {author} {\bibfnamefont {M.~J.}\ \bibnamefont {Manfra}}, \bibinfo {author} {\bibfnamefont {C.~M.}\ \bibnamefont {Marcus}},\ and\ \bibinfo {author} {\bibfnamefont {F.}~\bibnamefont {Kuemmeth}},\ }\bibfield  {title} {\bibinfo {title} {Noise suppression using symmetric exchange gates in spin qubits},\ }\href {https://doi.org/10.1103/PhysRevLett.116.116801} {\bibfield  {journal} {\bibinfo  {journal} {Phys. Rev. Lett.}\ }\textbf {\bibinfo {volume} {116}},\ \bibinfo {pages} {116801} (\bibinfo {year} {2016})}\BibitemShut {NoStop}%
\bibitem [{\citenamefont {Laflamme}\ \emph {et~al.}(1996)\citenamefont {Laflamme}, \citenamefont {Miquel}, \citenamefont {Paz},\ and\ \citenamefont {Zurek}}]{PerfectQECC}%
  \BibitemOpen
  \bibfield  {author} {\bibinfo {author} {\bibfnamefont {R.}~\bibnamefont {Laflamme}}, \bibinfo {author} {\bibfnamefont {C.}~\bibnamefont {Miquel}}, \bibinfo {author} {\bibfnamefont {J.~P.}\ \bibnamefont {Paz}},\ and\ \bibinfo {author} {\bibfnamefont {W.~H.}\ \bibnamefont {Zurek}},\ }\bibfield  {title} {\bibinfo {title} {Perfect quantum error correcting code},\ }\href {https://doi.org/10.1103/PhysRevLett.77.198} {\bibfield  {journal} {\bibinfo  {journal} {Phys. Rev. Lett.}\ }\textbf {\bibinfo {volume} {77}},\ \bibinfo {pages} {198} (\bibinfo {year} {1996})}\BibitemShut {NoStop}%
\bibitem [{\citenamefont {Knill}\ and\ \citenamefont {Laflamme}(1997)}]{TheoryofQEC}%
  \BibitemOpen
  \bibfield  {author} {\bibinfo {author} {\bibfnamefont {E.}~\bibnamefont {Knill}}\ and\ \bibinfo {author} {\bibfnamefont {R.}~\bibnamefont {Laflamme}},\ }\bibfield  {title} {\bibinfo {title} {Theory of quantum error-correcting codes},\ }\href {https://doi.org/10.1103/PhysRevA.55.900} {\bibfield  {journal} {\bibinfo  {journal} {Phys. Rev. A}\ }\textbf {\bibinfo {volume} {55}},\ \bibinfo {pages} {900} (\bibinfo {year} {1997})}\BibitemShut {NoStop}%
\bibitem [{\citenamefont {Ng}\ and\ \citenamefont {Preskill}(2009)}]{FTQCvsGaussiannoise}%
  \BibitemOpen
  \bibfield  {author} {\bibinfo {author} {\bibfnamefont {H.~K.}\ \bibnamefont {Ng}}\ and\ \bibinfo {author} {\bibfnamefont {J.}~\bibnamefont {Preskill}},\ }\bibfield  {title} {\bibinfo {title} {Fault-tolerant quantum computation versus gaussian noise},\ }\href {https://doi.org/10.1103/PhysRevA.79.032318} {\bibfield  {journal} {\bibinfo  {journal} {Phys. Rev. A}\ }\textbf {\bibinfo {volume} {79}},\ \bibinfo {pages} {032318} (\bibinfo {year} {2009})}\BibitemShut {NoStop}%
\bibitem [{\citenamefont {Preskill}(2013)}]{Preskill2012SufficientCO}%
  \BibitemOpen
  \bibfield  {author} {\bibinfo {author} {\bibfnamefont {J.}~\bibnamefont {Preskill}},\ }\bibfield  {title} {\bibinfo {title} {Sufficient condition on noise correlations for scalable quantum computing},\ }\href@noop {} {\bibfield  {journal} {\bibinfo  {journal} {Quantum Inf. Comput.}\ }\textbf {\bibinfo {volume} {13}},\ \bibinfo {pages} {181} (\bibinfo {year} {2013})}\BibitemShut {NoStop}%
\bibitem [{\citenamefont {Viola}\ \emph {et~al.}(1999)\citenamefont {Viola}, \citenamefont {Knill},\ and\ \citenamefont {Lloyd}}]{DDofOPS}%
  \BibitemOpen
  \bibfield  {author} {\bibinfo {author} {\bibfnamefont {L.}~\bibnamefont {Viola}}, \bibinfo {author} {\bibfnamefont {E.}~\bibnamefont {Knill}},\ and\ \bibinfo {author} {\bibfnamefont {S.}~\bibnamefont {Lloyd}},\ }\bibfield  {title} {\bibinfo {title} {Dynamical decoupling of open quantum systems},\ }\href {https://doi.org/10.1103/PhysRevLett.82.2417} {\bibfield  {journal} {\bibinfo  {journal} {Phys. Rev. Lett.}\ }\textbf {\bibinfo {volume} {82}},\ \bibinfo {pages} {2417} (\bibinfo {year} {1999})}\BibitemShut {NoStop}%
\bibitem [{\citenamefont {Khodjasteh}\ and\ \citenamefont {Lidar}(2005)}]{FTQDD}%
  \BibitemOpen
  \bibfield  {author} {\bibinfo {author} {\bibfnamefont {K.}~\bibnamefont {Khodjasteh}}\ and\ \bibinfo {author} {\bibfnamefont {D.~A.}\ \bibnamefont {Lidar}},\ }\bibfield  {title} {\bibinfo {title} {Fault-tolerant quantum dynamical decoupling},\ }\href {https://doi.org/10.1103/PhysRevLett.95.180501} {\bibfield  {journal} {\bibinfo  {journal} {Phys. Rev. Lett.}\ }\textbf {\bibinfo {volume} {95}},\ \bibinfo {pages} {180501} (\bibinfo {year} {2005})}\BibitemShut {NoStop}%
\bibitem [{\citenamefont {Paz-Silva}\ \emph {et~al.}(2016)\citenamefont {Paz-Silva}, \citenamefont {Lee}, \citenamefont {Green},\ and\ \citenamefont {Viola}}]{DD4multiqubit}%
  \BibitemOpen
  \bibfield  {author} {\bibinfo {author} {\bibfnamefont {G.~A.}\ \bibnamefont {Paz-Silva}}, \bibinfo {author} {\bibfnamefont {S.-W.}\ \bibnamefont {Lee}}, \bibinfo {author} {\bibfnamefont {T.~J.}\ \bibnamefont {Green}},\ and\ \bibinfo {author} {\bibfnamefont {L.}~\bibnamefont {Viola}},\ }\bibfield  {title} {\bibinfo {title} {Dynamical decoupling sequences for multi-qubit dephasing suppression and long-time quantum memory},\ }\href {https://doi.org/10.1088/1367-2630/18/7/073020} {\bibfield  {journal} {\bibinfo  {journal} {New Journal of Physics}\ }\textbf {\bibinfo {volume} {18}},\ \bibinfo {pages} {073020} (\bibinfo {year} {2016})}\BibitemShut {NoStop}%
\bibitem [{\citenamefont {Cywi\ifmmode~\acute{n}\else \'{n}\fi{}ski}\ \emph {et~al.}(2008)\citenamefont {Cywi\ifmmode~\acute{n}\else \'{n}\fi{}ski}, \citenamefont {Lutchyn}, \citenamefont {Nave},\ and\ \citenamefont {Das~Sarma}}]{PhysRevB.77.174509}%
  \BibitemOpen
  \bibfield  {author} {\bibinfo {author} {\bibfnamefont {L.}~\bibnamefont {Cywi\ifmmode~\acute{n}\else \'{n}\fi{}ski}}, \bibinfo {author} {\bibfnamefont {R.~M.}\ \bibnamefont {Lutchyn}}, \bibinfo {author} {\bibfnamefont {C.~P.}\ \bibnamefont {Nave}},\ and\ \bibinfo {author} {\bibfnamefont {S.}~\bibnamefont {Das~Sarma}},\ }\bibfield  {title} {\bibinfo {title} {How to enhance dephasing time in superconducting qubits},\ }\href {https://doi.org/10.1103/PhysRevB.77.174509} {\bibfield  {journal} {\bibinfo  {journal} {Phys. Rev. B}\ }\textbf {\bibinfo {volume} {77}},\ \bibinfo {pages} {174509} (\bibinfo {year} {2008})}\BibitemShut {NoStop}%
\bibitem [{\citenamefont {Nguyen}\ \emph {et~al.}(2026)\citenamefont {Nguyen}, \citenamefont {Rimbach-Russ},\ and\ \citenamefont {Bosco}}]{Nguyen2025ColorIC}%
  \BibitemOpen
  \bibfield  {author} {\bibinfo {author} {\bibfnamefont {M.~T.~P.}\ \bibnamefont {Nguyen}}, \bibinfo {author} {\bibfnamefont {M.}~\bibnamefont {Rimbach-Russ}},\ and\ \bibinfo {author} {\bibfnamefont {S.}~\bibnamefont {Bosco}},\ }\bibfield  {title} {\bibinfo {title} {Color it, code it, cancel it: k-local dynamical decoupling from classical additive codes},\ }\href@noop {} {\bibfield  {journal} {\bibinfo  {journal} {PRX Quantum}\ } (\bibinfo {year} {2026})},\ \bibinfo {note} {accepted for publication},\ \Eprint {https://arxiv.org/abs/arXiv:2508.08213} {arXiv:arXiv:2508.08213 [quant-ph]} \BibitemShut {NoStop}%
\bibitem [{\citenamefont {Brown}\ and\ \citenamefont {Lidar}(2025)}]{1d4l-73x6}%
  \BibitemOpen
  \bibfield  {author} {\bibinfo {author} {\bibfnamefont {A.~F.}\ \bibnamefont {Brown}}\ and\ \bibinfo {author} {\bibfnamefont {D.~A.}\ \bibnamefont {Lidar}},\ }\bibfield  {title} {\bibinfo {title} {Efficient chromatic-number-based multiqubit decoherence and crosstalk suppression},\ }\href {https://doi.org/10.1103/1d4l-73x6} {\bibfield  {journal} {\bibinfo  {journal} {PRX Quantum}\ }\textbf {\bibinfo {volume} {6}},\ \bibinfo {pages} {020354} (\bibinfo {year} {2025})}\BibitemShut {NoStop}%
\bibitem [{\citenamefont {Uhrig}(2007)}]{PhysRevLett.98.100504}%
  \BibitemOpen
  \bibfield  {author} {\bibinfo {author} {\bibfnamefont {G.~S.}\ \bibnamefont {Uhrig}},\ }\bibfield  {title} {\bibinfo {title} {Keeping a quantum bit alive by optimized $\ensuremath{\pi}$-pulse sequences},\ }\href {https://doi.org/10.1103/PhysRevLett.98.100504} {\bibfield  {journal} {\bibinfo  {journal} {Phys. Rev. Lett.}\ }\textbf {\bibinfo {volume} {98}},\ \bibinfo {pages} {100504} (\bibinfo {year} {2007})}\BibitemShut {NoStop}%
\bibitem [{\citenamefont {Vezvaee}\ \emph {et~al.}(2026)\citenamefont {Vezvaee}, \citenamefont {Tripathi}, \citenamefont {Morford-Oberst}, \citenamefont {Butt}, \citenamefont {Kasatkin},\ and\ \citenamefont {Lidar}}]{Vezvaee2026}%
  \BibitemOpen
  \bibfield  {author} {\bibinfo {author} {\bibfnamefont {A.}~\bibnamefont {Vezvaee}}, \bibinfo {author} {\bibfnamefont {V.}~\bibnamefont {Tripathi}}, \bibinfo {author} {\bibfnamefont {M.}~\bibnamefont {Morford-Oberst}}, \bibinfo {author} {\bibfnamefont {F.}~\bibnamefont {Butt}}, \bibinfo {author} {\bibfnamefont {V.}~\bibnamefont {Kasatkin}},\ and\ \bibinfo {author} {\bibfnamefont {D.~A.}\ \bibnamefont {Lidar}},\ }\bibfield  {title} {\bibinfo {title} {Demonstration of high-fidelity entangled logical qubits using transmons},\ }\href {https://doi.org/10.1038/s41467-026-70011-3} {\bibfield  {journal} {\bibinfo  {journal} {Nature Communications}\ }\textbf {\bibinfo {volume} {17}},\ \bibinfo {pages} {3281} (\bibinfo {year} {2026})}\BibitemShut {NoStop}%
\bibitem [{\citenamefont {Palao}\ and\ \citenamefont {Kosloff}(2002)}]{QCbyanOptimalControlAlgorithm}%
  \BibitemOpen
  \bibfield  {author} {\bibinfo {author} {\bibfnamefont {J.~P.}\ \bibnamefont {Palao}}\ and\ \bibinfo {author} {\bibfnamefont {R.}~\bibnamefont {Kosloff}},\ }\bibfield  {title} {\bibinfo {title} {Quantum computing by an optimal control algorithm for unitary transformations},\ }\href {https://doi.org/10.1103/PhysRevLett.89.188301} {\bibfield  {journal} {\bibinfo  {journal} {Phys. Rev. Lett.}\ }\textbf {\bibinfo {volume} {89}},\ \bibinfo {pages} {188301} (\bibinfo {year} {2002})}\BibitemShut {NoStop}%
\bibitem [{\citenamefont {Nielsen}\ \emph {et~al.}(2006)\citenamefont {Nielsen}, \citenamefont {Dowling}, \citenamefont {Gu},\ and\ \citenamefont {Doherty}}]{OptimalcontrolgeometryandQC}%
  \BibitemOpen
  \bibfield  {author} {\bibinfo {author} {\bibfnamefont {M.~A.}\ \bibnamefont {Nielsen}}, \bibinfo {author} {\bibfnamefont {M.~R.}\ \bibnamefont {Dowling}}, \bibinfo {author} {\bibfnamefont {M.}~\bibnamefont {Gu}},\ and\ \bibinfo {author} {\bibfnamefont {A.~C.}\ \bibnamefont {Doherty}},\ }\bibfield  {title} {\bibinfo {title} {Optimal control, geometry, and quantum computing},\ }\href {https://doi.org/10.1103/PhysRevA.73.062323} {\bibfield  {journal} {\bibinfo  {journal} {Phys. Rev. A}\ }\textbf {\bibinfo {volume} {73}},\ \bibinfo {pages} {062323} (\bibinfo {year} {2006})}\BibitemShut {NoStop}%
\bibitem [{\citenamefont {Li}\ \emph {et~al.}(2017)\citenamefont {Li}, \citenamefont {Yang}, \citenamefont {Peng},\ and\ \citenamefont {Sun}}]{HybridApproachtoQuantumOptimalControl}%
  \BibitemOpen
  \bibfield  {author} {\bibinfo {author} {\bibfnamefont {J.}~\bibnamefont {Li}}, \bibinfo {author} {\bibfnamefont {X.}~\bibnamefont {Yang}}, \bibinfo {author} {\bibfnamefont {X.}~\bibnamefont {Peng}},\ and\ \bibinfo {author} {\bibfnamefont {C.-P.}\ \bibnamefont {Sun}},\ }\bibfield  {title} {\bibinfo {title} {Hybrid quantum-classical approach to quantum optimal control},\ }\href {https://doi.org/10.1103/PhysRevLett.118.150503} {\bibfield  {journal} {\bibinfo  {journal} {Phys. Rev. Lett.}\ }\textbf {\bibinfo {volume} {118}},\ \bibinfo {pages} {150503} (\bibinfo {year} {2017})}\BibitemShut {NoStop}%
\bibitem [{\citenamefont {Hansen}\ \emph {et~al.}(2022)\citenamefont {Hansen}, \citenamefont {Seedhouse}, \citenamefont {Chan}, \citenamefont {Hudson}, \citenamefont {Itoh}, \citenamefont {Laucht}, \citenamefont {Saraiva}, \citenamefont {Yang},\ and\ \citenamefont {Dzurak}}]{hansen2022implementation}%
  \BibitemOpen
  \bibfield  {author} {\bibinfo {author} {\bibfnamefont {I.}~\bibnamefont {Hansen}}, \bibinfo {author} {\bibfnamefont {A.~E.}\ \bibnamefont {Seedhouse}}, \bibinfo {author} {\bibfnamefont {K.~W.}\ \bibnamefont {Chan}}, \bibinfo {author} {\bibfnamefont {F.~E.}\ \bibnamefont {Hudson}}, \bibinfo {author} {\bibfnamefont {K.~M.}\ \bibnamefont {Itoh}}, \bibinfo {author} {\bibfnamefont {A.}~\bibnamefont {Laucht}}, \bibinfo {author} {\bibfnamefont {A.}~\bibnamefont {Saraiva}}, \bibinfo {author} {\bibfnamefont {C.~H.}\ \bibnamefont {Yang}},\ and\ \bibinfo {author} {\bibfnamefont {A.~S.}\ \bibnamefont {Dzurak}},\ }\bibfield  {title} {\bibinfo {title} {Implementation of an advanced dressing protocol for global qubit control in silicon},\ }\href@noop {} {\bibfield  {journal} {\bibinfo  {journal} {Appl. Phys. Rev.}\ }\textbf {\bibinfo {volume} {9}},\ \bibinfo {pages} {031409} (\bibinfo {year} {2022})}\BibitemShut {NoStop}%
\bibitem [{\citenamefont {Hansen}\ \emph {et~al.}(2021)\citenamefont {Hansen}, \citenamefont {Seedhouse}, \citenamefont {Saraiva}, \citenamefont {Laucht}, \citenamefont {Dzurak},\ and\ \citenamefont {Yang}}]{PhysRevA.104.062415}%
  \BibitemOpen
  \bibfield  {author} {\bibinfo {author} {\bibfnamefont {I.}~\bibnamefont {Hansen}}, \bibinfo {author} {\bibfnamefont {A.~E.}\ \bibnamefont {Seedhouse}}, \bibinfo {author} {\bibfnamefont {A.}~\bibnamefont {Saraiva}}, \bibinfo {author} {\bibfnamefont {A.}~\bibnamefont {Laucht}}, \bibinfo {author} {\bibfnamefont {A.~S.}\ \bibnamefont {Dzurak}},\ and\ \bibinfo {author} {\bibfnamefont {C.~H.}\ \bibnamefont {Yang}},\ }\bibfield  {title} {\bibinfo {title} {Pulse engineering of a global field for robust and universal quantum computation},\ }\href {https://doi.org/10.1103/PhysRevA.104.062415} {\bibfield  {journal} {\bibinfo  {journal} {Phys. Rev. A}\ }\textbf {\bibinfo {volume} {104}},\ \bibinfo {pages} {062415} (\bibinfo {year} {2021})}\BibitemShut {NoStop}%
\bibitem [{\citenamefont {Bosco}\ \emph {et~al.}(2023)\citenamefont {Bosco}, \citenamefont {Geyer}, \citenamefont {Camenzind}, \citenamefont {Eggli}, \citenamefont {Fuhrer}, \citenamefont {Warburton}, \citenamefont {Zumb{\"u}hl}, \citenamefont {Egues}, \citenamefont {Kuhlmann},\ and\ \citenamefont {Loss}}]{Bosco2023PhaseDrivingHS}%
  \BibitemOpen
  \bibfield  {author} {\bibinfo {author} {\bibfnamefont {S.}~\bibnamefont {Bosco}}, \bibinfo {author} {\bibfnamefont {S.}~\bibnamefont {Geyer}}, \bibinfo {author} {\bibfnamefont {L.~C.}\ \bibnamefont {Camenzind}}, \bibinfo {author} {\bibfnamefont {R.~S.}\ \bibnamefont {Eggli}}, \bibinfo {author} {\bibfnamefont {A.}~\bibnamefont {Fuhrer}}, \bibinfo {author} {\bibfnamefont {R.~J.}\ \bibnamefont {Warburton}}, \bibinfo {author} {\bibfnamefont {D.~M.}\ \bibnamefont {Zumb{\"u}hl}}, \bibinfo {author} {\bibfnamefont {J.~C.}\ \bibnamefont {Egues}}, \bibinfo {author} {\bibfnamefont {A.~V.}\ \bibnamefont {Kuhlmann}},\ and\ \bibinfo {author} {\bibfnamefont {D.}~\bibnamefont {Loss}},\ }\bibfield  {title} {\bibinfo {title} {Phase-driving hole spin qubits},\ }\href {https://doi.org/10.1103/PhysRevLett.131.197001} {\bibfield  {journal} {\bibinfo  {journal} {Phys. Rev. Lett.}\ }\textbf {\bibinfo {volume} {131}},\ \bibinfo {pages} {197001} (\bibinfo {year} {2023})}\BibitemShut {NoStop}%
\bibitem [{\citenamefont {Rimbach-Russ}\ \emph {et~al.}(2023)\citenamefont {Rimbach-Russ}, \citenamefont {Philips}, \citenamefont {Xue},\ and\ \citenamefont {Vandersypen}}]{Rimbach-Russ_2023}%
  \BibitemOpen
  \bibfield  {author} {\bibinfo {author} {\bibfnamefont {M.}~\bibnamefont {Rimbach-Russ}}, \bibinfo {author} {\bibfnamefont {S.~G.~J.}\ \bibnamefont {Philips}}, \bibinfo {author} {\bibfnamefont {X.}~\bibnamefont {Xue}},\ and\ \bibinfo {author} {\bibfnamefont {L.~M.~K.}\ \bibnamefont {Vandersypen}},\ }\bibfield  {title} {\bibinfo {title} {Simple framework for systematic high-fidelity gate operations},\ }\href {https://doi.org/10.1088/2058-9565/acf786} {\bibfield  {journal} {\bibinfo  {journal} {Quantum Science and Technology}\ }\textbf {\bibinfo {volume} {8}},\ \bibinfo {pages} {045025} (\bibinfo {year} {2023})}\BibitemShut {NoStop}%
\bibitem [{\citenamefont {White}\ \emph {et~al.}(2020)\citenamefont {White}, \citenamefont {Hill}, \citenamefont {Pollock}, \citenamefont {Hollenberg},\ and\ \citenamefont {Modi}}]{DemonstrationofnonMarkovianprocesscharacterisation}%
  \BibitemOpen
  \bibfield  {author} {\bibinfo {author} {\bibfnamefont {G.~A.~L.}\ \bibnamefont {White}}, \bibinfo {author} {\bibfnamefont {C.~D.}\ \bibnamefont {Hill}}, \bibinfo {author} {\bibfnamefont {F.~A.}\ \bibnamefont {Pollock}}, \bibinfo {author} {\bibfnamefont {L.~C.~L.}\ \bibnamefont {Hollenberg}},\ and\ \bibinfo {author} {\bibfnamefont {K.}~\bibnamefont {Modi}},\ }\bibfield  {title} {\bibinfo {title} {Demonstration of non-markovian process characterisation and control on a quantum processor},\ }\href@noop {} {\bibfield  {journal} {\bibinfo  {journal} {Nature Communications}\ }\textbf {\bibinfo {volume} {11}},\ \bibinfo {pages} {6301} (\bibinfo {year} {2020})}\BibitemShut {NoStop}%
\bibitem [{\citenamefont {Kono}\ \emph {et~al.}(2024)\citenamefont {Kono}, \citenamefont {Pan}, \citenamefont {Chegnizadeh}, \citenamefont {Wang}, \citenamefont {Youssefi}, \citenamefont {Scigliuzzo},\ and\ \citenamefont {Kippenberg}}]{Mechanicallyinducedcorrelatederrorsonsuperconducting}%
  \BibitemOpen
  \bibfield  {author} {\bibinfo {author} {\bibfnamefont {S.}~\bibnamefont {Kono}}, \bibinfo {author} {\bibfnamefont {J.}~\bibnamefont {Pan}}, \bibinfo {author} {\bibfnamefont {M.}~\bibnamefont {Chegnizadeh}}, \bibinfo {author} {\bibfnamefont {X.}~\bibnamefont {Wang}}, \bibinfo {author} {\bibfnamefont {A.}~\bibnamefont {Youssefi}}, \bibinfo {author} {\bibfnamefont {M.}~\bibnamefont {Scigliuzzo}},\ and\ \bibinfo {author} {\bibfnamefont {T.~J.}\ \bibnamefont {Kippenberg}},\ }\bibfield  {title} {\bibinfo {title} {Mechanically induced correlated errors on superconducting qubits with relaxation times exceeding 0.4 ms},\ }\href@noop {} {\bibfield  {journal} {\bibinfo  {journal} {Nature Communications}\ }\textbf {\bibinfo {volume} {15}},\ \bibinfo {pages} {3950} (\bibinfo {year} {2024})}\BibitemShut {NoStop}%
\bibitem [{\citenamefont {Gul\'acsi}\ \emph {et~al.}(2025)\citenamefont {Gul\'acsi}, \citenamefont {Kattem\"olle},\ and\ \citenamefont {Burkard}}]{k2lx-vfqp}%
  \BibitemOpen
  \bibfield  {author} {\bibinfo {author} {\bibfnamefont {B.}~\bibnamefont {Gul\'acsi}}, \bibinfo {author} {\bibfnamefont {J.}~\bibnamefont {Kattem\"olle}},\ and\ \bibinfo {author} {\bibfnamefont {G.}~\bibnamefont {Burkard}},\ }\bibfield  {title} {\bibinfo {title} {Characterizing correlated noise with single-qubit operations},\ }\href {https://doi.org/10.1103/k2lx-vfqp} {\bibfield  {journal} {\bibinfo  {journal} {Phys. Rev. Res.}\ }\textbf {\bibinfo {volume} {7}},\ \bibinfo {pages} {L042067} (\bibinfo {year} {2025})}\BibitemShut {NoStop}%
\bibitem [{\citenamefont {Gul\'acsi}\ and\ \citenamefont {Burkard}(2023)}]{PhysRevB.107.174511}%
  \BibitemOpen
  \bibfield  {author} {\bibinfo {author} {\bibfnamefont {B.}~\bibnamefont {Gul\'acsi}}\ and\ \bibinfo {author} {\bibfnamefont {G.}~\bibnamefont {Burkard}},\ }\bibfield  {title} {\bibinfo {title} {Signatures of non-markovianity of a superconducting qubit},\ }\href {https://doi.org/10.1103/PhysRevB.107.174511} {\bibfield  {journal} {\bibinfo  {journal} {Phys. Rev. B}\ }\textbf {\bibinfo {volume} {107}},\ \bibinfo {pages} {174511} (\bibinfo {year} {2023})}\BibitemShut {NoStop}%
\bibitem [{\citenamefont {Contreras-Pulido}\ and\ \citenamefont {Aguado}(2008)}]{PhysRevB.77.155420}%
  \BibitemOpen
  \bibfield  {author} {\bibinfo {author} {\bibfnamefont {L.~D.}\ \bibnamefont {Contreras-Pulido}}\ and\ \bibinfo {author} {\bibfnamefont {R.}~\bibnamefont {Aguado}},\ }\bibfield  {title} {\bibinfo {title} {Entanglement between charge qubits induced by a common dissipative environment},\ }\href {https://doi.org/10.1103/PhysRevB.77.155420} {\bibfield  {journal} {\bibinfo  {journal} {Phys. Rev. B}\ }\textbf {\bibinfo {volume} {77}},\ \bibinfo {pages} {155420} (\bibinfo {year} {2008})}\BibitemShut {NoStop}%
\bibitem [{\citenamefont {Reina}\ \emph {et~al.}(2002)\citenamefont {Reina}, \citenamefont {Quiroga},\ and\ \citenamefont {Johnson}}]{PhysRevA.65.032326}%
  \BibitemOpen
  \bibfield  {author} {\bibinfo {author} {\bibfnamefont {J.~H.}\ \bibnamefont {Reina}}, \bibinfo {author} {\bibfnamefont {L.}~\bibnamefont {Quiroga}},\ and\ \bibinfo {author} {\bibfnamefont {N.~F.}\ \bibnamefont {Johnson}},\ }\bibfield  {title} {\bibinfo {title} {Decoherence of quantum registers},\ }\href {https://doi.org/10.1103/PhysRevA.65.032326} {\bibfield  {journal} {\bibinfo  {journal} {Phys. Rev. A}\ }\textbf {\bibinfo {volume} {65}},\ \bibinfo {pages} {032326} (\bibinfo {year} {2002})}\BibitemShut {NoStop}%
\bibitem [{\citenamefont {Mundada}\ \emph {et~al.}(2019)\citenamefont {Mundada}, \citenamefont {Zhang}, \citenamefont {Hazard},\ and\ \citenamefont {Houck}}]{corsstalk}%
  \BibitemOpen
  \bibfield  {author} {\bibinfo {author} {\bibfnamefont {P.}~\bibnamefont {Mundada}}, \bibinfo {author} {\bibfnamefont {G.}~\bibnamefont {Zhang}}, \bibinfo {author} {\bibfnamefont {T.}~\bibnamefont {Hazard}},\ and\ \bibinfo {author} {\bibfnamefont {A.}~\bibnamefont {Houck}},\ }\bibfield  {title} {\bibinfo {title} {Suppression of qubit crosstalk in a tunable coupling superconducting circuit},\ }\href {https://doi.org/10.1103/PhysRevApplied.12.054023} {\bibfield  {journal} {\bibinfo  {journal} {Phys. Rev. Appl.}\ }\textbf {\bibinfo {volume} {12}},\ \bibinfo {pages} {054023} (\bibinfo {year} {2019})}\BibitemShut {NoStop}%
\bibitem [{\citenamefont {Heinz}\ and\ \citenamefont {Burkard}(2022)}]{PhysRevB.105.085414}%
  \BibitemOpen
  \bibfield  {author} {\bibinfo {author} {\bibfnamefont {I.}~\bibnamefont {Heinz}}\ and\ \bibinfo {author} {\bibfnamefont {G.}~\bibnamefont {Burkard}},\ }\bibfield  {title} {\bibinfo {title} {Crosstalk analysis for simultaneously driven two-qubit gates in spin qubit arrays},\ }\href {https://doi.org/10.1103/PhysRevB.105.085414} {\bibfield  {journal} {\bibinfo  {journal} {Phys. Rev. B}\ }\textbf {\bibinfo {volume} {105}},\ \bibinfo {pages} {085414} (\bibinfo {year} {2022})}\BibitemShut {NoStop}%
\bibitem [{\citenamefont {Heinz}\ and\ \citenamefont {Burkard}(2021)}]{PhysRevB.104.045420}%
  \BibitemOpen
  \bibfield  {author} {\bibinfo {author} {\bibfnamefont {I.}~\bibnamefont {Heinz}}\ and\ \bibinfo {author} {\bibfnamefont {G.}~\bibnamefont {Burkard}},\ }\bibfield  {title} {\bibinfo {title} {Crosstalk analysis for single-qubit and two-qubit gates in spin qubit arrays},\ }\href {https://doi.org/10.1103/PhysRevB.104.045420} {\bibfield  {journal} {\bibinfo  {journal} {Phys. Rev. B}\ }\textbf {\bibinfo {volume} {104}},\ \bibinfo {pages} {045420} (\bibinfo {year} {2021})}\BibitemShut {NoStop}%
\bibitem [{\citenamefont {Rao}\ \emph {et~al.}(2025)\citenamefont {Rao}, \citenamefont {van Straaten}, \citenamefont {John}, \citenamefont {Yu}, \citenamefont {Oosterhout}, \citenamefont {Stehouwer}, \citenamefont {Scappucci}, \citenamefont {Stewart~Jr.}, \citenamefont {Veldhorst}, \citenamefont {Borsoi},\ and\ \citenamefont {Zwolak}}]{Rao2025TowardsAT}%
  \BibitemOpen
  \bibfield  {author} {\bibinfo {author} {\bibfnamefont {A.~S.}\ \bibnamefont {Rao}}, \bibinfo {author} {\bibfnamefont {B.}~\bibnamefont {van Straaten}}, \bibinfo {author} {\bibfnamefont {V.}~\bibnamefont {John}}, \bibinfo {author} {\bibfnamefont {C.~X.}\ \bibnamefont {Yu}}, \bibinfo {author} {\bibfnamefont {S.~D.}\ \bibnamefont {Oosterhout}}, \bibinfo {author} {\bibfnamefont {L.~E.~A.}\ \bibnamefont {Stehouwer}}, \bibinfo {author} {\bibfnamefont {G.}~\bibnamefont {Scappucci}}, \bibinfo {author} {\bibfnamefont {M.~D.}\ \bibnamefont {Stewart~Jr.}}, \bibinfo {author} {\bibfnamefont {M.}~\bibnamefont {Veldhorst}}, \bibinfo {author} {\bibfnamefont {F.}~\bibnamefont {Borsoi}},\ and\ \bibinfo {author} {\bibfnamefont {J.~P.}\ \bibnamefont {Zwolak}},\ }\href@noop {} {\bibinfo {title} {Towards autonomous time-calibration of large quantum-dot devices: Detection, real-time feedback, and noise spectroscopy}} (\bibinfo {year} {2025}),\ \Eprint {https://arxiv.org/abs/2512.24894} {arXiv:2512.24894 [cond-mat.mes-hall]}
  \BibitemShut {NoStop}%
\bibitem [{\citenamefont {Mokeev}\ \emph {et~al.}(2024)\citenamefont {Mokeev}, \citenamefont {Zhang},\ and\ \citenamefont {Dobrovitski}}]{mokeev2024modelingdecoherencefidelityenhancement}%
  \BibitemOpen
  \bibfield  {author} {\bibinfo {author} {\bibfnamefont {A.~S.}\ \bibnamefont {Mokeev}}, \bibinfo {author} {\bibfnamefont {Y.-N.}\ \bibnamefont {Zhang}},\ and\ \bibinfo {author} {\bibfnamefont {V.~V.}\ \bibnamefont {Dobrovitski}},\ }\href {https://arxiv.org/abs/2409.04404} {\bibinfo {title} {Modeling of decoherence and fidelity enhancement during transport of entangled qubits}} (\bibinfo {year} {2024}),\ \Eprint {https://arxiv.org/abs/2409.04404} {arXiv:2409.04404 [cond-mat.mes-hall]} \BibitemShut {NoStop}%
\bibitem [{\citenamefont {Zhang}\ \emph {et~al.}(2025)\citenamefont {Zhang}, \citenamefont {Mokeev},\ and\ \citenamefont {Dobrovitski}}]{78sg-ttth}%
  \BibitemOpen
  \bibfield  {author} {\bibinfo {author} {\bibfnamefont {Y.-N.}\ \bibnamefont {Zhang}}, \bibinfo {author} {\bibfnamefont {A.~S.}\ \bibnamefont {Mokeev}},\ and\ \bibinfo {author} {\bibfnamefont {V.~V.}\ \bibnamefont {Dobrovitski}},\ }\bibfield  {title} {\bibinfo {title} {Decoherence and fidelity enhancement during shuttling of entangled spin qubits},\ }\href {https://doi.org/10.1103/78sg-ttth} {\bibfield  {journal} {\bibinfo  {journal} {Phys. Rev. B}\ }\textbf {\bibinfo {volume} {112}},\ \bibinfo {pages} {205301} (\bibinfo {year} {2025})}\BibitemShut {NoStop}%
\bibitem [{\citenamefont {van Riggelen-Doelman}\ \emph {et~al.}(2024)\citenamefont {van Riggelen-Doelman}, \citenamefont {Wang}, \citenamefont {de~Snoo}, \citenamefont {Lawrie}, \citenamefont {Hendrickx}, \citenamefont {Rimbach-Russ}, \citenamefont {Sammak}, \citenamefont {Scappucci}, \citenamefont {D{\'e}prez},\ and\ \citenamefont {Veldhorst}}]{vanRiggelen-Doelman2024}%
  \BibitemOpen
  \bibfield  {author} {\bibinfo {author} {\bibfnamefont {F.}~\bibnamefont {van Riggelen-Doelman}}, \bibinfo {author} {\bibfnamefont {C.-A.}\ \bibnamefont {Wang}}, \bibinfo {author} {\bibfnamefont {S.~L.}\ \bibnamefont {de~Snoo}}, \bibinfo {author} {\bibfnamefont {W.~I.~L.}\ \bibnamefont {Lawrie}}, \bibinfo {author} {\bibfnamefont {N.~W.}\ \bibnamefont {Hendrickx}}, \bibinfo {author} {\bibfnamefont {M.}~\bibnamefont {Rimbach-Russ}}, \bibinfo {author} {\bibfnamefont {A.}~\bibnamefont {Sammak}}, \bibinfo {author} {\bibfnamefont {G.}~\bibnamefont {Scappucci}}, \bibinfo {author} {\bibfnamefont {C.}~\bibnamefont {D{\'e}prez}},\ and\ \bibinfo {author} {\bibfnamefont {M.}~\bibnamefont {Veldhorst}},\ }\bibfield  {title} {\bibinfo {title} {Coherent spin qubit shuttling through germanium quantum dots},\ }\href {https://doi.org/10.1038/s41467-024-49358-y} {\bibfield  {journal} {\bibinfo  {journal} {Nature Communications}\ }\textbf {\bibinfo {volume} {15}},\ \bibinfo {pages} {5716} (\bibinfo {year} {2024})}\BibitemShut
  {NoStop}%
\bibitem [{\citenamefont {Krzywda}\ \emph {et~al.}(2026)\citenamefont {Krzywda}, \citenamefont {Matsumoto}, \citenamefont {De~Smet}, \citenamefont {Tryputen}, \citenamefont {de~Snoo}, \citenamefont {Amitonov}, \citenamefont {van Nieuwenburg}, \citenamefont {Scappucci},\ and\ \citenamefont {Vandersypen}}]{krzywda2026coherence}%
  \BibitemOpen
  \bibfield  {author} {\bibinfo {author} {\bibfnamefont {J.~A.}\ \bibnamefont {Krzywda}}, \bibinfo {author} {\bibfnamefont {Y.}~\bibnamefont {Matsumoto}}, \bibinfo {author} {\bibfnamefont {M.}~\bibnamefont {De~Smet}}, \bibinfo {author} {\bibfnamefont {L.}~\bibnamefont {Tryputen}}, \bibinfo {author} {\bibfnamefont {S.~L.}\ \bibnamefont {de~Snoo}}, \bibinfo {author} {\bibfnamefont {S.~V.}\ \bibnamefont {Amitonov}}, \bibinfo {author} {\bibfnamefont {E.}~\bibnamefont {van Nieuwenburg}}, \bibinfo {author} {\bibfnamefont {G.}~\bibnamefont {Scappucci}},\ and\ \bibinfo {author} {\bibfnamefont {L.~M.~K.}\ \bibnamefont {Vandersypen}},\ }\href@noop {} {\bibinfo {title} {Coherence protection for mobile spin qubits in silicon}} (\bibinfo {year} {2026}),\ \Eprint {https://arxiv.org/abs/2602.09179} {arXiv:2602.09179 [cond-mat.mes-hall]} \BibitemShut {NoStop}%
\bibitem [{\citenamefont {Huang}\ and\ \citenamefont {Hu}(2013)}]{PhysRevB.88.075301}%
  \BibitemOpen
  \bibfield  {author} {\bibinfo {author} {\bibfnamefont {P.}~\bibnamefont {Huang}}\ and\ \bibinfo {author} {\bibfnamefont {X.}~\bibnamefont {Hu}},\ }\bibfield  {title} {\bibinfo {title} {Spin qubit relaxation in a moving quantum dot},\ }\href {https://doi.org/10.1103/PhysRevB.88.075301} {\bibfield  {journal} {\bibinfo  {journal} {Phys. Rev. B}\ }\textbf {\bibinfo {volume} {88}},\ \bibinfo {pages} {075301} (\bibinfo {year} {2013})}\BibitemShut {NoStop}%
\bibitem [{\citenamefont {Yoneda}\ \emph {et~al.}(2023)\citenamefont {Yoneda}, \citenamefont {Rojas-Arias}, \citenamefont {Stano}, \citenamefont {Takeda}, \citenamefont {Noiri}, \citenamefont {Nakajima}, \citenamefont {Loss},\ and\ \citenamefont {Tarucha}}]{Yoneda2022}%
  \BibitemOpen
  \bibfield  {author} {\bibinfo {author} {\bibfnamefont {J.}~\bibnamefont {Yoneda}}, \bibinfo {author} {\bibfnamefont {J.~S.}\ \bibnamefont {Rojas-Arias}}, \bibinfo {author} {\bibfnamefont {P.}~\bibnamefont {Stano}}, \bibinfo {author} {\bibfnamefont {K.}~\bibnamefont {Takeda}}, \bibinfo {author} {\bibfnamefont {A.}~\bibnamefont {Noiri}}, \bibinfo {author} {\bibfnamefont {T.}~\bibnamefont {Nakajima}}, \bibinfo {author} {\bibfnamefont {D.}~\bibnamefont {Loss}},\ and\ \bibinfo {author} {\bibfnamefont {S.}~\bibnamefont {Tarucha}},\ }\bibfield  {title} {\bibinfo {title} {{Noise-correlation spectrum for a pair of spin qubits in silicon}},\ }\href {https://doi.org/10.1038/s41567-023-02238-6} {\bibfield  {journal} {\bibinfo  {journal} {Nature Phys.}\ }\textbf {\bibinfo {volume} {19}},\ \bibinfo {pages} {1793} (\bibinfo {year} {2023})}\BibitemShut {NoStop}%
\bibitem [{\citenamefont {Rojas-Arias}\ \emph {et~al.}(2023)\citenamefont {Rojas-Arias}, \citenamefont {Noiri}, \citenamefont {Stano}, \citenamefont {Nakajima}, \citenamefont {Yoneda}, \citenamefont {Takeda}, \citenamefont {Kobayashi}, \citenamefont {Sammak}, \citenamefont {Scappucci}, \citenamefont {Loss},\ and\ \citenamefont {Tarucha}}]{AriasPRA2023}%
  \BibitemOpen
  \bibfield  {author} {\bibinfo {author} {\bibfnamefont {J.}~\bibnamefont {Rojas-Arias}}, \bibinfo {author} {\bibfnamefont {A.}~\bibnamefont {Noiri}}, \bibinfo {author} {\bibfnamefont {P.}~\bibnamefont {Stano}}, \bibinfo {author} {\bibfnamefont {T.}~\bibnamefont {Nakajima}}, \bibinfo {author} {\bibfnamefont {J.}~\bibnamefont {Yoneda}}, \bibinfo {author} {\bibfnamefont {K.}~\bibnamefont {Takeda}}, \bibinfo {author} {\bibfnamefont {T.}~\bibnamefont {Kobayashi}}, \bibinfo {author} {\bibfnamefont {A.}~\bibnamefont {Sammak}}, \bibinfo {author} {\bibfnamefont {G.}~\bibnamefont {Scappucci}}, \bibinfo {author} {\bibfnamefont {D.}~\bibnamefont {Loss}},\ and\ \bibinfo {author} {\bibfnamefont {S.}~\bibnamefont {Tarucha}},\ }\bibfield  {title} {\bibinfo {title} {Spatial noise correlations beyond nearest neighbors in ${}^{28}\mathrm{Si}/${{Si-Ge}} spin qubits},\ }\href {https://doi.org/10.1103/PhysRevApplied.20.054024} {\bibfield  {journal} {\bibinfo  {journal} {Phys. Rev. Appl.}\ }\textbf {\bibinfo {volume} {20}},\
  \bibinfo {pages} {054024} (\bibinfo {year} {2023})}\BibitemShut {NoStop}%
\bibitem [{\citenamefont {Rojas-Arias}\ \emph {et~al.}(2026{\natexlab{a}})\citenamefont {Rojas-Arias}, \citenamefont {Noiri}, \citenamefont {Yoneda}, \citenamefont {Stano}, \citenamefont {Nakajima}, \citenamefont {Takeda}, \citenamefont {Kobayashi}, \citenamefont {Scappucci}, \citenamefont {Tarucha},\ and\ \citenamefont {Loss}}]{RojasArias2025InferringCN}%
  \BibitemOpen
  \bibfield  {author} {\bibinfo {author} {\bibfnamefont {J.~S.}\ \bibnamefont {Rojas-Arias}}, \bibinfo {author} {\bibfnamefont {A.}~\bibnamefont {Noiri}}, \bibinfo {author} {\bibfnamefont {J.}~\bibnamefont {Yoneda}}, \bibinfo {author} {\bibfnamefont {P.}~\bibnamefont {Stano}}, \bibinfo {author} {\bibfnamefont {T.}~\bibnamefont {Nakajima}}, \bibinfo {author} {\bibfnamefont {K.}~\bibnamefont {Takeda}}, \bibinfo {author} {\bibfnamefont {T.}~\bibnamefont {Kobayashi}}, \bibinfo {author} {\bibfnamefont {G.}~\bibnamefont {Scappucci}}, \bibinfo {author} {\bibfnamefont {S.}~\bibnamefont {Tarucha}},\ and\ \bibinfo {author} {\bibfnamefont {D.}~\bibnamefont {Loss}},\ }\bibfield  {title} {\bibinfo {title} {Inferring charge-noise source locations from correlations in spin qubits},\ }\href {https://doi.org/10.1103/glp5-bb5b} {\bibfield  {journal} {\bibinfo  {journal} {Phys. Rev. Lett.}\ }\textbf {\bibinfo {volume} {136}},\ \bibinfo {pages} {027001} (\bibinfo {year} {2026}{\natexlab{a}})}\BibitemShut {NoStop}%
\bibitem [{\citenamefont {Rojas-Arias}\ \emph {et~al.}(2026{\natexlab{b}})\citenamefont {Rojas-Arias}, \citenamefont {Kojima}, \citenamefont {Takeda}, \citenamefont {Stano}, \citenamefont {Nakajima}, \citenamefont {Yoneda}, \citenamefont {Noiri}, \citenamefont {Kobayashi}, \citenamefont {Loss},\ and\ \citenamefont {Tarucha}}]{RojasArias2024TheOO}%
  \BibitemOpen
  \bibfield  {author} {\bibinfo {author} {\bibfnamefont {J.~S.}\ \bibnamefont {Rojas-Arias}}, \bibinfo {author} {\bibfnamefont {Y.}~\bibnamefont {Kojima}}, \bibinfo {author} {\bibfnamefont {K.}~\bibnamefont {Takeda}}, \bibinfo {author} {\bibfnamefont {P.}~\bibnamefont {Stano}}, \bibinfo {author} {\bibfnamefont {T.}~\bibnamefont {Nakajima}}, \bibinfo {author} {\bibfnamefont {J.}~\bibnamefont {Yoneda}}, \bibinfo {author} {\bibfnamefont {A.}~\bibnamefont {Noiri}}, \bibinfo {author} {\bibfnamefont {T.}~\bibnamefont {Kobayashi}}, \bibinfo {author} {\bibfnamefont {D.}~\bibnamefont {Loss}},\ and\ \bibinfo {author} {\bibfnamefont {S.}~\bibnamefont {Tarucha}},\ }\bibfield  {title} {\bibinfo {title} {The origins of noise in the zeeman splitting of spin qubits in natural-silicon devices},\ }\href {https://doi.org/10.1038/s41534-025-01150-6} {\bibfield  {journal} {\bibinfo  {journal} {npj Quantum Information}\ }\textbf {\bibinfo {volume} {12}},\ \bibinfo {pages} {9} (\bibinfo {year} {2026}{\natexlab{b}})}\BibitemShut
  {NoStop}%
\bibitem [{\citenamefont {Donnelly}\ \emph {et~al.}(2025)\citenamefont {Donnelly}, \citenamefont {Rowlands}, \citenamefont {Kranz}, \citenamefont {Hsueh}, \citenamefont {Chung}, \citenamefont {Timofeev}, \citenamefont {Geng}, \citenamefont {Singh-Gregory}, \citenamefont {Gorman}, \citenamefont {Keizer}, \citenamefont {Rahman},\ and\ \citenamefont {Simmons}}]{Donnelly2025}%
  \BibitemOpen
  \bibfield  {author} {\bibinfo {author} {\bibfnamefont {M.}~\bibnamefont {Donnelly}}, \bibinfo {author} {\bibfnamefont {J.}~\bibnamefont {Rowlands}}, \bibinfo {author} {\bibfnamefont {L.}~\bibnamefont {Kranz}}, \bibinfo {author} {\bibfnamefont {Y.}~\bibnamefont {Hsueh}}, \bibinfo {author} {\bibfnamefont {Y.}~\bibnamefont {Chung}}, \bibinfo {author} {\bibfnamefont {A.}~\bibnamefont {Timofeev}}, \bibinfo {author} {\bibfnamefont {H.}~\bibnamefont {Geng}}, \bibinfo {author} {\bibfnamefont {P.}~\bibnamefont {Singh-Gregory}}, \bibinfo {author} {\bibfnamefont {S.}~\bibnamefont {Gorman}}, \bibinfo {author} {\bibfnamefont {J.}~\bibnamefont {Keizer}}, \bibinfo {author} {\bibfnamefont {R.}~\bibnamefont {Rahman}},\ and\ \bibinfo {author} {\bibfnamefont {M.}~\bibnamefont {Simmons}},\ }\bibfield  {title} {\bibinfo {title} {Noise correlations in an atom-based quantum dot array},\ }\href {https://doi.org/10.1103/kr2l-c97c} {\bibfield  {journal} {\bibinfo  {journal} {Phys. Rev. Appl.}\ }\textbf {\bibinfo {volume} {23}},\
  \bibinfo {pages} {064058} (\bibinfo {year} {2025})}\BibitemShut {NoStop}%
\bibitem [{\citenamefont {von L\"upke}\ \emph {et~al.}(2020)\citenamefont {von L\"upke}, \citenamefont {Beaudoin}, \citenamefont {Norris}, \citenamefont {Sung}, \citenamefont {Winik}, \citenamefont {Qiu}, \citenamefont {Kjaergaard}, \citenamefont {Kim}, \citenamefont {Yoder}, \citenamefont {Gustavsson}, \citenamefont {Viola},\ and\ \citenamefont {Oliver}}]{superconductingqubits}%
  \BibitemOpen
  \bibfield  {author} {\bibinfo {author} {\bibfnamefont {U.}~\bibnamefont {von L\"upke}}, \bibinfo {author} {\bibfnamefont {F.}~\bibnamefont {Beaudoin}}, \bibinfo {author} {\bibfnamefont {L.~M.}\ \bibnamefont {Norris}}, \bibinfo {author} {\bibfnamefont {Y.}~\bibnamefont {Sung}}, \bibinfo {author} {\bibfnamefont {R.}~\bibnamefont {Winik}}, \bibinfo {author} {\bibfnamefont {J.~Y.}\ \bibnamefont {Qiu}}, \bibinfo {author} {\bibfnamefont {M.}~\bibnamefont {Kjaergaard}}, \bibinfo {author} {\bibfnamefont {D.}~\bibnamefont {Kim}}, \bibinfo {author} {\bibfnamefont {J.}~\bibnamefont {Yoder}}, \bibinfo {author} {\bibfnamefont {S.}~\bibnamefont {Gustavsson}}, \bibinfo {author} {\bibfnamefont {L.}~\bibnamefont {Viola}},\ and\ \bibinfo {author} {\bibfnamefont {W.~D.}\ \bibnamefont {Oliver}},\ }\bibfield  {title} {\bibinfo {title} {Two-qubit spectroscopy of spatiotemporally correlated quantum noise in superconducting qubits},\ }\href {https://doi.org/10.1103/PRXQuantum.1.010305} {\bibfield  {journal} {\bibinfo  {journal}
  {PRX Quantum}\ }\textbf {\bibinfo {volume} {1}},\ \bibinfo {pages} {010305} (\bibinfo {year} {2020})}\BibitemShut {NoStop}%
\bibitem [{\citenamefont {Wilen}\ \emph {et~al.}(2021)\citenamefont {Wilen}, \citenamefont {Abdullah}, \citenamefont {Kurinsky}, \citenamefont {Stanford}, \citenamefont {Cardani}, \citenamefont {D'Imperio}, \citenamefont {Tomei}, \citenamefont {Faoro}, \citenamefont {Ioffe}, \citenamefont {Liu}, \citenamefont {Opremcak}, \citenamefont {Christensen}, \citenamefont {DuBois},\ and\ \citenamefont {McDermott}}]{QECGoogle}%
  \BibitemOpen
  \bibfield  {author} {\bibinfo {author} {\bibfnamefont {C.~D.}\ \bibnamefont {Wilen}}, \bibinfo {author} {\bibfnamefont {S.}~\bibnamefont {Abdullah}}, \bibinfo {author} {\bibfnamefont {N.~A.}\ \bibnamefont {Kurinsky}}, \bibinfo {author} {\bibfnamefont {C.}~\bibnamefont {Stanford}}, \bibinfo {author} {\bibfnamefont {L.}~\bibnamefont {Cardani}}, \bibinfo {author} {\bibfnamefont {G.}~\bibnamefont {D'Imperio}}, \bibinfo {author} {\bibfnamefont {C.}~\bibnamefont {Tomei}}, \bibinfo {author} {\bibfnamefont {L.}~\bibnamefont {Faoro}}, \bibinfo {author} {\bibfnamefont {L.~B.}\ \bibnamefont {Ioffe}}, \bibinfo {author} {\bibfnamefont {C.~H.}\ \bibnamefont {Liu}}, \bibinfo {author} {\bibfnamefont {A.}~\bibnamefont {Opremcak}}, \bibinfo {author} {\bibfnamefont {B.~G.}\ \bibnamefont {Christensen}}, \bibinfo {author} {\bibfnamefont {J.~L.}\ \bibnamefont {DuBois}},\ and\ \bibinfo {author} {\bibfnamefont {R.}~\bibnamefont {McDermott}},\ }\bibfield  {title} {\bibinfo {title} {Correlated charge noise and relaxation errors in
  superconducting qubits},\ }\href@noop {} {\bibfield  {journal} {\bibinfo  {journal} {Nature}\ }\textbf {\bibinfo {volume} {594}},\ \bibinfo {pages} {369} (\bibinfo {year} {2021})}\BibitemShut {NoStop}%
\bibitem [{\citenamefont {Clemens}\ \emph {et~al.}(2004)\citenamefont {Clemens}, \citenamefont {Siddiqui},\ and\ \citenamefont {Gea-Banacloche}}]{QECcorr2004}%
  \BibitemOpen
  \bibfield  {author} {\bibinfo {author} {\bibfnamefont {J.~P.}\ \bibnamefont {Clemens}}, \bibinfo {author} {\bibfnamefont {S.}~\bibnamefont {Siddiqui}},\ and\ \bibinfo {author} {\bibfnamefont {J.}~\bibnamefont {Gea-Banacloche}},\ }\bibfield  {title} {\bibinfo {title} {Quantum error correction against correlated noise},\ }\href {https://doi.org/10.1103/PhysRevA.69.062313} {\bibfield  {journal} {\bibinfo  {journal} {Phys. Rev. A}\ }\textbf {\bibinfo {volume} {69}},\ \bibinfo {pages} {062313} (\bibinfo {year} {2004})}\BibitemShut {NoStop}%
\bibitem [{\citenamefont {Ahsan}\ and\ \citenamefont {Naqvi}(2018)}]{topoQECcorr2018}%
  \BibitemOpen
  \bibfield  {author} {\bibinfo {author} {\bibfnamefont {M.}~\bibnamefont {Ahsan}}\ and\ \bibinfo {author} {\bibfnamefont {S.~A.~Z.}\ \bibnamefont {Naqvi}},\ }\bibfield  {title} {\bibinfo {title} {Performance of topological quantum error correction in the presence of correlated noise},\ }\href {https://api.semanticscholar.org/CorpusID:52010690} {\bibfield  {journal} {\bibinfo  {journal} {Quantum Inf. Comput.}\ }\textbf {\bibinfo {volume} {18}},\ \bibinfo {pages} {743} (\bibinfo {year} {2018})}\BibitemShut {NoStop}%
\bibitem [{\citenamefont {Nickerson}\ and\ \citenamefont {Brown}(2019)}]{Nickerson2019analysingcorrelated}%
  \BibitemOpen
  \bibfield  {author} {\bibinfo {author} {\bibfnamefont {N.~H.}\ \bibnamefont {Nickerson}}\ and\ \bibinfo {author} {\bibfnamefont {B.~J.}\ \bibnamefont {Brown}},\ }\bibfield  {title} {\bibinfo {title} {Analysing correlated noise on the surface code using adaptive decoding algorithms},\ }\href {https://doi.org/10.22331/q-2019-04-08-131} {\bibfield  {journal} {\bibinfo  {journal} {{Quantum}}\ }\textbf {\bibinfo {volume} {3}},\ \bibinfo {pages} {131} (\bibinfo {year} {2019})}\BibitemShut {NoStop}%
\bibitem [{\citenamefont {Klesse}\ and\ \citenamefont {Frank}(2005)}]{QECinspatcorr2005}%
  \BibitemOpen
  \bibfield  {author} {\bibinfo {author} {\bibfnamefont {R.}~\bibnamefont {Klesse}}\ and\ \bibinfo {author} {\bibfnamefont {S.}~\bibnamefont {Frank}},\ }\bibfield  {title} {\bibinfo {title} {Quantum error correction in spatially correlated quantum noise},\ }\href {https://doi.org/10.1103/PhysRevLett.95.230503} {\bibfield  {journal} {\bibinfo  {journal} {Phys. Rev. Lett.}\ }\textbf {\bibinfo {volume} {95}},\ \bibinfo {pages} {230503} (\bibinfo {year} {2005})}\BibitemShut {NoStop}%
\bibitem [{\citenamefont {Hutter}\ and\ \citenamefont {Loss}(2014)}]{Breakdownofsurfacecode}%
  \BibitemOpen
  \bibfield  {author} {\bibinfo {author} {\bibfnamefont {A.}~\bibnamefont {Hutter}}\ and\ \bibinfo {author} {\bibfnamefont {D.}~\bibnamefont {Loss}},\ }\bibfield  {title} {\bibinfo {title} {Breakdown of surface-code error correction due to coupling to a bosonic bath},\ }\href {https://doi.org/10.1103/PhysRevA.89.042334} {\bibfield  {journal} {\bibinfo  {journal} {Phys. Rev. A}\ }\textbf {\bibinfo {volume} {89}},\ \bibinfo {pages} {042334} (\bibinfo {year} {2014})}\BibitemShut {NoStop}%
\bibitem [{\citenamefont {{Google Quantum AI and Collaborators}}(2025)}]{GoogleAI}%
  \BibitemOpen
  \bibfield  {author} {\bibinfo {author} {\bibnamefont {{Google Quantum AI and Collaborators}}},\ }\bibfield  {title} {\bibinfo {title} {Quantum error correction below the surface code threshold},\ }\href@noop {} {\bibfield  {journal} {\bibinfo  {journal} {Nature}\ }\textbf {\bibinfo {volume} {638}},\ \bibinfo {pages} {920} (\bibinfo {year} {2025})}\BibitemShut {NoStop}%
\bibitem [{\citenamefont {{Google Quantum AI and Collaborators}}(2021)}]{Chen2021}%
  \BibitemOpen
  \bibfield  {author} {\bibinfo {author} {\bibnamefont {{Google Quantum AI and Collaborators}}},\ }\bibfield  {title} {\bibinfo {title} {Exponential suppression of bit or phase errors with cyclic error correction},\ }\href@noop {} {\bibfield  {journal} {\bibinfo  {journal} {Nature}\ }\textbf {\bibinfo {volume} {595}},\ \bibinfo {pages} {383} (\bibinfo {year} {2021})}\BibitemShut {NoStop}%
\bibitem [{\citenamefont {Zou}\ \emph {et~al.}(2024)\citenamefont {Zou}, \citenamefont {Bosco},\ and\ \citenamefont {Loss}}]{Spatiallycorrelated}%
  \BibitemOpen
  \bibfield  {author} {\bibinfo {author} {\bibfnamefont {J.}~\bibnamefont {Zou}}, \bibinfo {author} {\bibfnamefont {S.}~\bibnamefont {Bosco}},\ and\ \bibinfo {author} {\bibfnamefont {D.}~\bibnamefont {Loss}},\ }\bibfield  {title} {\bibinfo {title} {Spatially correlated classical and quantum noise in driven qubits},\ }\href {https://doi.org/10.1038/s41534-024-00842-9} {\bibfield  {journal} {\bibinfo  {journal} {npj Quantum Information}\ }\textbf {\bibinfo {volume} {10}},\ \bibinfo {pages} {46} (\bibinfo {year} {2024})}\BibitemShut {NoStop}%
\bibitem [{\citenamefont {Kofman}\ and\ \citenamefont {Kurizki}(2001)}]{PhysRevLett.87.270405}%
  \BibitemOpen
  \bibfield  {author} {\bibinfo {author} {\bibfnamefont {A.~G.}\ \bibnamefont {Kofman}}\ and\ \bibinfo {author} {\bibfnamefont {G.}~\bibnamefont {Kurizki}},\ }\bibfield  {title} {\bibinfo {title} {Universal dynamical control of quantum mechanical decay: Modulation of the coupling to the continuum},\ }\href {https://doi.org/10.1103/PhysRevLett.87.270405} {\bibfield  {journal} {\bibinfo  {journal} {Phys. Rev. Lett.}\ }\textbf {\bibinfo {volume} {87}},\ \bibinfo {pages} {270405} (\bibinfo {year} {2001})}\BibitemShut {NoStop}%
\bibitem [{\citenamefont {Green}\ \emph {et~al.}(2012)\citenamefont {Green}, \citenamefont {Uys},\ and\ \citenamefont {Biercuk}}]{PhysRevLett.109.020501}%
  \BibitemOpen
  \bibfield  {author} {\bibinfo {author} {\bibfnamefont {T.}~\bibnamefont {Green}}, \bibinfo {author} {\bibfnamefont {H.}~\bibnamefont {Uys}},\ and\ \bibinfo {author} {\bibfnamefont {M.~J.}\ \bibnamefont {Biercuk}},\ }\bibfield  {title} {\bibinfo {title} {High-order noise filtering in nontrivial quantum logic gates},\ }\href {https://doi.org/10.1103/PhysRevLett.109.020501} {\bibfield  {journal} {\bibinfo  {journal} {Phys. Rev. Lett.}\ }\textbf {\bibinfo {volume} {109}},\ \bibinfo {pages} {020501} (\bibinfo {year} {2012})}\BibitemShut {NoStop}%
\bibitem [{\citenamefont {Paz-Silva}\ and\ \citenamefont {Viola}(2014)}]{PhysRevLett.113.250501}%
  \BibitemOpen
  \bibfield  {author} {\bibinfo {author} {\bibfnamefont {G.~A.}\ \bibnamefont {Paz-Silva}}\ and\ \bibinfo {author} {\bibfnamefont {L.}~\bibnamefont {Viola}},\ }\bibfield  {title} {\bibinfo {title} {General transfer-function approach to noise filtering in open-loop quantum control},\ }\href {https://doi.org/10.1103/PhysRevLett.113.250501} {\bibfield  {journal} {\bibinfo  {journal} {Phys. Rev. Lett.}\ }\textbf {\bibinfo {volume} {113}},\ \bibinfo {pages} {250501} (\bibinfo {year} {2014})}\BibitemShut {NoStop}%
\bibitem [{\citenamefont {Green}\ \emph {et~al.}(2013)\citenamefont {Green}, \citenamefont {Sastrawan}, \citenamefont {Uys},\ and\ \citenamefont {Biercuk}}]{Green_2013}%
  \BibitemOpen
  \bibfield  {author} {\bibinfo {author} {\bibfnamefont {T.~J.}\ \bibnamefont {Green}}, \bibinfo {author} {\bibfnamefont {J.}~\bibnamefont {Sastrawan}}, \bibinfo {author} {\bibfnamefont {H.}~\bibnamefont {Uys}},\ and\ \bibinfo {author} {\bibfnamefont {M.~J.}\ \bibnamefont {Biercuk}},\ }\bibfield  {title} {\bibinfo {title} {Arbitrary quantum control of qubits in the presence of universal noise},\ }\href {https://doi.org/10.1088/1367-2630/15/9/095004} {\bibfield  {journal} {\bibinfo  {journal} {New Journal of Physics}\ }\textbf {\bibinfo {volume} {15}},\ \bibinfo {pages} {095004} (\bibinfo {year} {2013})}\BibitemShut {NoStop}%
\bibitem [{\citenamefont {Paz-Silva}\ \emph {et~al.}(2017)\citenamefont {Paz-Silva}, \citenamefont {Norris},\ and\ \citenamefont {Viola}}]{PhysRevA.95.022121}%
  \BibitemOpen
  \bibfield  {author} {\bibinfo {author} {\bibfnamefont {G.~A.}\ \bibnamefont {Paz-Silva}}, \bibinfo {author} {\bibfnamefont {L.~M.}\ \bibnamefont {Norris}},\ and\ \bibinfo {author} {\bibfnamefont {L.}~\bibnamefont {Viola}},\ }\bibfield  {title} {\bibinfo {title} {Multiqubit spectroscopy of gaussian quantum noise},\ }\href {https://doi.org/10.1103/PhysRevA.95.022121} {\bibfield  {journal} {\bibinfo  {journal} {Phys. Rev. A}\ }\textbf {\bibinfo {volume} {95}},\ \bibinfo {pages} {022121} (\bibinfo {year} {2017})}\BibitemShut {NoStop}%
\bibitem [{\citenamefont {Cerfontaine}\ \emph {et~al.}(2021)\citenamefont {Cerfontaine}, \citenamefont {Hangleiter},\ and\ \citenamefont {Bluhm}}]{PhysRevLett.127.170403}%
  \BibitemOpen
  \bibfield  {author} {\bibinfo {author} {\bibfnamefont {P.}~\bibnamefont {Cerfontaine}}, \bibinfo {author} {\bibfnamefont {T.}~\bibnamefont {Hangleiter}},\ and\ \bibinfo {author} {\bibfnamefont {H.}~\bibnamefont {Bluhm}},\ }\bibfield  {title} {\bibinfo {title} {Filter functions for quantum processes under correlated noise},\ }\href {https://doi.org/10.1103/PhysRevLett.127.170403} {\bibfield  {journal} {\bibinfo  {journal} {Phys. Rev. Lett.}\ }\textbf {\bibinfo {volume} {127}},\ \bibinfo {pages} {170403} (\bibinfo {year} {2021})}\BibitemShut {NoStop}%
\bibitem [{\citenamefont {Hangleiter}\ \emph {et~al.}(2021)\citenamefont {Hangleiter}, \citenamefont {Cerfontaine},\ and\ \citenamefont {Bluhm}}]{PhysRevResearch.3.043047}%
  \BibitemOpen
  \bibfield  {author} {\bibinfo {author} {\bibfnamefont {T.}~\bibnamefont {Hangleiter}}, \bibinfo {author} {\bibfnamefont {P.}~\bibnamefont {Cerfontaine}},\ and\ \bibinfo {author} {\bibfnamefont {H.}~\bibnamefont {Bluhm}},\ }\bibfield  {title} {\bibinfo {title} {Filter-function formalism and software package to compute quantum processes of gate sequences for classical non-markovian noise},\ }\href {https://doi.org/10.1103/PhysRevResearch.3.043047} {\bibfield  {journal} {\bibinfo  {journal} {Phys. Rev. Res.}\ }\textbf {\bibinfo {volume} {3}},\ \bibinfo {pages} {043047} (\bibinfo {year} {2021})}\BibitemShut {NoStop}%
\bibitem [{\citenamefont {Loss}\ and\ \citenamefont {DiVincenzo}(1998)}]{QuantumcomputationwithQD}%
  \BibitemOpen
  \bibfield  {author} {\bibinfo {author} {\bibfnamefont {D.}~\bibnamefont {Loss}}\ and\ \bibinfo {author} {\bibfnamefont {D.~P.}\ \bibnamefont {DiVincenzo}},\ }\bibfield  {title} {\bibinfo {title} {Quantum computation with quantum dots},\ }\href {https://doi.org/10.1103/PhysRevA.57.120} {\bibfield  {journal} {\bibinfo  {journal} {Phys. Rev. A}\ }\textbf {\bibinfo {volume} {57}},\ \bibinfo {pages} {120} (\bibinfo {year} {1998})}\BibitemShut {NoStop}%
\bibitem [{\citenamefont {Burkard}\ \emph {et~al.}(1999)\citenamefont {Burkard}, \citenamefont {Loss},\ and\ \citenamefont {DiVincenzo}}]{Coupledquantumdotsasquantumgates}%
  \BibitemOpen
  \bibfield  {author} {\bibinfo {author} {\bibfnamefont {G.}~\bibnamefont {Burkard}}, \bibinfo {author} {\bibfnamefont {D.}~\bibnamefont {Loss}},\ and\ \bibinfo {author} {\bibfnamefont {D.~P.}\ \bibnamefont {DiVincenzo}},\ }\bibfield  {title} {\bibinfo {title} {Coupled quantum dots as quantum gates},\ }\href {https://doi.org/10.1103/PhysRevB.59.2070} {\bibfield  {journal} {\bibinfo  {journal} {Phys. Rev. B}\ }\textbf {\bibinfo {volume} {59}},\ \bibinfo {pages} {2070} (\bibinfo {year} {1999})}\BibitemShut {NoStop}%
\bibitem [{\citenamefont {Cirac}\ and\ \citenamefont {Zoller}(1995)}]{QCwithColdTrappedIons}%
  \BibitemOpen
  \bibfield  {author} {\bibinfo {author} {\bibfnamefont {J.~I.}\ \bibnamefont {Cirac}}\ and\ \bibinfo {author} {\bibfnamefont {P.}~\bibnamefont {Zoller}},\ }\bibfield  {title} {\bibinfo {title} {Quantum computations with cold trapped ions},\ }\href {https://doi.org/10.1103/PhysRevLett.74.4091} {\bibfield  {journal} {\bibinfo  {journal} {Phys. Rev. Lett.}\ }\textbf {\bibinfo {volume} {74}},\ \bibinfo {pages} {4091} (\bibinfo {year} {1995})}\BibitemShut {NoStop}%
\bibitem [{\citenamefont {Fern{\'a}ndez-Fern{\'a}ndez}\ \emph {et~al.}(2025)\citenamefont {Fern{\'a}ndez-Fern{\'a}ndez}, \citenamefont {Matsumoto}, \citenamefont {Vandersypen}, \citenamefont {Platero},\ and\ \citenamefont {Bosco}}]{SOenabledrealizationofarbitrary2qubitgates}%
  \BibitemOpen
  \bibfield  {author} {\bibinfo {author} {\bibfnamefont {D.}~\bibnamefont {Fern{\'a}ndez-Fern{\'a}ndez}}, \bibinfo {author} {\bibfnamefont {Y.}~\bibnamefont {Matsumoto}}, \bibinfo {author} {\bibfnamefont {L.~M.~K.}\ \bibnamefont {Vandersypen}}, \bibinfo {author} {\bibfnamefont {G.}~\bibnamefont {Platero}},\ and\ \bibinfo {author} {\bibfnamefont {S.}~\bibnamefont {Bosco}},\ }\bibfield  {title} {\bibinfo {title} {{Spin-orbit-enabled realization of arbitrary two-qubit gates on moving spins}},\ }\href@noop {} {\  (\bibinfo {year} {2025})},\ \Eprint {https://arxiv.org/abs/2508.08394} {arXiv:2508.08394 [cond-mat.mes-hall]} \BibitemShut {NoStop}%
\bibitem [{\citenamefont {Nielsen}(2002)}]{AGF}%
  \BibitemOpen
  \bibfield  {author} {\bibinfo {author} {\bibfnamefont {M.~A.}\ \bibnamefont {Nielsen}},\ }\bibfield  {title} {\bibinfo {title} {A simple formula for the average gate fidelity of a quantum dynamical operation},\ }\href {https://doi.org/https://doi.org/10.1016/S0375-9601(02)01272-0} {\bibfield  {journal} {\bibinfo  {journal} {Physics Letters A}\ }\textbf {\bibinfo {volume} {303}},\ \bibinfo {pages} {249} (\bibinfo {year} {2002})}\BibitemShut {NoStop}%
\bibitem [{\citenamefont {Abad}\ \emph {et~al.}(2022)\citenamefont {Abad}, \citenamefont {Fern\'andez-Pend\'as}, \citenamefont {Frisk~Kockum},\ and\ \citenamefont {Johansson}}]{Abad}%
  \BibitemOpen
  \bibfield  {author} {\bibinfo {author} {\bibfnamefont {T.}~\bibnamefont {Abad}}, \bibinfo {author} {\bibfnamefont {J.}~\bibnamefont {Fern\'andez-Pend\'as}}, \bibinfo {author} {\bibfnamefont {A.}~\bibnamefont {Frisk~Kockum}},\ and\ \bibinfo {author} {\bibfnamefont {G.}~\bibnamefont {Johansson}},\ }\bibfield  {title} {\bibinfo {title} {Universal fidelity reduction of quantum operations from weak dissipation},\ }\href {https://doi.org/10.1103/PhysRevLett.129.150504} {\bibfield  {journal} {\bibinfo  {journal} {Phys. Rev. Lett.}\ }\textbf {\bibinfo {volume} {129}},\ \bibinfo {pages} {150504} (\bibinfo {year} {2022})}\BibitemShut {NoStop}%
\bibitem [{SM()}]{SM}%
  \BibitemOpen
  \href@noop {} {\bibinfo {title} {See supplemental material at \url{URL_will_be_inserted_by_publisher} for additional details on the derivation of {Eqs.}~(3), (4), and (9), phase of the noise correlator, implementation of different gates in spin-qubit systems, the corresponding filter functions, and lorentzian noise infidelity results}}\BibitemShut {NoStop}%
\bibitem [{\citenamefont {Khodjasteh}\ and\ \citenamefont {Lidar}(2008)}]{PhysRevA.78.012355}%
  \BibitemOpen
  \bibfield  {author} {\bibinfo {author} {\bibfnamefont {K.}~\bibnamefont {Khodjasteh}}\ and\ \bibinfo {author} {\bibfnamefont {D.~A.}\ \bibnamefont {Lidar}},\ }\bibfield  {title} {\bibinfo {title} {Rigorous bounds on the performance of a hybrid dynamical-decoupling quantum-computing scheme},\ }\href {https://doi.org/10.1103/PhysRevA.78.012355} {\bibfield  {journal} {\bibinfo  {journal} {Phys. Rev. A}\ }\textbf {\bibinfo {volume} {78}},\ \bibinfo {pages} {012355} (\bibinfo {year} {2008})}\BibitemShut {NoStop}%
\bibitem [{\citenamefont {Jirovec}\ \emph {et~al.}(2025)\citenamefont {Jirovec}, \citenamefont {Fari\~na}, \citenamefont {Reale}, \citenamefont {Oosterhout}, \citenamefont {Zhang}, \citenamefont {de~Snoo}, \citenamefont {Sammak}, \citenamefont {Scappucci}, \citenamefont {Veldhorst},\ and\ \citenamefont {Vandersypen}}]{exchangecrosstalk}%
  \BibitemOpen
  \bibfield  {author} {\bibinfo {author} {\bibfnamefont {D.}~\bibnamefont {Jirovec}}, \bibinfo {author} {\bibfnamefont {P.~C.}\ \bibnamefont {Fari\~na}}, \bibinfo {author} {\bibfnamefont {S.}~\bibnamefont {Reale}}, \bibinfo {author} {\bibfnamefont {S.~D.}\ \bibnamefont {Oosterhout}}, \bibinfo {author} {\bibfnamefont {X.}~\bibnamefont {Zhang}}, \bibinfo {author} {\bibfnamefont {S.}~\bibnamefont {de~Snoo}}, \bibinfo {author} {\bibfnamefont {A.}~\bibnamefont {Sammak}}, \bibinfo {author} {\bibfnamefont {G.}~\bibnamefont {Scappucci}}, \bibinfo {author} {\bibfnamefont {M.}~\bibnamefont {Veldhorst}},\ and\ \bibinfo {author} {\bibfnamefont {L.~M.~K.}\ \bibnamefont {Vandersypen}},\ }\bibfield  {title} {\bibinfo {title} {Mitigation of exchange crosstalk in dense quantum dot arrays},\ }\href {https://doi.org/10.1103/qgnt-n527} {\bibfield  {journal} {\bibinfo  {journal} {Phys. Rev. Appl.}\ }\textbf {\bibinfo {volume} {24}},\ \bibinfo {pages} {034051} (\bibinfo {year} {2025})}\BibitemShut {NoStop}%
\bibitem [{\citenamefont {Burkard}\ \emph {et~al.}(2023)\citenamefont {Burkard}, \citenamefont {Ladd}, \citenamefont {Pan}, \citenamefont {Nichol},\ and\ \citenamefont {Petta}}]{RevModPhys.95.025003}%
  \BibitemOpen
  \bibfield  {author} {\bibinfo {author} {\bibfnamefont {G.}~\bibnamefont {Burkard}}, \bibinfo {author} {\bibfnamefont {T.~D.}\ \bibnamefont {Ladd}}, \bibinfo {author} {\bibfnamefont {A.}~\bibnamefont {Pan}}, \bibinfo {author} {\bibfnamefont {J.~M.}\ \bibnamefont {Nichol}},\ and\ \bibinfo {author} {\bibfnamefont {J.~R.}\ \bibnamefont {Petta}},\ }\bibfield  {title} {\bibinfo {title} {Semiconductor spin qubits},\ }\href {https://doi.org/10.1103/RevModPhys.95.025003} {\bibfield  {journal} {\bibinfo  {journal} {Rev. Mod. Phys.}\ }\textbf {\bibinfo {volume} {95}},\ \bibinfo {pages} {025003} (\bibinfo {year} {2023})}\BibitemShut {NoStop}%
\bibitem [{\citenamefont {Chatterjee}\ \emph {et~al.}(2021)\citenamefont {Chatterjee}, \citenamefont {Stevenson}, \citenamefont {De~Franceschi}, \citenamefont {Morello}, \citenamefont {de~Leon},\ and\ \citenamefont {Kuemmeth}}]{Chatterjee2021}%
  \BibitemOpen
  \bibfield  {author} {\bibinfo {author} {\bibfnamefont {A.}~\bibnamefont {Chatterjee}}, \bibinfo {author} {\bibfnamefont {P.}~\bibnamefont {Stevenson}}, \bibinfo {author} {\bibfnamefont {S.}~\bibnamefont {De~Franceschi}}, \bibinfo {author} {\bibfnamefont {A.}~\bibnamefont {Morello}}, \bibinfo {author} {\bibfnamefont {N.~P.}\ \bibnamefont {de~Leon}},\ and\ \bibinfo {author} {\bibfnamefont {F.}~\bibnamefont {Kuemmeth}},\ }\bibfield  {title} {\bibinfo {title} {Semiconductor qubits in practice},\ }\href {https://doi.org/10.1038/s42254-021-00283-9} {\bibfield  {journal} {\bibinfo  {journal} {Nature Reviews Physics}\ }\textbf {\bibinfo {volume} {3}},\ \bibinfo {pages} {157} (\bibinfo {year} {2021})}\BibitemShut {NoStop}%
\bibitem [{\citenamefont {Hanson}\ \emph {et~al.}(2007)\citenamefont {Hanson}, \citenamefont {Kouwenhoven}, \citenamefont {Petta}, \citenamefont {Tarucha},\ and\ \citenamefont {Vandersypen}}]{RevModPhys.79.1217}%
  \BibitemOpen
  \bibfield  {author} {\bibinfo {author} {\bibfnamefont {R.}~\bibnamefont {Hanson}}, \bibinfo {author} {\bibfnamefont {L.~P.}\ \bibnamefont {Kouwenhoven}}, \bibinfo {author} {\bibfnamefont {J.~R.}\ \bibnamefont {Petta}}, \bibinfo {author} {\bibfnamefont {S.}~\bibnamefont {Tarucha}},\ and\ \bibinfo {author} {\bibfnamefont {L.~M.~K.}\ \bibnamefont {Vandersypen}},\ }\bibfield  {title} {\bibinfo {title} {Spins in few-electron quantum dots},\ }\href {https://doi.org/10.1103/RevModPhys.79.1217} {\bibfield  {journal} {\bibinfo  {journal} {Rev. Mod. Phys.}\ }\textbf {\bibinfo {volume} {79}},\ \bibinfo {pages} {1217} (\bibinfo {year} {2007})}\BibitemShut {NoStop}%
\bibitem [{\citenamefont {Vandersypen}\ \emph {et~al.}(2017)\citenamefont {Vandersypen}, \citenamefont {Bluhm}, \citenamefont {Clarke}, \citenamefont {Dzurak}, \citenamefont {Ishihara}, \citenamefont {Morello}, \citenamefont {Reilly}, \citenamefont {Schreiber},\ and\ \citenamefont {Veldhorst}}]{Vandersypen2017}%
  \BibitemOpen
  \bibfield  {author} {\bibinfo {author} {\bibfnamefont {L.~M.~K.}\ \bibnamefont {Vandersypen}}, \bibinfo {author} {\bibfnamefont {H.}~\bibnamefont {Bluhm}}, \bibinfo {author} {\bibfnamefont {J.~S.}\ \bibnamefont {Clarke}}, \bibinfo {author} {\bibfnamefont {A.~S.}\ \bibnamefont {Dzurak}}, \bibinfo {author} {\bibfnamefont {R.}~\bibnamefont {Ishihara}}, \bibinfo {author} {\bibfnamefont {A.}~\bibnamefont {Morello}}, \bibinfo {author} {\bibfnamefont {D.~J.}\ \bibnamefont {Reilly}}, \bibinfo {author} {\bibfnamefont {L.~R.}\ \bibnamefont {Schreiber}},\ and\ \bibinfo {author} {\bibfnamefont {M.}~\bibnamefont {Veldhorst}},\ }\bibfield  {title} {\bibinfo {title} {Interfacing spin qubits in quantum dots and donors---hot, dense, and coherent},\ }\href {https://doi.org/10.1038/s41534-017-0038-y} {\bibfield  {journal} {\bibinfo  {journal} {npj Quantum Information}\ }\textbf {\bibinfo {volume} {3}},\ \bibinfo {pages} {34} (\bibinfo {year} {2017})}\BibitemShut {NoStop}%
\bibitem [{\citenamefont {DiVincenzo}\ \emph {et~al.}(2000)\citenamefont {DiVincenzo}, \citenamefont {Bacon}, \citenamefont {Kempe}, \citenamefont {Burkard},\ and\ \citenamefont {Whaley}}]{DiVincenzo2000}%
  \BibitemOpen
  \bibfield  {author} {\bibinfo {author} {\bibfnamefont {D.~P.}\ \bibnamefont {DiVincenzo}}, \bibinfo {author} {\bibfnamefont {D.}~\bibnamefont {Bacon}}, \bibinfo {author} {\bibfnamefont {J.}~\bibnamefont {Kempe}}, \bibinfo {author} {\bibfnamefont {G.}~\bibnamefont {Burkard}},\ and\ \bibinfo {author} {\bibfnamefont {K.~B.}\ \bibnamefont {Whaley}},\ }\bibfield  {title} {\bibinfo {title} {Universal quantum computation with the exchange interaction},\ }\href {https://doi.org/10.1038/35042541} {\bibfield  {journal} {\bibinfo  {journal} {Nature}\ }\textbf {\bibinfo {volume} {408}},\ \bibinfo {pages} {339} (\bibinfo {year} {2000})}\BibitemShut {NoStop}%
\bibitem [{\citenamefont {Petta}\ \emph {et~al.}(2005)\citenamefont {Petta}, \citenamefont {Johnson}, \citenamefont {Taylor}, \citenamefont {Laird}, \citenamefont {Yacoby}, \citenamefont {Lukin}, \citenamefont {Marcus}, \citenamefont {Hanson},\ and\ \citenamefont {Gossard}}]{doi:10.1126/science.1116955}%
  \BibitemOpen
  \bibfield  {author} {\bibinfo {author} {\bibfnamefont {J.~R.}\ \bibnamefont {Petta}}, \bibinfo {author} {\bibfnamefont {A.~C.}\ \bibnamefont {Johnson}}, \bibinfo {author} {\bibfnamefont {J.~M.}\ \bibnamefont {Taylor}}, \bibinfo {author} {\bibfnamefont {E.~A.}\ \bibnamefont {Laird}}, \bibinfo {author} {\bibfnamefont {A.}~\bibnamefont {Yacoby}}, \bibinfo {author} {\bibfnamefont {M.~D.}\ \bibnamefont {Lukin}}, \bibinfo {author} {\bibfnamefont {C.~M.}\ \bibnamefont {Marcus}}, \bibinfo {author} {\bibfnamefont {M.~P.}\ \bibnamefont {Hanson}},\ and\ \bibinfo {author} {\bibfnamefont {A.~C.}\ \bibnamefont {Gossard}},\ }\bibfield  {title} {\bibinfo {title} {Coherent manipulation of coupled electron spins in semiconductor quantum dots},\ }\href {https://doi.org/10.1126/science.1116955} {\bibfield  {journal} {\bibinfo  {journal} {Science}\ }\textbf {\bibinfo {volume} {309}},\ \bibinfo {pages} {2180} (\bibinfo {year} {2005})}\BibitemShut {NoStop}%
\bibitem [{\citenamefont {Saez-Mollejo}\ \emph {et~al.}(2025)\citenamefont {Saez-Mollejo}, \citenamefont {Jirovec}, \citenamefont {Schell}, \citenamefont {Kukucka}, \citenamefont {Calcaterra}, \citenamefont {Chrastina}, \citenamefont {Isella}, \citenamefont {Rimbach-Russ}, \citenamefont {Bosco},\ and\ \citenamefont {Katsaros}}]{ExchangeanisotropiesinmicrowavedrivenSTqubits}%
  \BibitemOpen
  \bibfield  {author} {\bibinfo {author} {\bibfnamefont {J.}~\bibnamefont {Saez-Mollejo}}, \bibinfo {author} {\bibfnamefont {D.}~\bibnamefont {Jirovec}}, \bibinfo {author} {\bibfnamefont {Y.}~\bibnamefont {Schell}}, \bibinfo {author} {\bibfnamefont {J.}~\bibnamefont {Kukucka}}, \bibinfo {author} {\bibfnamefont {S.}~\bibnamefont {Calcaterra}}, \bibinfo {author} {\bibfnamefont {D.}~\bibnamefont {Chrastina}}, \bibinfo {author} {\bibfnamefont {G.}~\bibnamefont {Isella}}, \bibinfo {author} {\bibfnamefont {M.}~\bibnamefont {Rimbach-Russ}}, \bibinfo {author} {\bibfnamefont {S.}~\bibnamefont {Bosco}},\ and\ \bibinfo {author} {\bibfnamefont {G.}~\bibnamefont {Katsaros}},\ }\bibfield  {title} {\bibinfo {title} {Exchange anisotropies in microwave-driven singlet-triplet qubits},\ }\href {https://doi.org/10.1038/s41467-025-58969-y} {\bibfield  {journal} {\bibinfo  {journal} {Nature Communications}\ }\textbf {\bibinfo {volume} {16}},\ \bibinfo {pages} {3862} (\bibinfo {year} {2025})}\BibitemShut {NoStop}%
\bibitem [{\citenamefont {Geyer}\ \emph {et~al.}(2024)\citenamefont {Geyer}, \citenamefont {Het{\'e}nyi}, \citenamefont {Bosco}, \citenamefont {Camenzind}, \citenamefont {Eggli}, \citenamefont {Fuhrer}, \citenamefont {Loss}, \citenamefont {Warburton}, \citenamefont {Zumb{\"u}hl},\ and\ \citenamefont {Kuhlmann}}]{Anisotropicexchangeinteractionoftwoholespinqubits}%
  \BibitemOpen
  \bibfield  {author} {\bibinfo {author} {\bibfnamefont {S.}~\bibnamefont {Geyer}}, \bibinfo {author} {\bibfnamefont {B.}~\bibnamefont {Het{\'e}nyi}}, \bibinfo {author} {\bibfnamefont {S.}~\bibnamefont {Bosco}}, \bibinfo {author} {\bibfnamefont {L.~C.}\ \bibnamefont {Camenzind}}, \bibinfo {author} {\bibfnamefont {R.~S.}\ \bibnamefont {Eggli}}, \bibinfo {author} {\bibfnamefont {A.}~\bibnamefont {Fuhrer}}, \bibinfo {author} {\bibfnamefont {D.}~\bibnamefont {Loss}}, \bibinfo {author} {\bibfnamefont {R.~J.}\ \bibnamefont {Warburton}}, \bibinfo {author} {\bibfnamefont {D.~M.}\ \bibnamefont {Zumb{\"u}hl}},\ and\ \bibinfo {author} {\bibfnamefont {A.~V.}\ \bibnamefont {Kuhlmann}},\ }\bibfield  {title} {\bibinfo {title} {Anisotropic exchange interaction of two hole-spin qubits},\ }\href {https://doi.org/10.1038/s41567-024-02481-5} {\bibfield  {journal} {\bibinfo  {journal} {Nature Physics}\ }\textbf {\bibinfo {volume} {20}},\ \bibinfo {pages} {1152} (\bibinfo {year} {2024})}\BibitemShut {NoStop}%
\bibitem [{\citenamefont {Meunier}\ \emph {et~al.}(2011)\citenamefont {Meunier}, \citenamefont {Calado},\ and\ \citenamefont {Vandersypen}}]{CPhase}%
  \BibitemOpen
  \bibfield  {author} {\bibinfo {author} {\bibfnamefont {T.}~\bibnamefont {Meunier}}, \bibinfo {author} {\bibfnamefont {V.~E.}\ \bibnamefont {Calado}},\ and\ \bibinfo {author} {\bibfnamefont {L.~M.~K.}\ \bibnamefont {Vandersypen}},\ }\bibfield  {title} {\bibinfo {title} {Efficient controlled-phase gate for single-spin qubits in quantum dots},\ }\href {https://doi.org/10.1103/PhysRevB.83.121403} {\bibfield  {journal} {\bibinfo  {journal} {Phys. Rev. B}\ }\textbf {\bibinfo {volume} {83}},\ \bibinfo {pages} {121403} (\bibinfo {year} {2011})}\BibitemShut {NoStop}%
\bibitem [{\citenamefont {Dutta}\ \emph {et~al.}(1979)\citenamefont {Dutta}, \citenamefont {Dimon},\ and\ \citenamefont {Horn}}]{Lor21f}%
  \BibitemOpen
  \bibfield  {author} {\bibinfo {author} {\bibfnamefont {P.}~\bibnamefont {Dutta}}, \bibinfo {author} {\bibfnamefont {P.}~\bibnamefont {Dimon}},\ and\ \bibinfo {author} {\bibfnamefont {P.~M.}\ \bibnamefont {Horn}},\ }\bibfield  {title} {\bibinfo {title} {Energy scales for noise processes in metals},\ }\href {https://doi.org/10.1103/PhysRevLett.43.646} {\bibfield  {journal} {\bibinfo  {journal} {Phys. Rev. Lett.}\ }\textbf {\bibinfo {volume} {43}},\ \bibinfo {pages} {646} (\bibinfo {year} {1979})}\BibitemShut {NoStop}%
\bibitem [{\citenamefont {Yoneda}\ \emph {et~al.}(2018)\citenamefont {Yoneda}, \citenamefont {Takeda}, \citenamefont {Otsuka}, \citenamefont {Nakajima}, \citenamefont {Delbecq}, \citenamefont {Allison}, \citenamefont {Honda}, \citenamefont {Kodera}, \citenamefont {Oda}, \citenamefont {Hoshi}, \citenamefont {Usami}, \citenamefont {Itoh},\ and\ \citenamefont {Tarucha}}]{Yoneda2018}%
  \BibitemOpen
  \bibfield  {author} {\bibinfo {author} {\bibfnamefont {J.}~\bibnamefont {Yoneda}}, \bibinfo {author} {\bibfnamefont {K.}~\bibnamefont {Takeda}}, \bibinfo {author} {\bibfnamefont {T.}~\bibnamefont {Otsuka}}, \bibinfo {author} {\bibfnamefont {T.}~\bibnamefont {Nakajima}}, \bibinfo {author} {\bibfnamefont {M.~R.}\ \bibnamefont {Delbecq}}, \bibinfo {author} {\bibfnamefont {G.}~\bibnamefont {Allison}}, \bibinfo {author} {\bibfnamefont {T.}~\bibnamefont {Honda}}, \bibinfo {author} {\bibfnamefont {T.}~\bibnamefont {Kodera}}, \bibinfo {author} {\bibfnamefont {S.}~\bibnamefont {Oda}}, \bibinfo {author} {\bibfnamefont {Y.}~\bibnamefont {Hoshi}}, \bibinfo {author} {\bibfnamefont {N.}~\bibnamefont {Usami}}, \bibinfo {author} {\bibfnamefont {K.~M.}\ \bibnamefont {Itoh}},\ and\ \bibinfo {author} {\bibfnamefont {S.}~\bibnamefont {Tarucha}},\ }\bibfield  {title} {\bibinfo {title} {A quantum-dot spin qubit with coherence limited by charge noise and fidelity higher than 99.9{\%}},\ }\href
  {https://doi.org/10.1038/s41565-017-0014-x} {\bibfield  {journal} {\bibinfo  {journal} {Nature Nanotechnology}\ }\textbf {\bibinfo {volume} {13}},\ \bibinfo {pages} {102} (\bibinfo {year} {2018})}\BibitemShut {NoStop}%
\bibitem [{\citenamefont {Petit}\ \emph {et~al.}(2018)\citenamefont {Petit}, \citenamefont {Boter}, \citenamefont {Eenink}, \citenamefont {Droulers}, \citenamefont {Tagliaferri}, \citenamefont {Li}, \citenamefont {Franke}, \citenamefont {Singh}, \citenamefont {Clarke}, \citenamefont {Schouten}, \citenamefont {Dobrovitski}, \citenamefont {Vandersypen},\ and\ \citenamefont {Veldhorst}}]{PhysRevLett.121.076801}%
  \BibitemOpen
  \bibfield  {author} {\bibinfo {author} {\bibfnamefont {L.}~\bibnamefont {Petit}}, \bibinfo {author} {\bibfnamefont {J.~M.}\ \bibnamefont {Boter}}, \bibinfo {author} {\bibfnamefont {H.~G.~J.}\ \bibnamefont {Eenink}}, \bibinfo {author} {\bibfnamefont {G.}~\bibnamefont {Droulers}}, \bibinfo {author} {\bibfnamefont {M.~L.~V.}\ \bibnamefont {Tagliaferri}}, \bibinfo {author} {\bibfnamefont {R.}~\bibnamefont {Li}}, \bibinfo {author} {\bibfnamefont {D.~P.}\ \bibnamefont {Franke}}, \bibinfo {author} {\bibfnamefont {K.~J.}\ \bibnamefont {Singh}}, \bibinfo {author} {\bibfnamefont {J.~S.}\ \bibnamefont {Clarke}}, \bibinfo {author} {\bibfnamefont {R.~N.}\ \bibnamefont {Schouten}}, \bibinfo {author} {\bibfnamefont {V.~V.}\ \bibnamefont {Dobrovitski}}, \bibinfo {author} {\bibfnamefont {L.~M.~K.}\ \bibnamefont {Vandersypen}},\ and\ \bibinfo {author} {\bibfnamefont {M.}~\bibnamefont {Veldhorst}},\ }\bibfield  {title} {\bibinfo {title} {Spin lifetime and charge noise in hot silicon quantum dot qubits},\ }\href
  {https://doi.org/10.1103/PhysRevLett.121.076801} {\bibfield  {journal} {\bibinfo  {journal} {Phys. Rev. Lett.}\ }\textbf {\bibinfo {volume} {121}},\ \bibinfo {pages} {076801} (\bibinfo {year} {2018})}\BibitemShut {NoStop}%
\bibitem [{\citenamefont {Ye}\ \emph {et~al.}(2024)\citenamefont {Ye}, \citenamefont {Ellaboudy}, \citenamefont {Albrecht}, \citenamefont {Vudatha}, \citenamefont {Jacobson},\ and\ \citenamefont {Nichol}}]{PhysRevB.110.235305}%
  \BibitemOpen
  \bibfield  {author} {\bibinfo {author} {\bibfnamefont {F.}~\bibnamefont {Ye}}, \bibinfo {author} {\bibfnamefont {A.}~\bibnamefont {Ellaboudy}}, \bibinfo {author} {\bibfnamefont {D.}~\bibnamefont {Albrecht}}, \bibinfo {author} {\bibfnamefont {R.}~\bibnamefont {Vudatha}}, \bibinfo {author} {\bibfnamefont {N.~T.}\ \bibnamefont {Jacobson}},\ and\ \bibinfo {author} {\bibfnamefont {J.~M.}\ \bibnamefont {Nichol}},\ }\bibfield  {title} {\bibinfo {title} {Characterization of individual charge fluctuators in {{Si/SiGe}} quantum dots},\ }\href {https://doi.org/10.1103/PhysRevB.110.235305} {\bibfield  {journal} {\bibinfo  {journal} {Phys. Rev. B}\ }\textbf {\bibinfo {volume} {110}},\ \bibinfo {pages} {235305} (\bibinfo {year} {2024})}\BibitemShut {NoStop}%
\bibitem [{\citenamefont {Müller}\ \emph {et~al.}(2019)\citenamefont {Müller}, \citenamefont {Cole},\ and\ \citenamefont {Lisenfeld}}]{Muller_2019}%
  \BibitemOpen
  \bibfield  {author} {\bibinfo {author} {\bibfnamefont {C.}~\bibnamefont {Müller}}, \bibinfo {author} {\bibfnamefont {J.~H.}\ \bibnamefont {Cole}},\ and\ \bibinfo {author} {\bibfnamefont {J.}~\bibnamefont {Lisenfeld}},\ }\bibfield  {title} {\bibinfo {title} {Towards understanding two-level-systems in amorphous solids: insights from quantum circuits},\ }\href {https://doi.org/10.1088/1361-6633/ab3a7e} {\bibfield  {journal} {\bibinfo  {journal} {Reports on Progress in Physics}\ }\textbf {\bibinfo {volume} {82}},\ \bibinfo {pages} {124501} (\bibinfo {year} {2019})}\BibitemShut {NoStop}%
\bibitem [{\citenamefont {Wang}\ \emph {et~al.}(2025)\citenamefont {Wang}, \citenamefont {Gholizadeh}, \citenamefont {Hu}, \citenamefont {Das~Sarma},\ and\ \citenamefont {Culcer}}]{1fnoiseplanarge}%
  \BibitemOpen
  \bibfield  {author} {\bibinfo {author} {\bibfnamefont {Z.}~\bibnamefont {Wang}}, \bibinfo {author} {\bibfnamefont {S.}~\bibnamefont {Gholizadeh}}, \bibinfo {author} {\bibfnamefont {X.}~\bibnamefont {Hu}}, \bibinfo {author} {\bibfnamefont {S.}~\bibnamefont {Das~Sarma}},\ and\ \bibinfo {author} {\bibfnamefont {D.}~\bibnamefont {Culcer}},\ }\bibfield  {title} {\bibinfo {title} {Dephasing of planar {Ge} hole spin qubits due to $1/f$ charge noise},\ }\href {https://doi.org/10.1103/PhysRevB.111.155403} {\bibfield  {journal} {\bibinfo  {journal} {Phys. Rev. B}\ }\textbf {\bibinfo {volume} {111}},\ \bibinfo {pages} {155403} (\bibinfo {year} {2025})}\BibitemShut {NoStop}%
\bibitem [{\citenamefont {Mickelsen}\ \emph {et~al.}(2023)\citenamefont {Mickelsen}, \citenamefont {Carruzzo},\ and\ \citenamefont {Yu}}]{Interactingtwolevelsystemsasasourceof1fchargenoiseinQD}%
  \BibitemOpen
  \bibfield  {author} {\bibinfo {author} {\bibfnamefont {D.~L.}\ \bibnamefont {Mickelsen}}, \bibinfo {author} {\bibfnamefont {H.~M.}\ \bibnamefont {Carruzzo}},\ and\ \bibinfo {author} {\bibfnamefont {C.~C.}\ \bibnamefont {Yu}},\ }\bibfield  {title} {\bibinfo {title} {Interacting two-level systems as a source of $1/f$ charge noise in quantum dot qubits},\ }\href {https://doi.org/10.1103/PhysRevB.108.195307} {\bibfield  {journal} {\bibinfo  {journal} {Phys. Rev. B}\ }\textbf {\bibinfo {volume} {108}},\ \bibinfo {pages} {195307} (\bibinfo {year} {2023})}\BibitemShut {NoStop}%
\bibitem [{\citenamefont {Kononovicius}\ and\ \citenamefont {Kaulakys}(2024)}]{1fnoiseinsemiconductorsarisingfromtheheterogeneousdetrappingprocess}%
  \BibitemOpen
  \bibfield  {author} {\bibinfo {author} {\bibfnamefont {A.}~\bibnamefont {Kononovicius}}\ and\ \bibinfo {author} {\bibfnamefont {B.}~\bibnamefont {Kaulakys}},\ }\bibfield  {title} {\bibinfo {title} {$1/f$ noise in semiconductors arising from the heterogeneous detrapping process of individual charge carriers},\ }\href {https://doi.org/10.1088/1742-5468/ad890b} {\bibfield  {journal} {\bibinfo  {journal} {Journal of Statistical Mechanics: Theory and Experiment}\ }\textbf {\bibinfo {volume} {2024}},\ \bibinfo {pages} {113201} (\bibinfo {year} {2024})}\BibitemShut {NoStop}%
\bibitem [{\citenamefont {Mehmandoost}\ and\ \citenamefont {Dobrovitski}(2024)}]{Decoherenceinducedbyasparsebathoftwolevelfluctuators}%
  \BibitemOpen
  \bibfield  {author} {\bibinfo {author} {\bibfnamefont {M.}~\bibnamefont {Mehmandoost}}\ and\ \bibinfo {author} {\bibfnamefont {V.~V.}\ \bibnamefont {Dobrovitski}},\ }\bibfield  {title} {\bibinfo {title} {Decoherence induced by a sparse bath of two-level fluctuators: Peculiar features of $1/f$ noise in high-quality qubits},\ }\href {https://doi.org/10.1103/PhysRevResearch.6.033175} {\bibfield  {journal} {\bibinfo  {journal} {Phys. Rev. Res.}\ }\textbf {\bibinfo {volume} {6}},\ \bibinfo {pages} {033175} (\bibinfo {year} {2024})}\BibitemShut {NoStop}%
\bibitem [{\citenamefont {Dutta}\ and\ \citenamefont {Horn}(1981)}]{RevModPhys.53.497}%
  \BibitemOpen
  \bibfield  {author} {\bibinfo {author} {\bibfnamefont {P.}~\bibnamefont {Dutta}}\ and\ \bibinfo {author} {\bibfnamefont {P.~M.}\ \bibnamefont {Horn}},\ }\bibfield  {title} {\bibinfo {title} {Low-frequency fluctuations in solids: $1/f$ noise},\ }\href {https://doi.org/10.1103/RevModPhys.53.497} {\bibfield  {journal} {\bibinfo  {journal} {Rev. Mod. Phys.}\ }\textbf {\bibinfo {volume} {53}},\ \bibinfo {pages} {497} (\bibinfo {year} {1981})}\BibitemShut {NoStop}%
\bibitem [{\citenamefont {Elsayed}\ \emph {et~al.}(2024)\citenamefont {Elsayed}, \citenamefont {Shehata}, \citenamefont {Godfrin}, \citenamefont {Kubicek}, \citenamefont {Massar}, \citenamefont {Canvel}, \citenamefont {Jussot}, \citenamefont {Simion}, \citenamefont {Mongillo}, \citenamefont {Wan}, \citenamefont {Govoreanu}, \citenamefont {Radu}, \citenamefont {Li}, \citenamefont {Van~Dorpe},\ and\ \citenamefont {De~Greve}}]{Elsayed2024}%
  \BibitemOpen
  \bibfield  {author} {\bibinfo {author} {\bibfnamefont {A.}~\bibnamefont {Elsayed}}, \bibinfo {author} {\bibfnamefont {M.~M.~K.}\ \bibnamefont {Shehata}}, \bibinfo {author} {\bibfnamefont {C.}~\bibnamefont {Godfrin}}, \bibinfo {author} {\bibfnamefont {S.}~\bibnamefont {Kubicek}}, \bibinfo {author} {\bibfnamefont {S.}~\bibnamefont {Massar}}, \bibinfo {author} {\bibfnamefont {Y.}~\bibnamefont {Canvel}}, \bibinfo {author} {\bibfnamefont {J.}~\bibnamefont {Jussot}}, \bibinfo {author} {\bibfnamefont {G.}~\bibnamefont {Simion}}, \bibinfo {author} {\bibfnamefont {M.}~\bibnamefont {Mongillo}}, \bibinfo {author} {\bibfnamefont {D.}~\bibnamefont {Wan}}, \bibinfo {author} {\bibfnamefont {B.}~\bibnamefont {Govoreanu}}, \bibinfo {author} {\bibfnamefont {I.~P.}\ \bibnamefont {Radu}}, \bibinfo {author} {\bibfnamefont {R.}~\bibnamefont {Li}}, \bibinfo {author} {\bibfnamefont {P.}~\bibnamefont {Van~Dorpe}},\ and\ \bibinfo {author} {\bibfnamefont {K.}~\bibnamefont {De~Greve}},\ }\bibfield  {title} {\bibinfo {title} {Low
  charge noise quantum dots with industrial {CMOS} manufacturing},\ }\href {https://doi.org/10.1038/s41534-024-00864-3} {\bibfield  {journal} {\bibinfo  {journal} {npj Quantum Information}\ }\textbf {\bibinfo {volume} {10}},\ \bibinfo {pages} {70} (\bibinfo {year} {2024})}\BibitemShut {NoStop}%
\bibitem [{\citenamefont {Freeman}\ \emph {et~al.}(2016)\citenamefont {Freeman}, \citenamefont {Schoenfield},\ and\ \citenamefont {Jiang}}]{10.1063/1.4954700}%
  \BibitemOpen
  \bibfield  {author} {\bibinfo {author} {\bibfnamefont {B.~M.}\ \bibnamefont {Freeman}}, \bibinfo {author} {\bibfnamefont {J.~S.}\ \bibnamefont {Schoenfield}},\ and\ \bibinfo {author} {\bibfnamefont {H.}~\bibnamefont {Jiang}},\ }\bibfield  {title} {\bibinfo {title} {Comparison of low frequency charge noise in identically patterned {{Si/SiO$_2$}} and {{Si/SiGe}} quantum dots},\ }\href {https://doi.org/10.1063/1.4954700} {\bibfield  {journal} {\bibinfo  {journal} {Applied Physics Letters}\ }\textbf {\bibinfo {volume} {108}},\ \bibinfo {pages} {253108} (\bibinfo {year} {2016})}\BibitemShut {NoStop}%
\bibitem [{\citenamefont {Paladino}\ \emph {et~al.}(2014)\citenamefont {Paladino}, \citenamefont {Galperin}, \citenamefont {Falci},\ and\ \citenamefont {Altshuler}}]{RevModPhys.86.361}%
  \BibitemOpen
  \bibfield  {author} {\bibinfo {author} {\bibfnamefont {E.}~\bibnamefont {Paladino}}, \bibinfo {author} {\bibfnamefont {Y.~M.}\ \bibnamefont {Galperin}}, \bibinfo {author} {\bibfnamefont {G.}~\bibnamefont {Falci}},\ and\ \bibinfo {author} {\bibfnamefont {B.~L.}\ \bibnamefont {Altshuler}},\ }\bibfield  {title} {\bibinfo {title} {1/$f$ noise: Implications for solid-state quantum information},\ }\href {https://doi.org/10.1103/RevModPhys.86.361} {\bibfield  {journal} {\bibinfo  {journal} {Rev. Mod. Phys.}\ }\textbf {\bibinfo {volume} {86}},\ \bibinfo {pages} {361} (\bibinfo {year} {2014})}\BibitemShut {NoStop}%
\bibitem [{CZc()}]{CZcomment}%
  \BibitemOpen
  \href@noop {} {\bibinfo {title} {Points along the $\delta=10$ line with $0\leq\theta <\pi/2$ can implement the {CZ} gate. {The} effective interaction strength is ${J}\cos\theta$, so the gate duration $\tau=\pi/({J}\cos\theta)$ increases with $\theta$. {The} suppression of correlated noise gained by increasing $\theta$ is offset by the larger infidelity arising from the longer gate duration. {Figure S1(d)} shows the infidelity of the {CZ} gate as a function of $\theta$; the orange dot at $\theta = 0$ has the lowest infidelity. {Therefore}, in what follows we choose $\theta = 0$ to realize the {CZ} gate.}}\BibitemShut {Stop}%
\bibitem [{fre()}]{frequencynoise}%
  \BibitemOpen
  \href@noop {} {\bibinfo {title} {The spatially correlated frequency noise {$S_{\Delta_1\Delta_2}$} makes a weak contribution to infidelity throughout the gate process. {However}, it could be relevant before the gate operation begins (i.e., before turning on the interaction), for instance, creating entanglement between idling qubits \cite{zou2022bell,driessen2026robusttripartiteentanglementgeneration}.}}\BibitemShut {Stop}%
\bibitem [{\citenamefont {Paquelet~Wuetz}\ \emph {et~al.}(2023)\citenamefont {Paquelet~Wuetz}, \citenamefont {Degli~Esposti}, \citenamefont {Zwerver}, \citenamefont {Amitonov}, \citenamefont {Botifoll}, \citenamefont {Arbiol}, \citenamefont {Sammak}, \citenamefont {Vandersypen}, \citenamefont {Russ},\ and\ \citenamefont {Scappucci}}]{PaqueletWuetz2023}%
  \BibitemOpen
  \bibfield  {author} {\bibinfo {author} {\bibfnamefont {B.}~\bibnamefont {Paquelet~Wuetz}}, \bibinfo {author} {\bibfnamefont {D.}~\bibnamefont {Degli~Esposti}}, \bibinfo {author} {\bibfnamefont {A.-M.~J.}\ \bibnamefont {Zwerver}}, \bibinfo {author} {\bibfnamefont {S.~V.}\ \bibnamefont {Amitonov}}, \bibinfo {author} {\bibfnamefont {M.}~\bibnamefont {Botifoll}}, \bibinfo {author} {\bibfnamefont {J.}~\bibnamefont {Arbiol}}, \bibinfo {author} {\bibfnamefont {A.}~\bibnamefont {Sammak}}, \bibinfo {author} {\bibfnamefont {L.~M.~K.}\ \bibnamefont {Vandersypen}}, \bibinfo {author} {\bibfnamefont {M.}~\bibnamefont {Russ}},\ and\ \bibinfo {author} {\bibfnamefont {G.}~\bibnamefont {Scappucci}},\ }\bibfield  {title} {\bibinfo {title} {Reducing charge noise in quantum dots by using thin silicon quantum wells},\ }\href {https://doi.org/10.1038/s41467-023-36951-w} {\bibfield  {journal} {\bibinfo  {journal} {Nature Communications}\ }\textbf {\bibinfo {volume} {14}},\ \bibinfo {pages} {1385} (\bibinfo {year}
  {2023})}\BibitemShut {NoStop}%
\bibitem [{\citenamefont {Kranz}\ \emph {et~al.}(2020)\citenamefont {Kranz}, \citenamefont {Gorman}, \citenamefont {Thorgrimsson}, \citenamefont {He}, \citenamefont {Keith}, \citenamefont {Keizer},\ and\ \citenamefont {Simmons}}]{https://doi.org/10.1002/adma.202003361}%
  \BibitemOpen
  \bibfield  {author} {\bibinfo {author} {\bibfnamefont {L.}~\bibnamefont {Kranz}}, \bibinfo {author} {\bibfnamefont {S.~K.}\ \bibnamefont {Gorman}}, \bibinfo {author} {\bibfnamefont {B.}~\bibnamefont {Thorgrimsson}}, \bibinfo {author} {\bibfnamefont {Y.}~\bibnamefont {He}}, \bibinfo {author} {\bibfnamefont {D.}~\bibnamefont {Keith}}, \bibinfo {author} {\bibfnamefont {J.~G.}\ \bibnamefont {Keizer}},\ and\ \bibinfo {author} {\bibfnamefont {M.~Y.}\ \bibnamefont {Simmons}},\ }\bibfield  {title} {\bibinfo {title} {Exploiting a single-crystal environment to minimize the charge noise on qubits in silicon},\ }\href {https://doi.org/https://doi.org/10.1002/adma.202003361} {\bibfield  {journal} {\bibinfo  {journal} {Advanced Materials}\ }\textbf {\bibinfo {volume} {32}},\ \bibinfo {pages} {2003361} (\bibinfo {year} {2020})}\BibitemShut {NoStop}%
\bibitem [{\citenamefont {Connors}\ \emph {et~al.}(2019)\citenamefont {Connors}, \citenamefont {Nelson}, \citenamefont {Qiao}, \citenamefont {Edge},\ and\ \citenamefont {Nichol}}]{PhysRevB.100.165305}%
  \BibitemOpen
  \bibfield  {author} {\bibinfo {author} {\bibfnamefont {E.~J.}\ \bibnamefont {Connors}}, \bibinfo {author} {\bibfnamefont {J.~J.}\ \bibnamefont {Nelson}}, \bibinfo {author} {\bibfnamefont {H.}~\bibnamefont {Qiao}}, \bibinfo {author} {\bibfnamefont {L.~F.}\ \bibnamefont {Edge}},\ and\ \bibinfo {author} {\bibfnamefont {J.~M.}\ \bibnamefont {Nichol}},\ }\bibfield  {title} {\bibinfo {title} {Low-frequency charge noise in {{Si/SiGe}} quantum dots},\ }\href {https://doi.org/10.1103/PhysRevB.100.165305} {\bibfield  {journal} {\bibinfo  {journal} {Phys. Rev. B}\ }\textbf {\bibinfo {volume} {100}},\ \bibinfo {pages} {165305} (\bibinfo {year} {2019})}\BibitemShut {NoStop}%
\bibitem [{ant()}]{anticorrelationcomment}%
  \BibitemOpen
  \href@noop {} {\bibinfo {title} {{The} hierarchy can be reversed in a special case: when the spatially correlated frequency noise ${S}_{\Delta_1\Delta_2}$ is anticorrelated, as arises when a two-level fluctuator is located between the two quantum dots~\cite{RojasArias2025InferringCN}, and is sufficiently strong compared with the interaction noise ${S}_{JJ}$. {For} anticorrelated noise the contribution of ${S}_{\Delta_1\Delta_2}$ to the infidelity is negative, and its magnitude decreases with detuning, since the corresponding filter function vanishes identically at the {CZ} point, {SM Sec. VI B}, spatially correlated frequency noise. {The} infidelity difference between the two gates is then a competition between the decrease of the ${S}_{JJ}$ contribution and the rise of the ${S}_{\Delta_1\Delta_2}$ contribution from negative toward zero as the detuning $\delta$ is increased. {When} the latter dominates, the {CZ} gate acquires a higher infidelity than the {SWAP} gate.}}\BibitemShut {Stop}%
\bibitem [{\citenamefont {Shehata}\ \emph {et~al.}(2023)\citenamefont {Shehata}, \citenamefont {Simion}, \citenamefont {Li}, \citenamefont {Mohiyaddin}, \citenamefont {Wan}, \citenamefont {Mongillo}, \citenamefont {Govoreanu}, \citenamefont {Radu}, \citenamefont {De~Greve},\ and\ \citenamefont {Van~Dorpe}}]{modelingnoiseenvir}%
  \BibitemOpen
  \bibfield  {author} {\bibinfo {author} {\bibfnamefont {M.~M. E.~K.}\ \bibnamefont {Shehata}}, \bibinfo {author} {\bibfnamefont {G.}~\bibnamefont {Simion}}, \bibinfo {author} {\bibfnamefont {R.}~\bibnamefont {Li}}, \bibinfo {author} {\bibfnamefont {F.~A.}\ \bibnamefont {Mohiyaddin}}, \bibinfo {author} {\bibfnamefont {D.}~\bibnamefont {Wan}}, \bibinfo {author} {\bibfnamefont {M.}~\bibnamefont {Mongillo}}, \bibinfo {author} {\bibfnamefont {B.}~\bibnamefont {Govoreanu}}, \bibinfo {author} {\bibfnamefont {I.}~\bibnamefont {Radu}}, \bibinfo {author} {\bibfnamefont {K.}~\bibnamefont {De~Greve}},\ and\ \bibinfo {author} {\bibfnamefont {P.}~\bibnamefont {Van~Dorpe}},\ }\bibfield  {title} {\bibinfo {title} {Modeling semiconductor spin qubits and their charge noise environment for quantum gate fidelity estimation},\ }\href {https://doi.org/10.1103/PhysRevB.108.045305} {\bibfield  {journal} {\bibinfo  {journal} {Phys. Rev. B}\ }\textbf {\bibinfo {volume} {108}},\ \bibinfo {pages} {045305} (\bibinfo {year}
  {2023})}\BibitemShut {NoStop}%
\bibitem [{\citenamefont {Samadi}\ \emph {et~al.}(2025)\citenamefont {Samadi}, \citenamefont {Łukasz Cywiński},\ and\ \citenamefont {Krzywda}}]{samadi2025statisticalstructurechargedisorder}%
  \BibitemOpen
  \bibfield  {author} {\bibinfo {author} {\bibfnamefont {S.}~\bibnamefont {Samadi}}, \bibinfo {author} {\bibnamefont {Łukasz Cywiński}},\ and\ \bibinfo {author} {\bibfnamefont {J.~A.}\ \bibnamefont {Krzywda}},\ }\href {https://arxiv.org/abs/2510.13578} {\bibinfo {title} {Statistical structure of charge disorder in si/sige quantum dots}} (\bibinfo {year} {2025}),\ \Eprint {https://arxiv.org/abs/2510.13578} {arXiv:2510.13578 [cond-mat.mes-hall]} \BibitemShut {NoStop}%
\bibitem [{\citenamefont {Breuer}\ and\ \citenamefont {Petruccione}(2007)}]{thetheoryofopenquantumsystem}%
  \BibitemOpen
  \bibfield  {author} {\bibinfo {author} {\bibfnamefont {H.-P.}\ \bibnamefont {Breuer}}\ and\ \bibinfo {author} {\bibfnamefont {F.}~\bibnamefont {Petruccione}},\ }\href {https://doi.org/10.1093/acprof:oso/9780199213900.001.0001} {\emph {\bibinfo {title} {The Theory of Open Quantum Systems}}}\ (\bibinfo  {publisher} {Oxford University Press},\ \bibinfo {year} {2007})\BibitemShut {NoStop}%
\bibitem [{\citenamefont {Mele}(2024)}]{Haarmeasure}%
  \BibitemOpen
  \bibfield  {author} {\bibinfo {author} {\bibfnamefont {A.~A.}\ \bibnamefont {Mele}},\ }\bibfield  {title} {\bibinfo {title} {Introduction to haar measure tools in quantum information: A beginner's tutorial},\ }\href {https://quantum-journal.org/wp-content/uploads/2024/05/q-2024-05-08-1340-1.pdf} {\bibfield  {journal} {\bibinfo  {journal} {Quantum}\ }\textbf {\bibinfo {volume} {8}},\ \bibinfo {pages} {1340} (\bibinfo {year} {2024})}\BibitemShut {NoStop}%
\bibitem [{\citenamefont {Clerk}\ \emph {et~al.}(2010)\citenamefont {Clerk}, \citenamefont {Devoret}, \citenamefont {Girvin}, \citenamefont {Marquardt},\ and\ \citenamefont {Schoelkopf}}]{Clerk2008IntroductionTQ}%
  \BibitemOpen
  \bibfield  {author} {\bibinfo {author} {\bibfnamefont {A.~A.}\ \bibnamefont {Clerk}}, \bibinfo {author} {\bibfnamefont {M.~H.}\ \bibnamefont {Devoret}}, \bibinfo {author} {\bibfnamefont {S.~M.}\ \bibnamefont {Girvin}}, \bibinfo {author} {\bibfnamefont {F.}~\bibnamefont {Marquardt}},\ and\ \bibinfo {author} {\bibfnamefont {R.~J.}\ \bibnamefont {Schoelkopf}},\ }\bibfield  {title} {\bibinfo {title} {Introduction to quantum noise, measurement, and amplification},\ }\href {https://doi.org/10.1103/RevModPhys.82.1155} {\bibfield  {journal} {\bibinfo  {journal} {Rev. Mod. Phys.}\ }\textbf {\bibinfo {volume} {82}},\ \bibinfo {pages} {1155} (\bibinfo {year} {2010})}\BibitemShut {NoStop}%
\bibitem [{\citenamefont {Nielsen}\ and\ \citenamefont {Chuang}(2010)}]{Nielsen_Chuang_2010}%
  \BibitemOpen
  \bibfield  {author} {\bibinfo {author} {\bibfnamefont {M.~A.}\ \bibnamefont {Nielsen}}\ and\ \bibinfo {author} {\bibfnamefont {I.~L.}\ \bibnamefont {Chuang}},\ }\href@noop {} {\emph {\bibinfo {title} {Quantum Computation and Quantum Information: 10th Anniversary Edition}}}\ (\bibinfo  {publisher} {Cambridge University Press},\ \bibinfo {year} {2010})\BibitemShut {NoStop}%
\bibitem [{\citenamefont {Bergonzoni}\ \emph {et~al.}(2025)\citenamefont {Bergonzoni}, \citenamefont {Jandura},\ and\ \citenamefont {Pupillo}}]{iswapgate}%
  \BibitemOpen
  \bibfield  {author} {\bibinfo {author} {\bibfnamefont {M.}~\bibnamefont {Bergonzoni}}, \bibinfo {author} {\bibfnamefont {S.}~\bibnamefont {Jandura}},\ and\ \bibinfo {author} {\bibfnamefont {G.}~\bibnamefont {Pupillo}},\ }\bibfield  {title} {\bibinfo {title} {iswap gate with polar molecules: Robustness criteria for entangling operations},\ }\href {https://doi.org/10.1103/q9sd-rfp6} {\bibfield  {journal} {\bibinfo  {journal} {Phys. Rev. A}\ }\textbf {\bibinfo {volume} {112}},\ \bibinfo {pages} {032621} (\bibinfo {year} {2025})}\BibitemShut {NoStop}%
\bibitem [{\citenamefont {Zou}\ \emph {et~al.}(2022)\citenamefont {Zou}, \citenamefont {Zhang},\ and\ \citenamefont {Tserkovnyak}}]{zou2022bell}%
  \BibitemOpen
  \bibfield  {author} {\bibinfo {author} {\bibfnamefont {J.}~\bibnamefont {Zou}}, \bibinfo {author} {\bibfnamefont {S.}~\bibnamefont {Zhang}},\ and\ \bibinfo {author} {\bibfnamefont {Y.}~\bibnamefont {Tserkovnyak}},\ }\bibfield  {title} {\bibinfo {title} {Bell-state generation for spin qubits via dissipative coupling},\ }\href@noop {} {\bibfield  {journal} {\bibinfo  {journal} {Physical Review B}\ }\textbf {\bibinfo {volume} {106}},\ \bibinfo {pages} {L180406} (\bibinfo {year} {2022})}\BibitemShut {NoStop}%
\bibitem [{\citenamefont {Driessen}\ \emph {et~al.}(2026)\citenamefont {Driessen}, \citenamefont {Zou}, \citenamefont {Thingstad}, \citenamefont {Klinovaja},\ and\ \citenamefont {Loss}}]{driessen2026robusttripartiteentanglementgeneration}%
  \BibitemOpen
  \bibfield  {author} {\bibinfo {author} {\bibfnamefont {S.}~\bibnamefont {Driessen}}, \bibinfo {author} {\bibfnamefont {J.}~\bibnamefont {Zou}}, \bibinfo {author} {\bibfnamefont {E.}~\bibnamefont {Thingstad}}, \bibinfo {author} {\bibfnamefont {J.}~\bibnamefont {Klinovaja}},\ and\ \bibinfo {author} {\bibfnamefont {D.}~\bibnamefont {Loss}},\ }\bibfield  {title} {\bibinfo {title} {Robust tripartite entanglement generation via correlated noise in spin qubits},\ }\href {https://doi.org/10.1103/5mr9-84qh} {\bibfield  {journal} {\bibinfo  {journal} {Phys. Rev. Res.}\ }\textbf {\bibinfo {volume} {8}},\ \bibinfo {pages} {033153} (\bibinfo {year} {2026})}\BibitemShut {NoStop}%
\end{thebibliography}%
\end{document}


\title{Supplementary Material for \\
``Resilience of Entangling Gates to Spatiotemporally Correlated Noise"}

\author{Xinru Tang}
\affiliation{Department of Physics, University of Basel, Klingelbergstrasse 82, 4056 Basel, Switzerland}
\author{Sander Driessen}
\affiliation{Department of Physics, University of Basel, Klingelbergstrasse 82, 4056 Basel, Switzerland}
\affiliation{Department of Applied Physics, Eindhoven University of Technology, Eindhoven 5612 AZ, The Netherlands}
\author{Ji Zou}
\email{ji.zou@kfupm.edu.sa}
\affiliation{Physics Department, King Fahd University of Petroleum and Minerals, 31261, Dhahran, Saudi Arabia}
\affiliation{Quantum Center, KFUPM, Dhahran, Saudi Arabia}
\author{Stefano Bosco}
\affiliation{QuTech, Delft University of Technology, Lorentzweg 1, 2628 CJ Delft, The Netherlands}
\author{Jelena Klinovaja} 
\affiliation{Department of Physics, University of Basel, Klingelbergstrasse 82, 4056 Basel, Switzerland}
\author{Daniel Loss} 
\affiliation{Department of Physics, University of Basel, Klingelbergstrasse 82, 4056 Basel, Switzerland}
\affiliation{Physics Department, King Fahd University of Petroleum and Minerals, 31261, Dhahran, Saudi Arabia}
\affiliation{Quantum Center, KFUPM, Dhahran, Saudi Arabia}
\affiliation{RDIA Chair in Quantum Computing}

\maketitle

\section{Derivation of the average gate fidelity expression in the time domain}\label{secI}
Here we derive the average gate infidelity in Eq. (3). Throughout we set $\hbar=1$, so that energies are measured in units of angular frequency. The quantum map $\Phi_\tau=\mathcal{T}\exp{\int_0^\tau ds\,\mathcal{L}(s) }$ in Eq. (2) can be expanded to second order:
\begin{align}\label{EqS1}
    \Phi_\tau&=I+\int_0^\tau ds\,\mathcal{L}(s)+\int_0^\tau dt_{1}\int_0^{t_{1}} dt_{2}\,\mathcal{L}(t_{1})\mathcal{L}(t_{2})+\mathcal{O}((||H_{\text{SE}}||_{\infty}\tau)^{3})\notag\\
    &=I-i\int_0^\tau ds\,[H_{\text{SE}}(s),\star]+(-i)^{2}\int_0^\tau dt_{1}\int_0^{t_{1}} dt_{2}\,[H_{\text{SE}}(t_{1}),[H_{\text{SE}}(t_{2}),\star]]+\mathcal{O}((||H_{\text{SE}}||_{\infty}\tau)^{3}),
\end{align}
where we have used the definition of the Liouville superoperator $\mathcal{L}(t)=(-i/\hbar)[H_{\text{SE}}(t),\star]$. The operator norm is defined as $||A||_{\infty}\equiv \sup_{\|\ket{\psi}\|=1} \|A\ket{\psi}\|$. We justify this truncation at the end of this section.\\

Using the general form of the coupling Hamiltonian $H_{\text{SE}}=\sum_{\alpha}\hat{A
}_{\alpha}\otimes\hat{B}_{\alpha}$, the interaction-picture error channel of Eq. (2) becomes, to second order:
\begin{align}\label{channel}
    \mathcal{E}_I(\ket{\psi}\bra{\psi}) &=\ket{\psi}\bra{\psi}+(-i)^{2}\int_0^\tau dt_{1}\int_0^{t_{1}} dt_{2}\,\Tr_E\{[H_{\text{SE}}(t_{1}),[H_{\text{SE}}(t_{2}),\ket{\psi}\bra{\psi}\otimes \rho_E]]\}\\
    &=\ket{\psi}\bra{\psi}+(-i)^{2}{\sum_{\alpha,\beta}}\int_0^\tau dt_{1}\int_0^{t_{1}} \!dt_{2}\!\left[[\hat{A}_{\alpha}(t_{1})\hat{A}_{\beta}(t_{2})\ket{\psi}\bra{\psi}-\hat{A}_{\beta}(t_{2})\ket{\psi}\bra{\psi}\hat{A}_{\alpha}(t_{1})]S_{\alpha\beta}(t_1-t_2)+\text{H.c.}\right],
\end{align}
where we have used $\Tr_E\{[H_{\text{SE}}(s),\ket{\psi}\bra{\psi}\otimes \rho_E]\}=0$, which holds for centered noise \cite{thetheoryofopenquantumsystem}. We have introduced the two-point noise correlation function $S_{\alpha\beta}(t_1-t_2)=\tr_{E}[\rho_{E}\hat{B}_{\alpha}(t_1)\hat{B}_{\beta}(t_2)]\equiv \langle \hat{B}_{\alpha}(t_1)\hat{B}_{\beta}(t_2)\rangle$.\\
Then the average gate infidelity $\overline{\mathcal{I}}=1-\overline{\mathcal{F}}$, which carries an overall minus sign relative to the second-order correction of $\overline{\mathcal{F}}$, becomes:
\begin{align}
    \overline{\mathcal{I}}
    &={\sum_{\alpha,\beta}}\int_0^\tau \!\! dt_{1}\int_0^{t_{1}} \!\!dt_{2} S_{\alpha\beta}(t_1-t_2)\Tr_{S}\left[\int d\psi \ket{\psi}\bra{\psi}
    \hat{A}_{\alpha}(t_{1})\hat{A}_{\beta}(t_{2})-\int d\psi \ket{\psi}\bra{\psi}\hat{A}_{\beta}(t_{2})\ket{\psi}\bra{\psi}\hat{A}_{\alpha}(t_{1})\right]+\text{H.c.}\\
    &={\sum_{\alpha,\beta}}\int_0^\tau dt_{1}\int_0^{t_{1}} \!dt_{2} S_{\alpha\beta}(t_1-t_2)\left\{\frac{\Tr_S[\hat{A}_{\alpha}(t_{1})\hat{A}_{\beta}(t_{2})]}{d}-\frac{\Tr_{S\otimes S}[(\mathbb{I}_{d^2}+\mathbb{F}_{d^2})(\hat{A}_{\beta}(t_{2})\otimes\hat{A}_{\alpha}(t_{1}))]}{d(d+1)}\right\}+\text{H.c.}\\
    &={\sum_{\alpha,\beta}}\int_0^\tau dt_{1}\int_0^{t_{1}} \!dt_{2} S_{\alpha\beta}(t_1-t_2)\left\{\frac{\Tr_{S}[\hat{A}_{\alpha}(t_{1})\hat{A}_{\beta}(t_{2})]}{d+1}-\frac{\Tr_{S}[\hat{A}_{\beta}(t_{2})]\Tr_{S}[\hat{A}_{\alpha}(t_{1})]}{d(d+1)}\right\}+\text{H.c.}.
\end{align}
Here we have used the following properties of the Haar measure \cite{Haarmeasure},
\( \int d\psi\ket{\psi}\!\bra{\psi}=\frac{\mathbb{I}_d}{d},\;\;\;\text{and} \;\;\;   \int d\psi \bigl(\ket{\psi}\!\bra{\psi}\bigr)^{\otimes2} = \frac{\mathbb{I}_{d^2} + \mathbb{F}_{d^2}}{d(d+1)}, \)
where $\mathbb{I}_{d^2}$ and $\mathbb{F}_{d^2}$ are the identity and flip operators, respectively, defined in the computational basis as $\mathbb{I}_{d^2}=\sum_{i,j=1}^{d}\ket{i,j}\bra{i,j}$ and $\mathbb{F}_{d^2}=\sum_{i,j=1}^{d}\ket{i,j}\bra{j,i}$. These operators satisfy the trace relations $\Tr((A\otimes B)\mathbb{I}_{d^2})=\Tr(A)\Tr(B)$ and $\Tr((A\otimes B)\mathbb{F}_{d^2})=\Tr(AB)$. The identity $\mathbb{I}_d$ acts on the single-copy system Hilbert space; $\mathbb{I}_{d^2}$ and $\mathbb{F}_{d^2}$ act on its tensor square.

We conclude this section by justifying the truncation in Eq.~\eqref{EqS1} for a bounded coupling Hamiltonian $H_{SE}$.
The operator norm satisfies $\|AB\|_{\infty}\le\|A\|_{\infty}\|B\|_{\infty}$,
the triangle inequality $\|A+B\|_{\infty}\le\|A\|_{\infty}+\|B\|_{\infty}$,
and invariance under unitary conjugation,
$\|UAU^{\dagger}\|_{\infty}=\|A\|_{\infty}$~\cite{PhysRevA.78.012355}.
Using the first two, for any operator $X$,
\begin{align}\label{EqS1b}
\|\mathcal{L}(t)X\|_{\infty}=\|[H_{\text{SE}}(t),X]\|_{\infty}
\le\|H_{\text{SE}}(t)X\|_{\infty}+\|XH_{\text{SE}}(t)\|_{\infty}
\le 2\|H_{\text{SE}}\|_{\infty}\|X\|_{\infty},
\end{align}
where the last step uses unitary invariance, which gives
$\|H_{\text{SE}}(t)\|_{\infty}=\|H_{\text{SE}}\|_{\infty}$ since the
interaction picture is a unitary conjugation. For a density matrix,
$\|\rho\|_{\infty}\le1$: the operator norm of a Hermitian operator is its
largest eigenvalue in magnitude, and the eigenvalues of $\rho$ are
nonnegative and sum to unity, so none can exceed one. Applying
Eq.~\eqref{EqS1b} $n$ times and using that the ordered time domain has
volume $\tau^{n}/n!$, the $n$-th term of Eq.~\eqref{EqS1} is bounded by
$\|\Phi^{(n)}_{\tau}\|_{\infty}\le(2\|H_{\text{SE}}\|_{\infty}\tau)^{n}/n!$.
With $y\equiv\|H_{\text{SE}}\|_{\infty}\tau$ and $(k+3)!\ge 6\,k!$,
\begin{align}
\Big\|\Phi_{\tau}-\sum_{n=0}^{2}\Phi^{(n)}_{\tau}\Big\|_{\infty}
\le\sum_{n\ge3}\frac{(2y)^{n}}{n!}
\le\frac{(2y)^{3}}{6}\sum_{k\ge0}\frac{(2y)^{k}}{k!}
=\frac{4}{3}\,y^{3}e^{2y},
\end{align}
so the truncation error is
$\mathcal{O}\big((\|H_{\text{SE}}\|_{\infty}\tau)^{3}\big)$ and is controlled
once $\|H_{\text{SE}}\|_{\infty}\tau\ll1$.

\section{Derivation of the average gate fidelity expression in the frequency domain}
Here we present the details of the derivation of the average gate fidelity in the frequency domain, as given in Eq. (4). Using $\Tr_S\hat{A}_{\alpha}=0$, we can write Eq. (3) as:
\begin{align}
\overline{\mathcal{I}}=\sum_{\alpha,\beta}\int^\tau_0&\!\!\!dt_1\!\int^{t_1}_0\!\!\!dt_2\, \bigl[S_{\alpha\beta}(t_1-t_2)+S_{\beta\alpha}(t_2-t_1)\bigr]\frac{\Tr_S[\hat A_\alpha(t_1)\hat A_\beta(t_2)]}{d+1}.
\end{align}
Due to cyclicity of the trace $\Tr_S[\hat A_\alpha(t_1)\hat A_\beta(t_2)]=\Tr_S[\hat A_\alpha(t_1-t_2)\hat A_\beta]$, the time dependence in the integrand appears only through the time difference $s=t_1-t_2$. We can exploit the symmetry of the integration domain by performing the change of variables $t_1=\tilde{t}+s/2$ and $t_2=\tilde{t}-s/2$, which allows one to write the average gate infidelity as:
\begin{align}
    \overline{\mathcal{I}}\!&=\!\frac{1}{d+1}\sum_{\alpha,\beta}\int_{-\tau}^{\tau}ds(\tau-|s|)\frac{S_{\alpha\beta}(s)+S_{\beta\alpha}(-s)}{2}\Tr_S[\hat A_\alpha(s)\hat A_\beta].
\end{align}
One can immediately recognize $[S_{\alpha\beta}(s)+S_{\beta\alpha}(-s)]/2=S^{C}_{\alpha\beta}(s)$ to be the symmetrized correlator, which encodes the classical component of the noise \cite{Clerk2008IntroductionTQ}. 
To gain further insight, we now transform the expression to the frequency domain. Using the convention $K_{\alpha\beta}({\omega})= \int dt\,e^{i{\omega}t}\,\mathrm{Tr}_S[\hat{A}_\alpha(t)\hat{A}_\beta]$ and $S^C(\omega)= \int dt\, e^{i\omega t} S^C(t)$, and applying the identity $\int_{-\tau}^{\tau} (\tau - |s|)\, e^{-i \tilde{\omega} s} \, ds = 4\sin^2\left(\tilde{\omega}\tau/2\right)/\tilde{\omega}^{2}$, we obtain Eq. (4).

\section{Phase of the noise correlator}
For spatially correlated frequency noise $\alpha\neq\beta$, the noise spectrum can be complex-valued: \newline$S_{\alpha\beta}(\omega)=|S_{\alpha\beta}(\omega)|e^{i\theta_{\alpha\beta}(\omega)}$. The corresponding classical noise spectrum in Eq. (4) is then $S^{C}_{\alpha\beta}(\omega)=(|S_{\alpha\beta}(\omega)|e^{i\theta_{\alpha\beta}(\omega)}+|S_{\beta\alpha}(-\omega)|e^{i\theta_{\beta\alpha}(-\omega)})/2$. Depending on the phase $\theta_{\alpha\beta}(\omega)$, the contribution of spatially correlated frequency noise to the infidelity can be negative, for instance, when a charge trap is located in between two qubits \cite{Yoneda2022}. However, this contribution is bounded by that of local noise, which ensures the total infidelity remains positive.

We now demonstrate this bound. First, the spatially correlated noise spectrum is constrained by $|S_{\alpha\beta}^{C}(\omega)|^{2}\leq S_{\alpha\alpha}^{C}(\omega)S_{\beta\beta}^{C}(\omega)$ \cite{Clerk2008IntroductionTQ}. Second, a similar bound exists for the effective filter function: $|F_{\alpha\beta}(\omega,\tau)|^{2}\leq F_{\alpha\alpha}(\omega,\tau)F_{\beta\beta}(\omega,\tau)$. Applying the Cauchy–Schwarz inequality \cite{Nielsen_Chuang_2010} to the filter function in Eq. (6) yields:
\begin{align}
|\sum_{nm}\hat{A}_{\alpha}^{nm}\hat{A}_{\beta}^{mn} F_{\text{FID}}(\omega-\omega_{nm},\tau)|^{2}\leq (\sum_{nm}|\hat{A}_{\alpha}^{nm}|^{2}F_{\text{FID}}(\omega-\omega_{nm},\tau))(\sum_{nm}|\hat{A}_{\beta}^{mn}|^{2}F_{\text{FID}}(\omega-\omega_{nm},\tau)).
\end{align}
The above inequalities indicate that both the magnitude of the correlated noise spectrum $|S_{\alpha\beta}^{C}(\omega)|$ and its filter function $|F_{\alpha\beta}(\omega,\tau)|$ are bounded by their local counterparts. These inequalities, in fact, are the $2\times2$ minor conditions of a stronger statement: both $S^{C}_{\alpha\beta}(\omega)$ and $F_{\alpha\beta}(\omega,\tau)$ are positive-semidefinite matrices in the channel indices for every $\omega$, the former by spectral positivity of stationary noise \cite{Clerk2008IntroductionTQ}, the latter because Eq.~(6) expresses it as a sum of rank-one Gram matrices, $F_{\alpha\beta}(\omega,\tau)=\sum_{nm}\hat{A}^{nm}_{\alpha}(\hat{A}^{nm}_{\beta})^{*}F_{\mathrm{FID}}(\omega-\omega_{nm},\tau)$, with nonnegative weights $F_{\mathrm{FID}}\geq0$. The integrand of Eq.~(4), $\sum_{\alpha\beta}S^{C}_{\alpha\beta}F_{\alpha\beta}=\Tr[S^{C}(\omega)F^{*}(\omega,\tau)]$, is therefore nonnegative at every frequency, guaranteeing a positive total infidelity.

\section{Implementation of different gates in spin-qubit systems}

In this section, we describe how to implement the SWAP, CZ, and $\sqrt{i\text{SWAP}}$ gates in spin-qubit systems by tuning two key parameters in Eq. (8): the detuning $\delta=|\Delta_{1}-\Delta_{2}|/J$ and the rotation angle $\theta$, in the regime $\Delta_1,\Delta_2\gg J$.

\subsection{\text{SWAP} and \text{CZ} gates}
We first consider the case $\theta = 0$, for which the system Hamiltonian reads: 
\begin{equation}
H_{S}
=
\frac{\Delta_{1}}{2}\sigma_{1}^{z}
+
\frac{\Delta_{2}}{2}\sigma_{2}^{z}
+
\frac{J}{4}\,\boldsymbol{\sigma}_{1}\cdot\boldsymbol{\sigma}_{2}.
\end{equation}
We move to the rotating frame defined by $R(t)
=
\exp\!\left(i\Delta_{1}\sigma_{1}^{z}t/2\right)
\otimes
\exp\!\left(i\Delta_{2}\sigma_{2}^{z}t/2\right)$. We use $\sigma_i^{\pm}=(\sigma_i^x\pm i\sigma_i^y)/2$, so $(\sigma_i^+)^{\dagger}=\sigma_i^-$; Pauli operators on distinct qubits act on distinct tensor factors and commute.

In this frame, the Hamiltonian becomes:
\begin{equation}
\tilde{H}_{S}
=
\frac{J}{4}
\left[
2\!\left(
e^{i(\Delta_{1}-\Delta_{2})t}\sigma_{1}^{+}\sigma_{2}^{-}
+
e^{-i(\Delta_{1}-\Delta_{2})t}\sigma_{1}^{-}\sigma_{2}^{+}
\right)
+
\sigma_{1}^{z}\sigma_{2}^{z}
\right].
\end{equation}

When the detuning vanishes, $\delta=0$, the Hamiltonian reduces to the isotropic Heisenberg interaction:
\begin{equation}
\tilde{H}_{S}
=
\frac{J}{4}\,\boldsymbol{\sigma}_{1}\cdot\boldsymbol{\sigma}_{2}.
\end{equation}
In contrast, for large detuning $\delta \gg 1$, the fast-oscillating transverse terms average out under the rotating-wave approximation, yielding an effective Ising interaction
\begin{equation}
\tilde{H}_{S}
=
\frac{J}{4}\sigma_{1}^{z}\sigma_{2}^{z}.
\end{equation}

The SWAP gate can be expressed in terms of Pauli operators as $\mathrm{SWAP}
=
\frac{1}{2}
\left(
I + \boldsymbol{\sigma}_{1}\cdot\boldsymbol{\sigma}_{2}
\right)$ \cite{Nielsen_Chuang_2010}. Using this identity, the Hamiltonian at zero detuning can be rewritten as 
$\tilde{H}_{S}
=
\frac{J}{4}
\left(
2\,\mathrm{SWAP} - I
\right)$. The corresponding time-evolution operator is:
\begin{equation}
U_{S}(\tau)=
\exp\!\left(-i\tilde{H}_{S}\tau\right)
=
\exp\!\left(iJ\tau I/4\right)
\exp\!\left(-iJ\tau\text{SWAP}/2\right),
\end{equation}
where the first factor represents a global phase and can be neglected.  Using the identity $\exp(-i\theta\,\text{SWAP})=\cos\theta\,I-i\sin\theta\,\text{SWAP}$, which is a direct result of $\mathrm{SWAP}^{2}=I$, we obtain
\begin{equation}
U_{S}(\tau)
\sim
\cos\!\left(\frac{J\tau}{2}\right) I
-
i\sin\!\left(\frac{J\tau}{2}\right)\text{SWAP},
\end{equation}
where the symbol $\sim$ denotes equality up to a global phase. Choosing $\frac{J\tau}{2}=\frac{\pi}{2}$, $\tau=\pi/J$, yields the $\mathrm{SWAP}$ gate: $U_{S}(\tau){\sim -i\,\text{SWAP}}\sim\text{SWAP}$. Hence, the SWAP gate can be implemented using an isotropic exchange interaction and a gate time $\tau=\pi/J$.

The CZ gate can be implemented through the Ising interaction: $\tilde{H}_{S}=\frac{J}{4}\sigma_{1}^{z}\sigma_{2}^{z}=\text{diag}\{J/4,-J/4,-J/4,J/4\}$. The corresponding time-evolution operator is:
\begin{align}
U_{S}(\tau)&=\text{diag}\{\exp(-iJ\tau/4),\exp(iJ\tau/4),\exp(iJ\tau/4),\exp(-iJ\tau/4)\}\notag\\
&\sim\text{diag}\{1,\exp(iJ\tau/2),\exp(iJ\tau/2),1\}.
\end{align}
Apply the single-qubit rotations: $Z_{1}(-J\tau/2)Z_{2}(-J\tau/2)=\text{diag}\{1,\exp(-iJ\tau/2),\exp(-iJ\tau/2),\exp(-iJ\tau)\}$. Then:
\begin{align}
    Z_{1}(-J\tau/2)Z_{2}(-J\tau/2)U_{S}(\tau)\sim\text{diag}\{1,1,1,\exp(-iJ\tau)\}.
\end{align}
Choosing $J\tau=\pi$, $\tau=\pi/J$, we obtain: 
\begin{equation}
Z_{1}(-J\tau/2)Z_{2}(-J\tau/2)U_{S}(\tau)\sim\text{diag}\{1,1,1,-1\}=\text{CZ}.
\end{equation}
Therefore, the CZ gate can be implemented up to single-qubit rotations with the Ising interaction and gate time $\tau=\pi/J$.

\subsection{$\sqrt{\text{iSWAP}}$ gate}
We next consider the case $\theta=\pi/2$ and $\delta=0$, for which the system Hamiltonian reads:
\begin{equation}
    H_{S}=\frac{\Delta_{1}}{2}\sigma_{1}^{z}+\frac{\Delta_{2}}{2}\sigma_{2}^{z}+
\frac{J}{4}[\sigma_{1}^{y}\sigma_{2}^{y}+(\sigma_{1}^{z}\sigma_{2}^{x}-\sigma_{1}^{x}\sigma_{2}^{z})].
\end{equation}
We move to the rotating frame defined by $R(t)
=
\exp\!\left(i\Delta_{1}\sigma_{1}^{z}t/2\right)
\otimes
\exp\!\left(i\Delta_{2}\sigma_{2}^{z}t/2\right)$.
In this frame, and after applying the rotating-wave approximation, the effective Hamiltonian becomes:
\begin{equation}
\tilde{H}_{S}
=
\frac{J}{8}(\sigma_{1}^{x}\sigma_{2}^{x}+\sigma_{1}^{y}\sigma_{2}^{y})=\frac{J}{4}(\sigma_{1}^{-}\sigma_{2}^{+}+\sigma_{1}^{+}\sigma_{2}^{-}).
\end{equation}
Restricting to the subspace
$\mathcal{H}=\mathrm{span}\{\ket{10},\ket{01}\}$,
the effective Hamiltonian takes the form
\begin{equation}
\tilde{H}_{S}^{\mathrm{sub}}
=
\frac{J}{4} X,
\end{equation}
where $X$ is the Pauli-$X$ operator in this subspace. The corresponding time-evolution operator is \newline$U_{S}^{\text{sub}}(\tau)=\exp(-iJX\tau/4)=\cos(J\tau/4)I-i\sin(J\tau/4)X$.
In the full two-qubit Hilbert space, the evolution operator is therefore:
\begin{align}
U_{S}(\tau) =
\begin{pmatrix}
1 & 0 & 0 & 0 \\
0 & \cos(J\tau/4) & -i\sin(J\tau/4) & 0 \\
0 & -i\sin(J\tau/4) & \cos(J\tau/4) & 0 \\
0 & 0 & 0 & 1
\end{pmatrix}.
\end{align}
Choosing $J\tau/4=\pi/4$ (i.e., $\tau=\pi/J$) realizes the $\sqrt{\mathrm{iSWAP}}$ gate up to single-qubit $z$ rotations: the resulting evolution equals $(\sqrt{\mathrm{iSWAP}})^{\dagger}$, which is mapped to $\sqrt{\mathrm{iSWAP}}$ by conjugation with $\sigma^{z}$ on either qubit.
We note that $J\tau/4=3\pi/4$ yields $Z_{1}(\pi)Z_{2}(\pi)\,\sqrt{\mathrm{iSWAP}}$, i.e., again the $\sqrt{\mathrm{iSWAP}}$ operation up to single-qubit $z$ rotations. Here we choose the shorter positive bare evolution, $J\tau/4=\pi/4$.

\section{Derivation of the analytical expression for the infidelity under $1/f$ noise}
The truncation in Eq.~\eqref{EqS1} was justified in Sec.~\ref{secI}
for a bounded coupling Hamiltonian, where the error is controlled by
$\|H_{\text{SE}}\|_{\infty}\tau$. 
For the idealized Gaussian power-law model below, this operator-norm criterion does not provide a finite control parameter. We therefore give a separate bound using integrated noise correlations.

For this higher-order estimate, we additionally assume centered stationary Gaussian noise with $0< \eta< 1$. Many independent weak fluctuators with no dominant contribution can motivate this approximation~\cite{RevModPhys.53.497,RevModPhys.86.361}. 
With $H_{\text{SE}}(t)=\sum_{\alpha}\hat A_{\alpha}(t)\otimes\hat B_{\alpha}(t)$, the
$m$-th term of Eq.~\eqref{EqS1} acting on the initial product state is
\begin{align}\label{EqS1c}
\Phi^{(m)}_{\tau}(\rho_{S}\otimes\rho_{E})
=(-i)^{m}\!\!\!\int\limits_{0<t_{m}<\cdots<t_{1}<\tau}\!\!\!\!\!d^{m}t\;
\bigl[H_{\text{SE}}(t_{1}),\bigl[H_{\text{SE}}(t_{2}),\cdots
[H_{\text{SE}}(t_{m}),\rho_{S}\otimes\rho_{E}]\cdots\bigr]\bigr].
\end{align}
Each commutator is the difference of a left and a right multiplication, so the
nested commutator expands into $2^{m}$ terms, each carrying one channel index
per time slot; since operators on $\mathcal{H}_{S}\otimes\mathcal{H}_{E}$
multiply factor by factor, every term is a product of a system and an
environment factor,
\begin{align}\label{EqS1d}
&\bigl[H_{\text{SE}}(t_{1}),\cdots
[H_{\text{SE}}(t_{m}),\rho_{S}\otimes\rho_{E}]\cdots\bigr]
\nonumber\\
&\quad=\sum_{\alpha_{1}\cdots\alpha_{m}}\;\sum_{r=1}^{2^{m}}\pm\,
\bigl(\hat A_{\alpha_{i_{1}}}\!(t_{i_{1}})\!\cdots\hat A_{\alpha_{i_{p}}}(t_{i_{p}})\,\rho_{S}\,
\hat A_{\alpha_{j_{q}}}\!(t_{j_{q}})\!\cdots\hat A_{\alpha_{j_{1}}}(t_{j_{1}})\bigr)\otimes
\bigl(\hat B_{\alpha_{i_{1}}}\!(t_{i_{1}})\cdots
\hat B_{\alpha_{i_{p}}}\!(t_{i_{p}})\,\rho_{E}\,
\hat B_{\alpha_{j_{q}}}\!(t_{j_{q}})\cdots
\hat B_{\alpha_{j_{1}}}\!(t_{j_{1}})\bigr),
\end{align}
where $\{i\}$ and $\{j\}$ collect the left- and right-acting slots of the
$r$-th term, $p+q=m$, and the signs play no role below. The channel index is
tied to its time slot, appearing on the same $\hat A$ and $\hat B$, since both
originate from the same factor of $H_{\text{SE}}$. Tracing out the environment
and using cyclicity of the trace, each environment factor becomes a single
$m$-time correlator,
\begin{align}\label{EqS1e}
\mathcal{E}^{(m)}(\rho_{S})=(-i)^{m}\!\!\int\!\!d^{m}t
\sum_{\alpha_{1}\cdots\alpha_{m}}\;\sum_{r=1}^{2^{m}}\pm\,
\hat A_{\alpha_{i_{1}}}\!(t_{i_{1}})\!\cdots\rho_{S}\cdots
\hat A_{\alpha_{j_{1}}}\!(t_{j_{1}})\;
\bigl\langle\hat B_{\alpha_{j_{q}}}\!(t_{j_{q}})\cdots
\hat B_{\alpha_{i_{p}}}\!(t_{i_{p}})\bigr\rangle,
\end{align}
with $\mathcal{E}^{(m)}$ the $m$-th order term of the reduced channel of
Eq.~\eqref{channel}. The unbounded operators $\hat B_{\alpha}$ now appear only
inside correlators, so only the $\hat A_{\alpha}$, which are bounded system operators formed from sums of Pauli strings, are
normed; writing
$\|\hat A\|_{\infty}\equiv\max_{\alpha}\|\hat A_{\alpha}\|_{\infty}$, which is
time independent because the interaction picture acts by unitary conjugation,
and using $\|\rho_{S}\|_{\infty}\le1$,
\begin{align}\label{EqS1f}
\bigl\|\mathcal{E}^{(m)}\bigr\|_{\infty}\le
\|\hat A\|_{\infty}^{m}\sum_{\alpha_{1}\cdots\alpha_{m}}\;\sum_{r=1}^{2^{m}}
\int\!\!d^{m}t\;
\Bigl|\bigl\langle\hat B_{\alpha_{j_{q}}}\!(t_{j_{q}})\cdots
\hat B_{\alpha_{i_{p}}}\!(t_{i_{p}})\bigr\rangle\Bigr| .
\end{align}
Wick's theorem determines all correlators from the two-point functions.
Correlators with an odd number of operators vanish, since they admit no
complete pairing, and for $m=2n$
\begin{align}\label{EqS1g}
\bigl|\langle\hat B_{\alpha_{1}}\!(t_{1})\cdots
\hat B_{\alpha_{2n}}\!(t_{2n})\rangle\bigr|
\le\sum_{\text{pairings}}\;\prod_{(i,j)\in P}
\bigl|S_{\alpha_{i}\alpha_{j}}(t_{i}-t_{j})\bigr|,
\end{align}
a sum of $(2n-1)!!$ terms. As $S_{\alpha\beta}(-t)=S_{\beta\alpha}(t)^{*}$,
these magnitudes are unchanged under exchange of the two slots of a pair, so
the same bound applies to every one of the $2^{2n}$ correlators in
Eq.~\eqref{EqS1f}. 
On the full time hypercube, each pairing assigns the $2n$ times and channel indices to $n$ disjoint pairs. The integrals and channel sums then factorize into $n$ identical factors,
\begin{align}\label{EqS1h}
\kappa\equiv\sum_{\alpha\beta}\int_{0}^{\tau}\!\!\int_{0}^{\tau}\!dt_{1}dt_{2}\,
\bigl|S_{\alpha\beta}(t_{1}-t_{2})\bigr|
=\Bigl(\sum_{\alpha\beta}|c_{\alpha\beta}|\Bigr)
\int\!\frac{d\omega}{2\pi}\,S(\omega)\,F_{\text{FID}}(\omega,\tau),
\end{align}
the second form using $S_{\alpha\beta}(t)=c_{\alpha\beta}S(t)$ with $S(t)\ge0$,
as holds here. Since $F_{\text{FID}}=4\sin^{2}(\omega\tau/2)/\omega^{2}$ decays
as $2/\omega^{2}$, the integrand is $|\omega|^{\eta-1}$ at small $\omega$ and
$|\omega|^{\eta-3}$ at large $\omega$, so $\kappa$ is finite for all $0<\eta<2$.

After summing over all channel indices,
the Wick bound is symmetric under permutations of the time slots, and the hypercube is the union of $(2n)!$ congruent ordered
simplices, so the time-ordered domain of Eq.~\eqref{EqS1f} supplies a factor
$1/(2n)!$. With $(2n-1)!!=(2n)!/(2^{n}n!)$ the factorials cancel,
\begin{align}\label{EqS1i}
\bigl\|\mathcal{E}^{(2n)}\bigr\|_{\infty}
\le\frac{(2\|\hat A\|_{\infty})^{2n}}{(2n)!}\,(2n-1)!!\;\kappa^{n}
=\frac{K^{n}}{n!},\qquad K\equiv2\|\hat A\|_{\infty}^{2}\,\kappa .
\end{align}
Everything beyond second order is therefore bounded by
$\sum_{n\ge2}K^{n}/n!=e^{K}-1-K\simeq K^{2}/2$, so the truncation is
controlled once $K\ll1$, exactly as $\|H_{\text{SE}}\|_{\infty}\tau\ll1$
controls the bounded case. Where $\|H_{\text{SE}}\|_{\infty}\tau$ counts one
power of the coupling per order, $K$ counts one Wick pair, which is why it
remains finite where $\|H_{\text{SE}}\|_{\infty}\tau$ does not.

Evaluating both $\kappa$ and the infidelity itself requires the scalar integral
$X(\tau,\omega_{nm})$, which is distinct from the Pauli operator $X$ used
above:
\begin{equation}
X(\tau,\omega_{nm})
= \int_{-\infty}^{\infty} d\omega \;
\frac{1}{|\omega|^{1-\eta}}\left\{
\frac{\sin^2\!\left(\frac{\omega-\omega_{nm}}{2} \tau \right)}{\left( \frac{\omega-\omega_{nm}}{2} \right)^2}+\frac{\sin^2\!\left(\frac{\omega+\omega_{nm}}{2} \tau \right)}{\left( \frac{\omega+\omega_{nm}}{2} \right)^2}\right\}.
\end{equation}
The integrand is even in $\omega$, so we can write
\begin{equation}
X(\tau,\omega_{nm})
= 2\int_{0}^{\infty} d\omega \;
\frac{1}{\omega^{1-\eta}}\left\{
\frac{\sin^2\!\left(\frac{\omega-\omega_{nm}}{2} \tau \right)}{\left( \frac{\omega-\omega_{nm}}{2} \right)^2}+\frac{\sin^2\!\left(\frac{\omega+\omega_{nm}}{2} \tau \right)}{\left( \frac{\omega+\omega_{nm}}{2} \right)^2}\right\}.
\end{equation}
Then, we can apply the identity
\begin{equation}
\frac{\sin^2(ax)}{x^2}
= 2 \int_{0}^{a} du\; (a-u)\cos(2ux)
\end{equation}
to rewrite the integrand:
\begin{align}
X(\tau,\omega_{nm})
&= 4 \int_{0}^{\infty} d\omega\;\frac{1}{\omega^{1-\eta}}
\int_{0}^{\tau} du\; (\tau-u)
\bigl[
    \cos\!\left((\omega-\omega_{nm})u\right)
    + \cos\!\left((\omega+\omega_{nm})u\right)
\bigr] \\
&= 8 \int_{0}^{\infty} d\omega\;\frac{1}{\omega^{1-\eta}}
\int_{0}^{\tau} du\; (\tau-u)
\cos\!\left(\omega u\right)\cos\!\left(\omega_{nm}u\right),
\end{align}
after which we can switch the order of integration to obtain
\begin{equation}
X(\tau,\omega_{nm})
= 8 \int_{0}^{\tau} du\; (\tau-u)\cos(\omega_{nm} u)
\int_{0}^{\infty} d\omega\;
\frac{\cos(\omega u)}{\omega^{1-\eta}}.
\end{equation}
For $0<\eta<1$, evaluating the integral over $\omega$ using the identity
\begin{equation}
\int_{0}^{\infty} d\omega\;
\frac{\cos(\omega u)}{\omega^{1-\eta}}
= |u|^{-\eta} \cos(\pi\eta/2)\, \Gamma[\eta]
\end{equation}
gives
\begin{equation}
X(\tau,\omega_{nm})
= 8 \cos(\pi\eta/2)\Gamma[\eta]
\int_{0}^{\tau} du\; (\tau-u)u^{-\eta}\cos(\omega_{nm} u).
\end{equation}
Finally, expanding $\cos(\omega_{nm} u)$ in its Taylor series and performing
the integral term by term yields
\begin{align}
X(\tau,\omega_{nm})
&=8 \cos(\pi\eta/2)\Gamma[\eta]
\sum_{k=0}^{\infty}
\frac{(-1)^k {\omega_{nm}^{2k}}}{(2k)!}
\int_{0}^{\tau} du\; (\tau-u)u^{2k-\eta}\\
&=8 \cos(\pi\eta/2)\Gamma[\eta]\,
\tau^{2-\eta}
\sum_{k=0}^{\infty}
\frac{(-1)^k}{(2k)!}
\frac{(\omega_{nm} \tau)^{2k}}
{[\eta - (2k+1)][\eta - (2k+2)]}.
\label{eq:X-series}
\end{align}
Note that $X(\tau,\omega_{nm})$ as defined above contains both frequency
components $\pm\omega_{nm}$, i.e., the contributions of both orderings $(n,m)$
and $(m,n)$ of each transition pair: since the single-channel integral
$\int_{-\infty}^{\infty}d\omega\,|\omega|^{\eta-1}\,
\sin^{2}[(\omega-\omega_{nm})\tau/2]/[(\omega-\omega_{nm})/2]^{2}$ is even in
$\omega_{nm}$, $X(\tau,\omega_{nm})$ equals twice this integral. Because
Eq.~(9) is written with the unrestricted double sum $\sum_{nm}$, the
compensating factor $1/2$ must therefore be included. Restoring the constants
from Eq.~(4),
$\overline{\mathcal{I}}=\frac{T^{\eta-2}}{2(d+1)}\sum_{\alpha\beta}\sum_{nm}
\hat{A}_{\alpha}^{nm}\hat{A}_{\beta}^{mn}\,X(\tau,\omega_{nm})$, which for
$d=4$ yields the prefactor $8/[2(d+1)]=4/5$ of Eq.~(9).

We can now evaluate $K$. For the covariance used here all $\hat B_{\alpha}$
coincide, so the three channels combine into the single channel
$H_{\text{SE}}(t)=\hat A_{\theta}(t)\otimes\hat B(t)$ with
$\hat A_{\theta}(t)=\sum_{\alpha}\hat A_{\alpha}(t)$. The sum in
Eq.~\eqref{EqS1h} then contains a single term with $c=1$, so
$\kappa=\int(d\omega/2\pi)S(\omega)F_{\text{FID}}(\omega,\tau)$ and
$\|\hat A\|_{\infty}=\|\hat A_{\theta}\|_{\infty}$. This is
$X(\tau,\omega_{nm})$ at $\omega_{nm}=0$, where every term of
Eq.~\eqref{eq:X-series} with $k\ge1$ vanishes; restoring the spectral
amplitude and the factor $1/2$ of Eq.~(9),
\begin{equation}\label{eq:kappavalue}
\kappa=\tfrac{1}{2}T^{\eta-2}X(\tau,0)
=\frac{4\,\Gamma(\eta)\cos(\pi\eta/2)}{(1-\eta)(2-\eta)}
\left(\frac{\tau}{T}\right)^{2-\eta}.
\end{equation}
For the parameters of Fig.~2, namely $J=1$, $\tau=\pi/J$, $T=200\pi\tau$ and
$\eta=0.01$, Eq.~\eqref{eq:kappavalue} gives $\kappa=5.45\times10^{-4}$; since
$\kappa$ depends only on the spectrum and the gate duration, and all three
operating points use $\tau=\pi/J$, this single value applies to each of them,
which differ only through $\|\hat A_{\theta}\|_{\infty}$, the largest
eigenvalue of $\hat A_{\theta}$ in magnitude. At the SWAP and CZ points, which
both have $\theta=0$, the operator
$\hat A_{0}=\sigma^{z}_{1}+\sigma^{z}_{2}+\vec\sigma_{1}\!\cdot\!\vec\sigma_{2}$
is already diagonal in the singlet--triplet basis, with spectrum
$\{3,1,-1,-3\}$, so $\|\hat A_{0}\|_{\infty}=3$. At the $\sqrt{i\text{SWAP}}$
point, $\theta=\pi/2$, diagonalizing
$\hat A_{\pi/2}=\sigma^{z}_{1}+\sigma^{z}_{2}+\sigma^{y}_{1}\sigma^{y}_{2}
+\sigma^{z}_{1}\sigma^{x}_{2}-\sigma^{x}_{1}\sigma^{z}_{2}$ gives
$\{2.604,\,1,\,-0.110,\,-3.494\}$, so $\|\hat A_{\pi/2}\|_{\infty}=3.494$.
Hence $K=9.8\times10^{-3}$ and $1.33\times10^{-2}$ respectively, so the
neglected orders are bounded by $e^{K}-1-K\le9\times10^{-5}$ at these three operating points, subject to the stated Gaussian assumptions. Note that this is an absolute remainder bound, not a uniform relative-error estimate over all parameter sweeps.

\section{Filter functions of different gates in spin-qubit systems}
Here we derive the filter functions of SWAP, CZ, and $\sqrt{\text{iSWAP}}$ gates in spin-qubit systems.

\subsection{SWAP gate}
We first consider the SWAP gate. The system Hamiltonian is
\begin{equation}
H_{S}
=
\frac{\Delta}{2}\sigma_{1}^{z}
+
\frac{\Delta}{2}\sigma_{2}^{z}
+
\frac{J}{4}\,\boldsymbol{\sigma}_{1}\cdot\boldsymbol{\sigma}_{2}.
\end{equation}
The eigenstates of this Hamiltonian consist of the triplet states
$\ket{T_{+1}}=\ket{\uparrow\uparrow}$,
$\ket{T_{0}}=(\ket{\uparrow\downarrow}+\ket{\downarrow\uparrow})/\sqrt{2}$,
$\ket{T_{-1}}=\ket{\downarrow\downarrow}$,
and the singlet state
$\ket{S}=(\ket{\uparrow\downarrow}-\ket{\downarrow\uparrow})/\sqrt{2}$.
The transition frequency between $\ket{T_{0}}$ and $\ket{S}$ is $J$.
The relevant system operators are
\begin{equation}
\hat{A}\in
\left\{
\sigma_{1}^{z},\,
\sigma_{2}^{z},\,
\sigma_{1}^{z}\sigma_{2}^{z}+\,
\sigma_{1}^{x}\sigma_{2}^{x}+\,
\sigma_{1}^{y}\sigma_{2}^{y}
\right\}.
\end{equation}

\begin{figure*}[htbp]
\centering
\includegraphics[width=\textwidth]{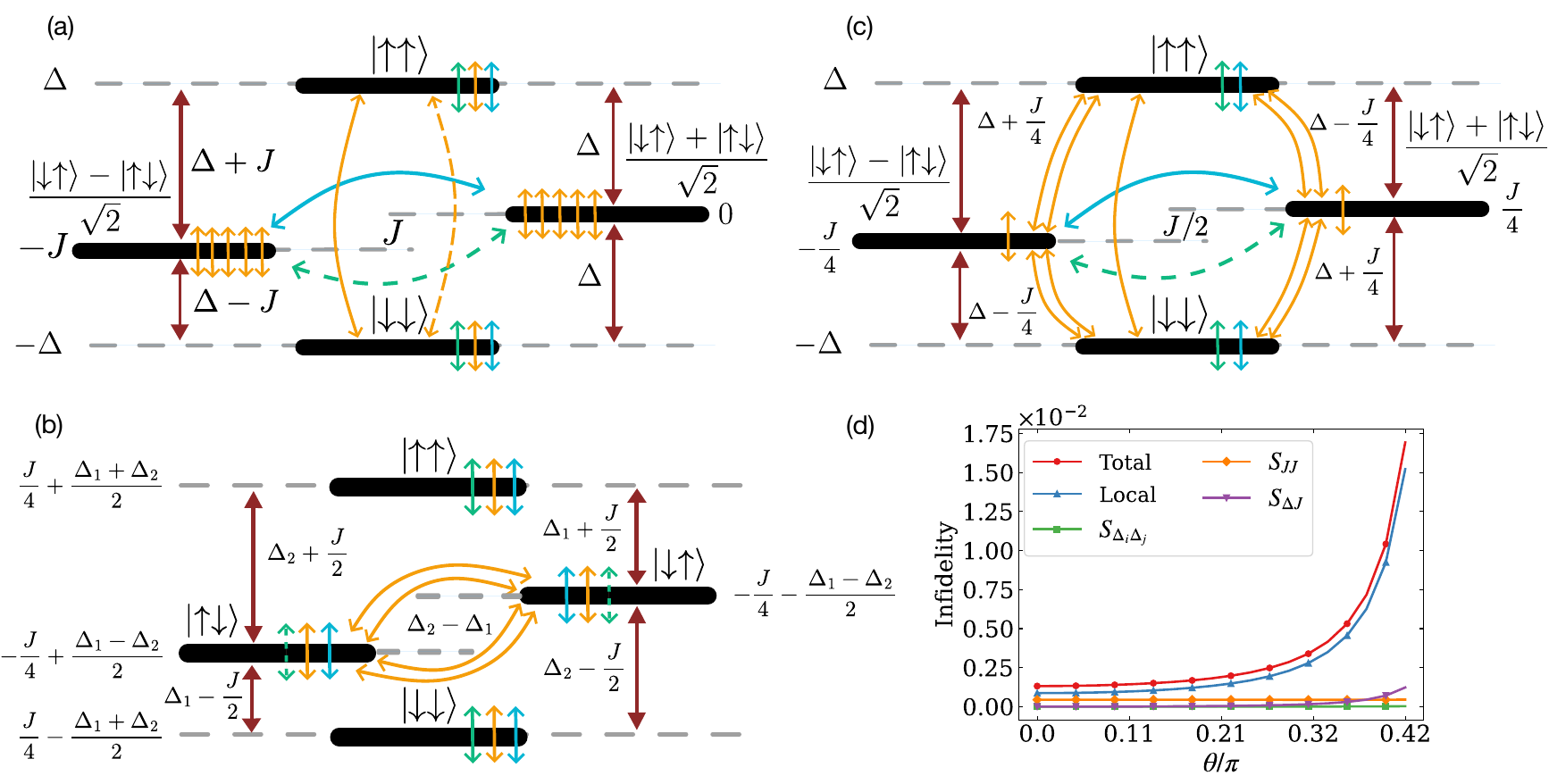}
\caption{Schematic energy diagrams for the (a) SWAP, (b) CZ, and (c) $\sqrt{\text{iSWAP}}$ gates. Blue double-arrow lines indicate dephasing and transition channels due to local noise. Green double-arrow lines represent dephasing and transition channels arising from spatially correlated frequency noise, while orange lines correspond to dephasing and transition channels induced by interaction noise. Solid lines denote a weight of $1$, and dashed lines denote a weight of $-1$. The system operators responsible for each channel and the calculation of the corresponding weights are presented in Sec. VI. (d) Infidelity of the CZ gate as a function of $\theta$. For $0\leq\theta<\pi/2$, the effective interaction strength is $J\cos\theta$, and the gate time $\tau$ required to implement the CZ gate is $\pi/(J\cos\theta)$,  which increases as $\theta$ increases. The parameters used in (d) are: $J=1$, $\Delta_{1}=10$, $\Delta_{2}=\Delta_{1}+\delta\cross J$, $\delta=10$, $\tau=\pi/(J\cos\theta)$, $T=2\pi\cross 100 \cross(\pi/J)$, $\eta=0.01$.}
\label{fig:s1}
\end{figure*}

\paragraph*{(1) Local noise.}
Local noise is described by the system operators
$\hat{A}_{\alpha}=\hat{A}_{\beta}=\sigma_{1,2}^{z}$.
These operators induce
(i) pure dephasing of the states $\ket{T_{\pm1}}$ at zero frequency, with weight $\bra{T_{\pm1}}\sigma_{1,2}^{z}\ket{T_{\pm1}}
\bra{T_{\pm1}}\sigma_{1,2}^{z}\ket{T_{\pm1}}=1$,
and
(ii) transitions between $\ket{T_{0}}$ and $\ket{S}$ at frequency $J$, with weight $\bra{T_{0}}\sigma_{1,2}^{z}\ket{S}
\bra{S}\sigma_{1,2}^{z}\ket{T_{0}}=1$. Figure~\ref{fig:s1} shows the dephasing and transition channels within the energy diagrams. The corresponding effective filter function is
\begin{equation}
F_{\mathrm{local}}(\omega,\tau)
=\sum_{i=1}^{2}(F_{\mathrm{FID}}(\omega\pm0,\tau)
+
F_{\mathrm{FID}}(\omega\pm J,\tau)),
\end{equation}
where for any $x$ we use the shorthand $F_{\mathrm{FID}}(\omega\pm x,\tau)\equiv F_{\mathrm{FID}}(\omega+x,\tau)+F_{\mathrm{FID}}(\omega-x,\tau)$; in particular, $F_{\mathrm{FID}}(\omega\pm0,\tau)=2F_{\mathrm{FID}}(\omega,\tau)$.

\paragraph*{\color{black}(2) Spatially correlated frequency \color{black}noise.}
Spatially correlated frequency noise is characterized by the system operators
$\hat{A}_{\alpha}=\sigma_{1,2}^{z}$ and
$\hat{A}_{\beta}=\sigma_{2,1}^{z}$.
As in the local-noise case, these operators give rise to
(i) dephasing of the states $\ket{T_{\pm1}}$ at zero frequency,
and
(ii) transitions between $\ket{T_{0}}$ and $\ket{S}$ at frequency $J$.
However, the latter process carries an opposite sign,
$\bra{T_{0}}\sigma_{1,2}^{z}\ket{S}
\bra{S}\sigma_{2,1}^{z}\ket{T_{0}}=-1$. As a result, the effective filter function becomes
\begin{equation}
F_{S_{\Delta_{1}\Delta_{2}}}(\omega,\tau)
=
\sum_{i=1}^{2}(F_{\mathrm{FID}}(\omega\pm0,\tau)
-
F_{\mathrm{FID}}(\omega\pm J,\tau)).
\end{equation}

\paragraph*{(3) Interaction noise.}
Interaction noise originates from fluctuations in the exchange coupling and is described by the system operator
$\hat{A}=\sigma_{1}^{z}\sigma_{2}^{z}+\,
\sigma_{1}^{x}\sigma_{2}^{x}+\,
\sigma_{1}^{y}\sigma_{2}^{y}$.
This operator induces the following processes:
(i) for $\hat{A}_{\alpha}=\hat{A}_{\beta}=\sigma_{1}^{z}\sigma_{2}^{z}$,
pure dephasing of the states $\ket{T_{\pm1}}$ and $\ket{T_{0},S}$ at zero frequency, with unit weight;
(ii) for $\hat{A}_{\alpha}=\hat{A}_{\beta}=\sigma_{1}^{x}\sigma_{2}^{x}$ or $\sigma_{1}^{y}\sigma_{2}^{y}$,
pure dephasing of $\ket{T_{0}}$ and $\ket{S}$ at zero frequency, and transitions between $\ket{T_{+1}}$ and $\ket{T_{-1}}$ at frequency $2\Delta$, with unit weight;
(iii) for $\hat{A}_{\alpha}=\sigma_{1}^{x}\sigma_{2}^{x}$ and $\hat{A}_{\beta}=\sigma_{1}^{y}\sigma_{2}^{y}$ (or vice versa),
the same transitions occur but the transitions between $\ket{T_{+1}}$ and $\ket{T_{-1}}$ occur with the opposite sign, yielding a weight of $-1$.
All other combinations of system operators yield zero net weight. Collecting all nonvanishing contributions, the effective filter function for interaction noise is
\begin{equation}
F_{S_{JJ}}(\omega,\tau)
=
6\,F_{\mathrm{FID}}(\omega\pm0,\tau).
\end{equation}

\paragraph*{(4) Cross correlations.}
Cross-correlated noise is described by the system operators
$\hat{A}_{\alpha}\in\{\sigma_{1,2}^{z}\}$ and
$\hat{A}_{\beta}=\sigma_{1}^{z}\sigma_{2}^{z}+\,
\sigma_{1}^{x}\sigma_{2}^{x}+\,
\sigma_{1}^{y}\sigma_{2}^{y}$.
All such combinations yield zero net weight.
Therefore, the effective filter function for cross-correlated noise vanishes.

\subsection{CZ gate}
We next consider the CZ gate. In this regime, the effective system Hamiltonian is
\begin{equation}
H_{S}
=
\frac{\Delta_{1}}{2}\sigma_{1}^{z}
+
\frac{\Delta_{2}}{2}\sigma_{2}^{z}
+
\frac{J}{4}\,\sigma_{1}^{z}\sigma_{2}^{z},
\end{equation}
where $|\Delta_{1}-\Delta_{2}|\gg J$.
The eigenstates of this Hamiltonian are
$\{\ket{\uparrow\uparrow},\ket{\uparrow\downarrow},\ket{\downarrow\uparrow},\ket{\downarrow\downarrow}\}$.
The transition frequency between $\ket{\uparrow\downarrow}$ and $\ket{\downarrow\uparrow}$ is $|\Delta_{1}-\Delta_{2}|$.
The relevant system operators are
\begin{equation}
\hat{A}\in
\left\{
\sigma_{1}^{z},\,
\sigma_{2}^{z},\,
\sigma_{1}^{z}\sigma_{2}^{z}+\,
\sigma_{1}^{x}\sigma_{2}^{x}+\,
\sigma_{1}^{y}\sigma_{2}^{y}
\right\}.
\end{equation}

\paragraph*{(1) Local noise.}
Local noise induces
(i) pure dephasing of the states $\ket{\uparrow\uparrow}$ and $\ket{\downarrow\downarrow}$ at zero frequency, with unit weight, and
(ii) pure dephasing of the states $\ket{\uparrow\downarrow}$ and $\ket{\downarrow\uparrow}$ at zero frequency, also with unit weight.
The corresponding effective filter function is therefore
\begin{equation}
F_{\mathrm{local}}(\omega,\tau)
=
\sum_{i=1}^{2}(\,F_{\mathrm{FID}}(\omega\pm0,\tau)+F_{\mathrm{FID}}(\omega\pm0,\tau)).
\end{equation}

\paragraph*{\color{black}(2) Spatially correlated frequency \color{black}noise.}
Spatially correlated frequency noise induces pure dephasing of all computational basis states at zero frequency. The two parallel-spin states contribute with weight $+1$ each, and the two antiparallel-spin states with weight $-1$ each, leading to a complete cancellation.
As a result, the effective filter function vanishes,
\begin{equation}
F_{S_{\Delta_{1}\Delta_{2}}}(\omega,\tau)
=
\sum_{i=1}^{2}(\,F_{\mathrm{FID}}(\omega\pm0,\tau)
-
F_{\mathrm{FID}}(\omega\pm0,\tau))
=0.
\end{equation}

\paragraph*{(3) Interaction noise.}
Interaction noise arises from fluctuations in the exchange coupling and is described by the system operator
\begin{equation}
\hat{A}=\sigma_{1}^{z}\sigma_{2}^{z}+\,
\sigma_{1}^{x}\sigma_{2}^{x}+\,
\sigma_{1}^{y}\sigma_{2}^{y}.
\end{equation}
This operator induces the following processes:
(i) for $\hat{A}_{\alpha}=\hat{A}_{\beta}=\sigma_{1}^{z}\sigma_{2}^{z}$,
pure dephasing of all eigenstates at zero frequency, with unit weight;
(ii) for $\hat{A}_{\alpha}=\hat{A}_{\beta}=\sigma_{1}^{x}\sigma_{2}^{x}$ or $\sigma_{1}^{y}\sigma_{2}^{y}$,
transitions between $\ket{\uparrow\downarrow}$ and $\ket{\downarrow\uparrow}$ at frequency $|\Delta_{1}-\Delta_{2}|$, with unit weight;
(iii) for $\hat{A}_{\alpha}=\sigma_{1}^{x}\sigma_{2}^{x}$ and
$\hat{A}_{\beta}=\sigma_{1}^{y}\sigma_{2}^{y}$ (or vice versa),
the same transitions occur with unit weight.
All other combinations of system operators yield zero net weight.
Collecting all nonvanishing contributions, the effective filter function for interaction noise is
\begin{equation}
F_{S_{JJ}}(\omega,\tau)
=
2\,F_{\mathrm{FID}}(\omega\pm0,\tau)
+
4\,F_{\mathrm{FID}}(\omega\pm|\Delta_{1}-\Delta_{2}|,\tau).
\end{equation}

\paragraph*{(4) Cross correlations.}
Cross-correlated noise is described by the system operators
$\hat{A}_{\alpha}\in\{\sigma_{1,2}^{z}\}$ and
$\hat{A}_{\beta}=\sigma_{1}^{z}\sigma_{2}^{z}+\,
\sigma_{1}^{x}\sigma_{2}^{x}+\,
\sigma_{1}^{y}\sigma_{2}^{y}$.
All such combinations yield zero net weight.
Therefore, the effective filter function for cross-correlated noise vanishes.

\subsection{$\sqrt{\text{iSWAP}}$ gate}
We finally consider the $\sqrt{\text{iSWAP}}$ gate. In this regime, the effective system Hamiltonian is
\begin{align}
    H_{S}=\frac{\Delta}{2}\sigma_{1}^{z}+\frac{\Delta}{2}\sigma_{2}^{z}+\frac{J}{4}[\sigma_{1}^{y}\sigma_{2}^{y}+(\sigma_{1}^{z}\sigma_{2}^{x}-\sigma_{1}^{x}\sigma_{2}^{z})].
\end{align} 
The eigenstates of this Hamiltonian are, to leading order, in $J/\Delta$ the triplet states
$\ket{T_{+1}}=\ket{\uparrow\uparrow}$,
$\ket{T_{0}}=(\ket{\uparrow\downarrow}+\ket{\downarrow\uparrow})/\sqrt{2}$,
$\ket{T_{-1}}=\ket{\downarrow\downarrow}$,
and the singlet state
$\ket{S}=(\ket{\uparrow\downarrow}-\ket{\downarrow\uparrow})/\sqrt{2}$.
The transition frequency between $\ket{T_{0}}$ and $\ket{S}$ is $J/2$.  
Weak mixing between the $\ket{T_{\pm1}}$ and $\ket{T_{0},S}$ states allows existing cross correlations to contribute to the infidelity.
The relevant system operators are
\begin{equation}
\hat{A}\in
\left\{
\sigma_{1}^{z},\,
\sigma_{2}^{z},\,
\sigma_{1}^{y}\sigma_{2}^{y}\,
-\sigma_{1}^{x}\sigma_{2}^{z}+\,
\sigma_{1}^{z}\sigma_{2}^{x}
\right\}.
\end{equation}

\paragraph*{(1) Local noise.}
Local noise is described by the system operators
$\hat{A}_{\alpha}=\hat{A}_{\beta}=\sigma_{1,2}^{z}$.
These operators induce
(i) pure dephasing of the states $\ket{T_{\pm1}}$ at zero frequency, with unit weight and
(ii) transitions between $\ket{T_{0}}$ and $\ket{S}$ at frequency $J/2$, with unit weight.
The corresponding effective filter function to leading order in $J/\Delta$ is
\begin{equation}
F_{\mathrm{local}}(\omega,\tau)
=
\sum_{i=1}^{2}(\,F_{\mathrm{FID}}(\omega\pm0,\tau)
+
F_{\mathrm{FID}}(\omega\pm J/2,\tau)).
\end{equation}

\paragraph*{\color{black}(2) Spatially correlated frequency \color{black}noise.}
Spatially correlated frequency noise induces the same transition processes as local noise, but with the second process carrying an opposite sign. The effective filter function to leading order in $J/\Delta$ becomes
\begin{equation}
F_{S_{\Delta_{1}\Delta_{2}}}(\omega,\tau)
=
\sum_{i=1}^{2}(\,F_{\mathrm{FID}}(\omega\pm0,\tau)
-
F_{\mathrm{FID}}(\omega\pm J/2,\tau)).
\end{equation}

\paragraph*{(3) Interaction noise.}
Interaction noise in this case is described by the system operator
$\hat{A}=
\sigma_{1}^{y}\sigma_{2}^{y}-\,\sigma_{1}^{x}\sigma_{2}^{z}+\,
\sigma_{1}^{z}\sigma_{2}^{x}
$.
This operator induces the following processes:
(i) for $\hat{A}_{\alpha}=\hat{A}_{\beta}=\sigma_{1}^{y}\sigma_{2}^{y}$,
pure dephasing of $\ket{T_{0}}$ and $\ket{S}$ at zero frequency, and transitions between $\ket{T_{+1}}$ and $\ket{T_{-1}}$ at frequency $2\Delta$, with unit weight;
(ii) for $\hat{A}_{\alpha}=\hat{A}_{\beta}=-\sigma_{1}^{x}\sigma_{2}^{z}$ or $\sigma_{1}^{z}\sigma_{2}^{x}$, transitions between $\ket{T_{\pm1}}$ and $\ket{S,T_{0}}$  with weight $\frac{1}{2}$, and frequencies $\Delta\pm J/4$.
All other combinations of system operators yield zero net weight. Collecting all nonvanishing contributions, the effective filter function for interaction noise to leading order in $J/\Delta$ is
\begin{equation}
F_{S_{JJ}}(\omega,\tau)
=
F_{\mathrm{FID}}(\omega\pm0,\tau)+F_{\mathrm{FID}}(\omega\pm 2\Delta,\tau)+4(\frac{1}{2}F_{\mathrm{FID}}(\omega\pm (\Delta+\frac{J}{4}),\tau)+\frac{1}{2}F_{\mathrm{FID}}(\omega\pm (\Delta-\frac{J}{4}),\tau)).
\end{equation}

\paragraph*{(4) Cross correlations.}
Due to the small mixing between $\ket{T_{\pm1}}$ and $\ket{T_{0},S}$ states, the cross-correlated noise has nonzero but small contributions to the average gate infidelity at order $J/\Delta$. We neglect it in the analytic filter functions above; it is retained in the numerical results.

\section{Infidelity under Lorentzian noise}
We derive the analytical expression for the infidelity under Lorentzian noise and then present the resulting gate comparison. As in the $1/f$ case, we take all noise correlations to share the same spectrum and amplitude, $S_{\alpha\beta}(\omega)=S(\omega)$.

The Lorentzian noise spectrum is given by
\begin{equation}
    S(\omega)=\frac{1}{T_{2}^{*}}\frac{1}{1+\omega^{2}\tau_{c}^{2}},
\end{equation}
where $T_{2}^{*}$ is a reference time scale that sets the overall noise strength and $\tau_{c}$ is the noise correlation time.

The evaluation of the infidelity requires the calculation of the following integral:
\begin{align}
    Y(\omega_{nm},\tau,\tau_{c})=
    \int_{-\infty}^{\infty}\frac{d\omega}{2\pi}
    \frac{1}{1 + \omega^2 \tau_c^2}
    \frac{ \sin^2\!\left[\frac{(\omega-\omega_{nm}) \tau}{2}\right]}
    {\left(\frac{\omega -\omega_{nm}}{2}\right)^2}.
    \label{Ydef}
\end{align}
Introducing the dimensionless variables
\begin{equation}
    x=\frac{\omega\tau}{2}, \qquad
    a=\frac{\tau}{2\tau_{c}}, \qquad
    b=\frac{\omega_{nm}\tau}{2},
\end{equation}
the integral~\eqref{Ydef} can be rewritten as
\begin{align}
    Y(\omega_{nm},\tau,\tau_{c})
    =\frac{1}{2\pi}
    \int_{-\infty}^{\infty}dx
    \frac{2\tau a^{2}}{a^{2}+x^{2}}
    \frac{\sin^{2}(x-b)}{(x-b)^{2}}.
    \label{Ydimensionless}
\end{align}
This integral can be evaluated analytically, yielding
\begin{align}
   Y(\omega_{nm},\tau,\tau_{c})
    =
    \frac{
       a \tau e^{-2(a + i b)}
        \left[
            (a - i b)^2
            + 2\left( -a^2 + 2a^{3} + b^2 + 2ab^2 \right) e^{2a + 2i b}
            + (a + i b)^2 e^{4i b}
        \right]
    }{
        4  (a^2+b^2)^2
    }.
    \label{Yresult}
\end{align}
Using the above result, the average gate infidelity under Lorentzian noise can be written as
\begin{align}
    \overline{\mathcal{I}}
    =\frac{1}{5T_{2}^{*}}
\sum_{\alpha\beta}\sum_{nm}\hat{A}_{\alpha}^{nm}\hat{A}_{\beta}^{mn} 
    Y(\omega_{nm},\tau,\tau_{c}).
\end{align}

\begin{figure*}[htbp]
\centering
\includegraphics[width=\textwidth]{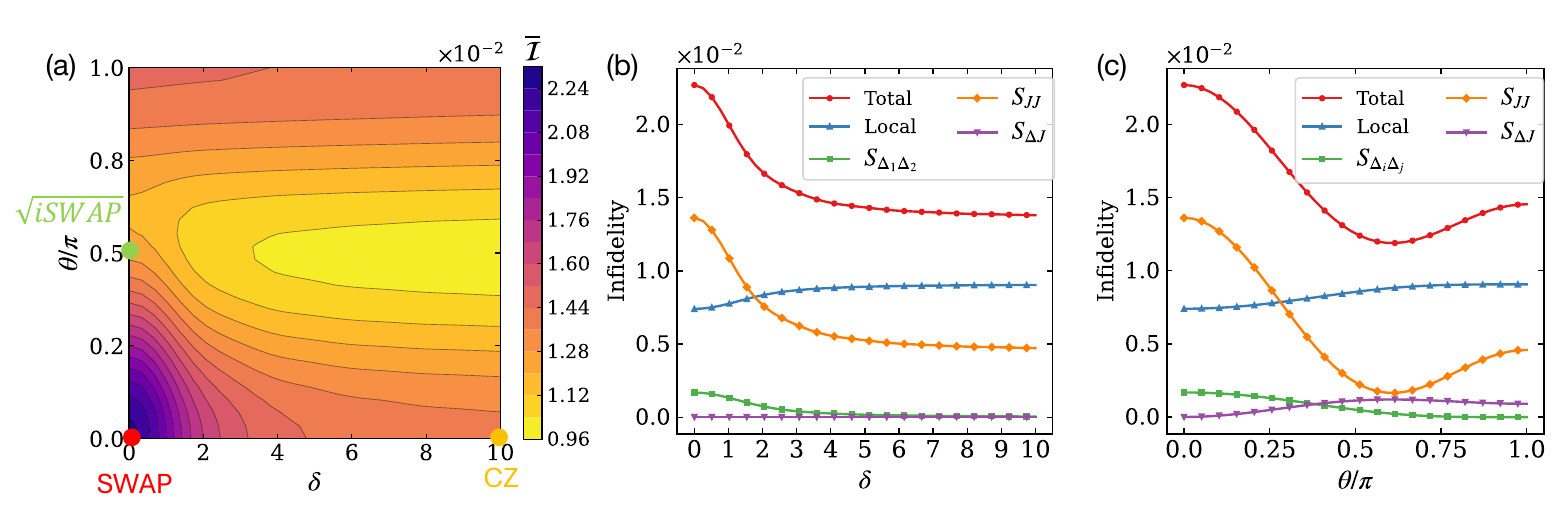}
\caption{Infidelity of two-qubit gates under Lorentzian noise. (a)\ Total infidelity across the $(\delta,\theta)$ plane. (b) Line cuts of the infidelity at $\theta = 0$, showing dependence of infidelity induced by each type of noise on detuning. (c) Line cuts at $\delta = 0$, showing infidelity induced by each type of noise as a function of rotation angle $\theta$. Parameters used in the plots: $J=1$, $\Delta_{1}=10$, $\Delta_{2}=\Delta_{1}+\delta\cross J$, $\tau=\pi/J$, $T_{2}^{*}=100\tau$, and $\tau_c=\tau/2$.}
\label{fig:S2}
\end{figure*}

For the equal-amplitude, common-spectrum covariance $S_{\alpha\beta}(\omega)=S(\omega)$ used here, Fig. \ref{fig:S2} shows the same gate ordering under Lorentzian noise as under  $1/f$ noise. This comparison extends the result to a second low-frequency spectral shape for this covariance. 

\bibliography{bib}